\documentclass[12pt]{article}
\usepackage{amsmath}
\usepackage{graphicx,psfrag,epsf}
\usepackage{enumerate}
\usepackage{psfrag,epsf}
\usepackage{url} 
\usepackage{amsmath,amssymb,amsfonts,amsthm,mathtools}
\usepackage{bm,bbm}
\usepackage{color}
\usepackage{subfigure}
\usepackage{algorithm,algpseudocode}
\usepackage{scalefnt}
\usepackage{multirow}

\usepackage{natbib}
\usepackage{dsfont}
\usepackage{setspace}
\usepackage{authblk}
\usepackage{adjustbox}

\usepackage[left=1in,top=1.2in,right=0.7in]{geometry}

\theoremstyle{definition}

\numberwithin{equation}{section} 

\makeatletter
\def\@seccntformat#1{\@ifundefined{#1@cntformat}%
	{\csname the#1\endcsname\quad}
	{\csname #1@cntformat\endcsname}
}
\makeatother

\newif\ifShowComments
\ShowCommentstrue
\def\strutdepth{\dp\strutbox}
\def\druk#1{\strut\vadjust{\kern-\strutdepth
        {\vtop to \strutdepth{%
                \baselineskip\strutdepth\vss
                        \llap{\hbox{#1}\quad}\null}}}}

\title{\bf  Parametric quantile regression for income data}
\author{
\normalsize
\textbf{Helton Saulo}$^{1}$, \textbf{Roberto Vila}$^{1}$, \textbf{Giovanna V. Borges}$^{1}$ and \textbf{Marcelo Bourguignon}$^{2}$\\
{\small $^{1}$Department of Statistics, University of Brasilia, Brasilia, Brazil}\\
{\small $^{2}$Department of Statistics, Federal University of Rio Grande do Norte, Natal, Brazil}\\
}

\begin{document}
\maketitle

\begin{abstract}

Univariate normal regression models are statistical tools widely applied in many areas of economics. Nevertheless, income data have asymmetric behavior and are best modeled by non-normal distributions. The modeling of income plays an important role in determining workers' earnings, as well as being an important research topic in labor economics. Thus, the objective of this work is to propose parametric quantile regression models based on two important asymmetric income distributions, namely, Dagum and Singh-Maddala distributions. The proposed quantile models are based on reparameterizations of the original distributions by inserting a quantile parameter.  We present the reparameterizations, some properties of the distributions, and the quantile regression models with their inferential aspects. We proceed with Monte Carlo simulation studies, considering the maximum likelihood estimation performance evaluation and an analysis of the empirical distribution of two residuals. The Monte Carlo results show that both models meet the expected outcomes. We apply the proposed quantile regression models to a household income data set provided by the National Institute of Statistics of Chile. We showed that both proposed models had a good performance both in terms of model fitting.
Thus, we conclude that results were favorable to the use of Singh-Maddala and Dagum quantile regression models for positive asymmetric data, such as income data.
\end{abstract}


\onehalfspacing

\section{Introduction}
\noindent

Income modeling plays an important role in determining workers' earnings, as well as being an important research topic in labor economics. In general, income data are modeled using mean-based regression models based on the normality assumption. Nevertheless, income is often unequally distributed, hence why this type of data usually has an asymmetric behavior and then the mean is not an appropriate central tendency measure. Therefore, quantile regression models are usually more useful in this context; see \cite{galarza2020}, \cite{sanchez2021a} and \cite{sauloetal:21}.

Quantile regression models are robust alternatives to traditional mean-based models. That is because instead of focusing on the conditional mean, these models are based on the conditional quantile, such as median; see \cite{koenker:05}. The quantile approach has the advantage of providing flexibility in modeling, as it allows considering the effects of explanatory variables throughout the spectrum of the dependent variable, thus also including the effect on the median, which is a measure of central tendency better than the mean in the asymmetric context.

Income modelling begins with \cite{pareto1897} propositions, establishing a law on how income distribution works. Later on, this suggested a distribution -- known as Pareto distribution -- and it has set a reference for other distributions, such as log-normal and gamma, to show their potential as for describing income distribution; see \cite{shirras1935, reed2003}. Even though the Pareto, log-normal and gamma are the most frequently distributions applied to income data because of their abilities to describe this type of data, they have limitations. On the one hand, the Pareto model is appropriate to describe only the upper tail of the distribution. On the other hand, the log-normal and gamma distributions perform poor in describing both the upper and lower tails of the actual distributions. Income distributions such as Dagum and Singh-Maddala have outperformed the Pareto, log-normal and gamma distributions in terms of model fitting; see \cite{cramer1971, dagum2008}.

Originally proposed by \cite{dagum1973,dagum1975}, the Dagum distribution has flexibility to deal with distribution changes, nil and negative income, income range with non-predetermined positive minimum income start, and strictly decreasing and unimodal density functions. This distribution also shows good goodness of fit to income data and obeys a weak version of the Pareto law, i.e. it asymptotically approaches the Pareto distribution. The Dagum model accommodates both heavy tails and an interior mode, characteristics commonly found in income data, and not found singly in well-known distributions -- such as log-normal and Pareto; see \cite{kramer2002,dagum2008,kleiber2008}.

The Singh-Maddala distribution was derived from the concept of hazard rate, an approach widely used in the reliability literature; see \cite{singh1976}. This model also obeys the weak Pareto law, and one of its advantages is to be more flexible than other income distributions. The Dagum and Singh-Maddala distributions are special cases of the generalized beta distribution of the second kind (GB2); for more details on these models, one may refer to the works by \cite{kleiber1996,kleiber2008,kumar2017,hajargasht2012}.

This work explores a parametric quantile regression approach for the Dagum and Singh-Maddala distributions. We first introduce reparameterizations of the Dagum and Singh-Maddala model by inserting quantile parameters, and then develop the new regression models. We then demonstrate that the proposed models outperform the recently proposed Birnbaum-Saunders quantile regression model \citep{sanchez2021b} in terms of model fitting.

The rest of this paper proceeds as follows. In Section \ref{sec:2}, we describe the usual Dagum and Singh-Maddala distributions and propose reparameterizations of these distributions in terms of a quantile parameter. In this section, we also present some properties including mode, real and truncated moments. In Section \ref{sec:3}, we introduce the quantile regression models and also describe the parameter estimation by the maximum likelihood (ML) method. In Section~\ref{sec:4}, we carry out a Monte Carlo simulation study to evaluate the performance of the estimators and generalized Cox-Snell (GCS) and random quantile (RQ) residuals. In Section \ref{sec:5}, we apply the Dagum and Singh-Maddala quantile regression models to a household income data set provided by the National Institute of Statistics of Chile, and finally in Section \ref{sec:6}, we provide some concluding remarks.

\section{Classical and quantile-based income distributions}\label{sec:2}
\noindent

In this section, we describe the classical Singh-Maddala and Dagum distributions along with the proposed quantile-based reparameterizations of these distributions, which will be useful subsequently for developing the parametric quantile regression models. We also present some properties for each model, including mode, real and truncated moments.


\subsection{Classical Singh-Maddala distribution}
\label{Singh-Maddala distribution}
\noindent

If a random variable $Y$ follows a Singh-Maddala distribution with shape parameters $a,q>0$ and scale parameter $b>0$, denoted by $Y\sim \text{SM}(a, b, q)$, then the corresponding probability density function (PDF) and cumulative distribution function (CDF) are given by
\begin{equation}\label{sm:01}
f_{\rm SM}(y; a, b, q)
=
\frac{a\,q (y/b)^{a-1}}{b[1+(y/b)^a]^{1+q}}, \quad {y}>0,
\end{equation}
and
\begin{equation}\label{sm:02}
F_{\rm SM}(y; a, b, q)
=
1 - \left[1 + (y/b)^a\right]^{-q}, \quad {y}>0,
\end{equation}
respectively. The Singh-Maddala distribution includes as special cases the Lomax distribution when $a = 1$, and the log-logistic distribution when $q = 1$.  If $Y$ follows a Singh-Maddala distribution, then $1/Y$ follows a Dagum distribution, and vice-versa.

The $\tau$-th quantile of $Y\sim \text{SM}(a, b, q)$ is obtained by inverting Equation \eqref{sm:02}, which yields
\begin{equation}\label{sm:03}
q(\tau; a, b, q) = b{ c_q}^{1/a}
\quad
{
\text{where} \, c_q=(1-\tau)^{-1/q} - 1 \ \text{for} \
}
0 < \tau < 1.
\end{equation}

\subsection{Quantile-based Singh-Maddala distribution}
\label{Quantile-Singh-Maddala-distribution}
\noindent

From the quantile function \eqref{sm:03}, we find that the most parsimonious way of conducting the reparametrization is using the scale parameter $b$, where we can then write
\begin{equation*}\label{sm:04}
b
=
{\gamma}\,{{ c_q}^{-1/a}},
\end{equation*}
where $\gamma = q(\tau; a, b, q) > 0$. Then, the quantile-based Singh-Maddala PDF is given by
\begin{equation*}\label{sm:05}
f_{\rm QSM}(y; a, \gamma, q)
=
\frac{aq{ c_q} (y/\gamma)^{a-1}}{\gamma[1+{ c_q}(y/\gamma)^a]^{1+q}}, \quad {y}>0,
\end{equation*}
with notation $Y\sim {\rm QSM}(a,\gamma,q)$.

If $Y\sim {\rm QSM}(a,\gamma,q)$, then the following properties hold:
\begin{itemize}
	\item[(QSM1)]
	Mode \citep{kd:17,kpw:2019}:
	\begin{align*}
	\biggl({a-1\over a\,q+1}\biggr)^{1/ a}, \quad a>1, \ \text{else} \ 0.
	\end{align*}
	\item[(QSM2)]
	Real moments \citep{kd:17,kpw:2019}:
	\begin{align*}
	\mathbb{E}(Y^r)
	=
	{q\gamma^r\over c_q^{r/a}} \,
	{\rm B}\biggl(1+{r\over a},q-{r\over a}\biggr),
	\quad -a<{r}<aq,
	\end{align*}
	where  ${\rm B}(x,y)$ denotes the beta function.
	\item[(QSM3)]
	Truncated moments \citep{kd:17}:
	\begin{align*}
	\mathbb{E}(Y^r\mathds{1}_{\{Y>x\}})
	=
	{aq\gamma^{r} ({\gamma/ x})^{aq-r}\over (
	aq-{r})c_q^{q}}
	\, _2F_1
	\biggl(
	1+q, q-{r\over a};q-{r\over a}+1;-{(\gamma/x)^a\over  c_q}
	\biggr),
	\quad aq>{r},
	\end{align*}
	where $\, _2F_1(a,b;c;x)$ denotes the Gauss hypergeometric function.


\end{itemize}

\subsection{Classical Dagum distribution}\label{Dagum-distribution}
\noindent

The PDF and CDF of a random variable $Y$ following a classical Dagum distribution with shape parameters $a,p>0$ and scale parameter $b>0$, denoted by $Y\sim \text{DA}(a, b, p)$, are given by
\begin{equation*}\label{da:01}
f_{\rm DA}(y; a, b, p)
=
\frac{ap(y/b)^{ap-1}}{b[1+(y/b)^a]^{1+p}}, \quad {y}>0,
\end{equation*}
and
\begin{equation}\label{da:02}
F_{\rm DA}(y; a, b, p)
=
\left[1 + (y/b)^{-a}\right]^{-p}, \quad {y}>0,
\end{equation}
respectively. It is simple to observe that
$
f_{\rm DA}(y; a, b, p)
=
(y/b)^{a(p-1)}
f_{\rm SM}(y; a, b, p)
$
and that when $p=1$ both densities coincide with the log-logistic distribution.

The $\tau$-th quantile of $Y\sim \text{DA}(a, b, p)$ is given by
\begin{equation*}\label{da:03}
q(\tau; a, b, p) = b\,{ e_p}^{-1/a}
\quad
{
	\text{where} \,\, e_p=\tau^{-1/p} - 1 \ \text{for} \
}
0 < \tau < 1.
\end{equation*}

\subsection{Quantile-based Dagum distribution}\label{Quantile-Dagum-distribution}
\noindent

By observing the three parameters of the classical Dagum distribution, the isolation of the scale according to the quantile would produce the simplest form of the new quantile-based Dagum distribution; it is represented as follows:
\begin{equation*}\label{da:04}
b = \gamma{ e_p}^{1/a},
\end{equation*}
where $\gamma = q(\tau; a, b, p) > 0$. Then, the quantile-based Dagum PDF can be written as
\begin{equation*}\label{da:05}
f_{\rm QDA}(y; a, \gamma, p)
=
\frac{a\,p (y/\gamma)^{ap-1}}{
\gamma
{ e_p}^{p}
[1+{ e_p^{-1}}{(y/\gamma)^a}]^{1+p}},
\quad {y}>0.
\end{equation*}

If $Y\sim {\rm QDA}(a,\gamma,p)$, then the following properties hold:
	\begin{itemize}
	\item[(QDA1)]
	Mode \citep{kpw:2019}:
	\begin{align*}
	\gamma \, e_p^{1/a}
	\biggl({a\,p-1\over p+1}\biggr)^{1/ p} , \quad ap>1, \ \text{else} \ 0.
	\end{align*}
	\item[(QDA2)]
	Real moments \citep{kpw:2019}:
	\begin{align*}
		\mathbb{E}(Y^r)
		=
		\gamma^r e_p^{r/a}
		{\rm B}\biggl(a+{r\over p}, 1-{r\over p}\biggr),
		\quad -a\,p<r<p.
	\end{align*}
	\item[(QDA3)]
	Truncated moments (see Appendix \ref{Apendix-A}):
	\begin{align*}
	\mathbb{E}(Y^r\mathds{1}_{\{Y>x\}})
	=
{p\gamma^r
	(\gamma/x)^{a(1-ar)} \over   (1-ar)e_p^{-p}}
\, _2F_1\biggl(1+p,1-ar;2-ar;-{(\gamma/x)^a\over e_p^{-1}}\biggr), \quad ar<1.
	\end{align*}
\end{itemize}



\subsection{Summary table and density plots}
\noindent

\begin{figure}[!ht]
\vspace{-0.25cm}
\centering
\subfigure[$a$ and $q$ fixed]{\includegraphics[height=5.2cm,width=5.2cm]{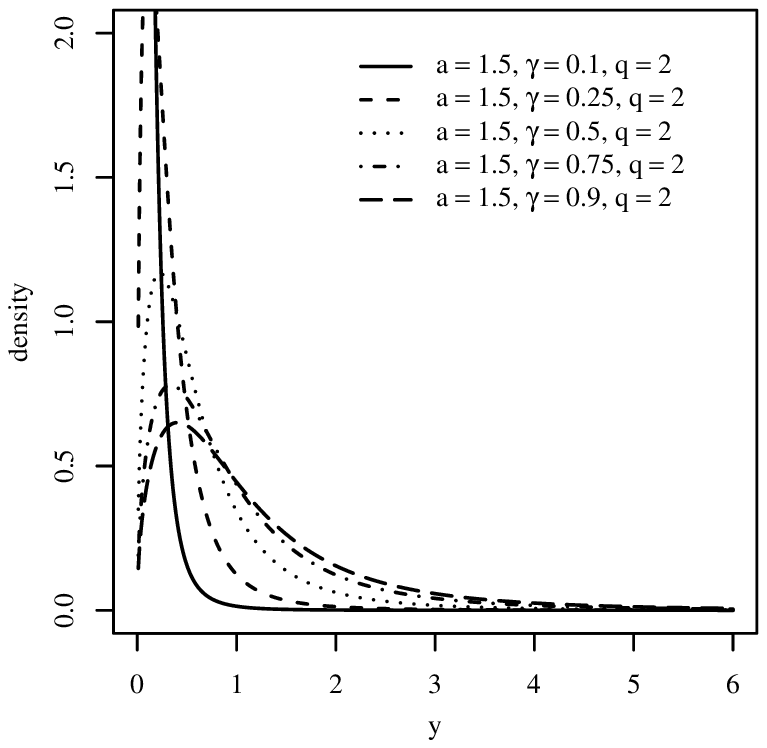}}
\subfigure[$\gamma$ and $q$ fixed]{\includegraphics[height=5.2cm,width=5.2cm]{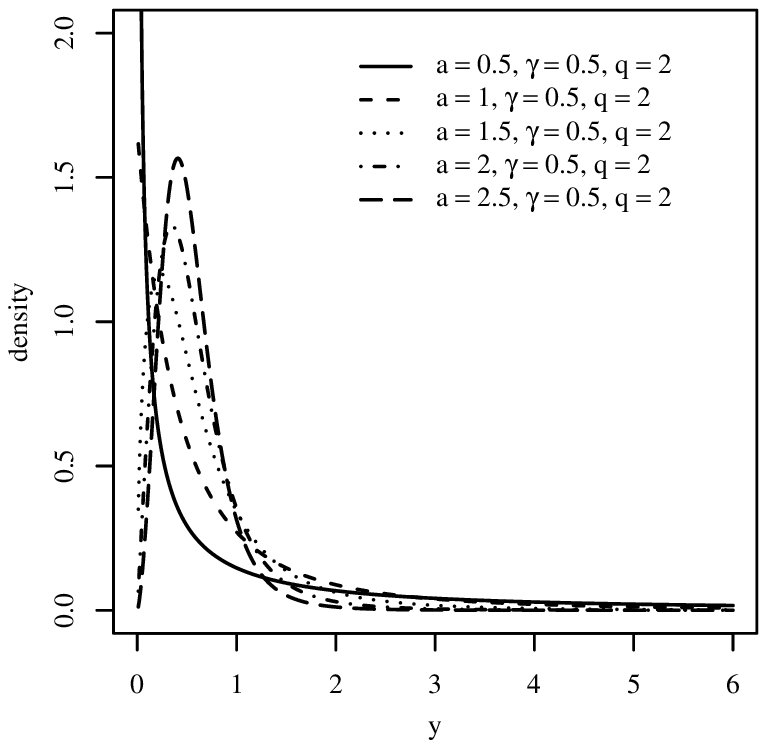}}
\subfigure[$\gamma$ and $a$ fixed]{\includegraphics[height=5.2cm,width=5.2cm]{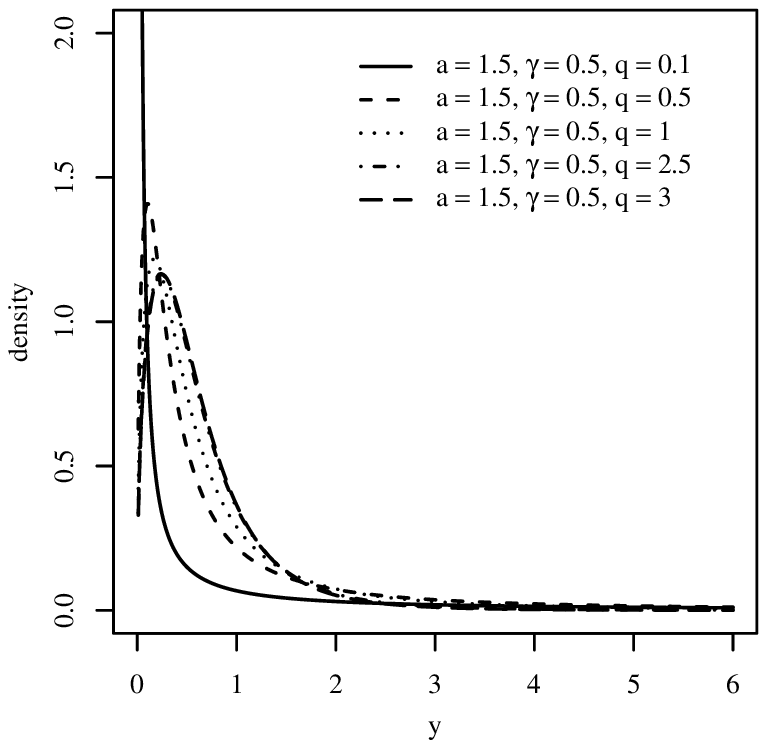}}\\
\vspace{-0.2cm}
\caption{Quantile-based Singh-Maddala PDFs for some choices of parameters.}
\label{figpdf:1}
\end{figure}

\begin{figure}[!ht]
\vspace{-0.25cm}
\centering
\subfigure[$a$ and $p$ fixed]{\includegraphics[height=5.2cm,width=5.2cm]{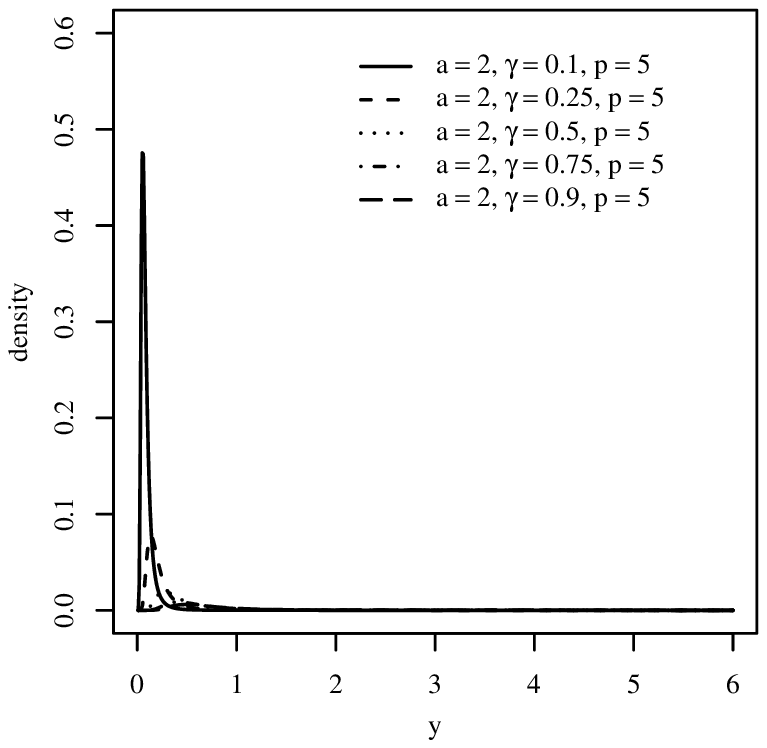}}
\subfigure[$\gamma$ and $p$ fixed]{\includegraphics[height=5.2cm,width=5.2cm]{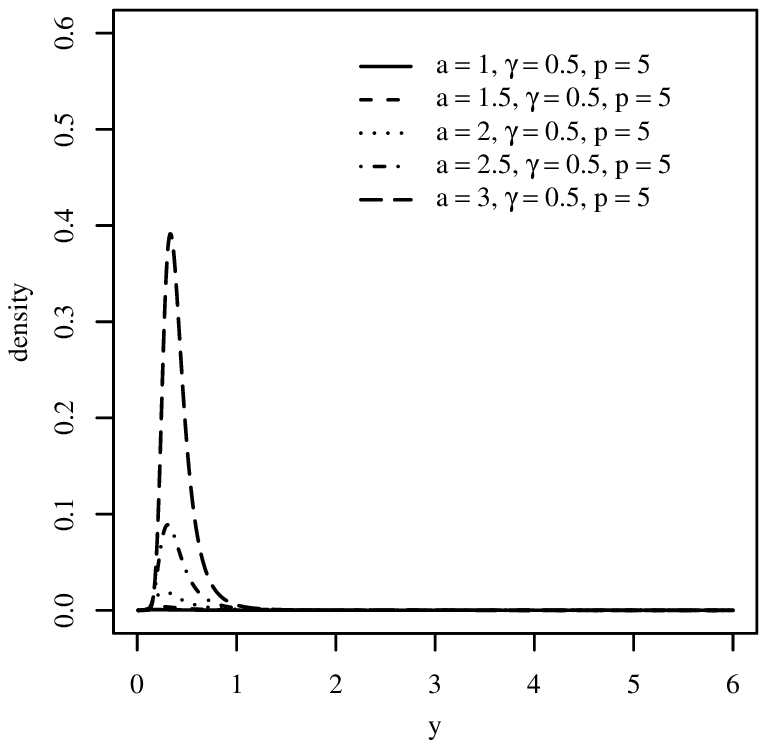}}
\subfigure[$\gamma$ and $a$ fixed]{\includegraphics[height=5.2cm,width=5.2cm]{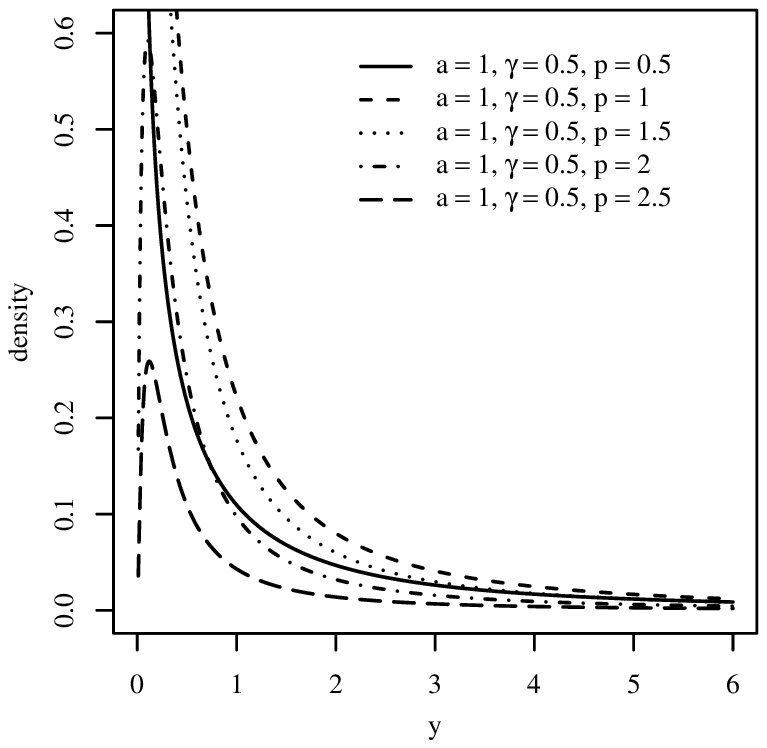}}\\
\vspace{-0.2cm}
\caption{Quantile-based Dagum PDFs for some choices of parameters.}
\label{figpdf:2}
\end{figure}

Table \ref{tab:gfuncions} presents the Singh-Maddala and Dagum distributions in their original and quantile-based versions. Figures \ref{figpdf:1} and \ref{figpdf:2} display different shapes of the quantile-based income distributions for different combinations of parameters, considering scenarios where $a$, $p$, $q$ and $\gamma$ are fixed. For Singh-Maddala, we can see that $a$ influences the kurtosis and skewness, while $q$ changes the kurtosis, as it decreases when $q$ increases. For Dagum, we see a similar pattern for $a$, changing both kurtosis and skewness, while $p$ affects the kurtosis.

\begin{table}[!ht]
\centering
\normalsize
\caption{Income distributions for the original and quantile parameterizations.}\label{tab:gfuncions}
\vspace{0.15cm}
	\adjustbox{max height=\dimexpr\textheight-0.5cm\relax,
		max width=\textwidth}{
\begin{tabular}{llllll}
\hline Distribution         &    Classical density     &    $\gamma$: $\tau$-th quantile  &   Substitution  & Quantile-based density                                                   \\ 
\hline
Singh-Maddala
&
$\frac{aq (y/b)^{a-1}}{b[1+(y/b)^a]^{1+q}}$
&
$\gamma=b { c_q}^{1/a}$
&
$b = \frac{\gamma}{{ c_q}^{1/a}}$
&
$\frac{a\,q { c_q} (y/\gamma)^{a-1}}{\gamma [1+{ c_q}(y/\gamma)^a]^{1+q}}$ \\
&                            &                  &          &\\
Dagum
&
$\frac{ap (y/b)^{ap-1}}{b[1+(y/b)^a]^{1+p}}$
&
$\gamma=b \,{ e_p}^{-1/a}$
&
$b = \gamma \,{ e_p}^{1/a}$
&
$\frac{a\,p (y/\gamma)^{ap-1}}{\gamma
{ e_p}^{p}
[1+
{ e_p^{-1}}{(y/\gamma)^a}
]^{1+p}}$
\\
\hline
\end{tabular}
}
\end{table}

\section{Income quantile regression models}\label{sec:3}
\noindent

Let $ Y_1, \ldots, Y_n $ be independent random
variables such that each $ Y_i $, for $ i = 1,\ldots,n $, has PDF given by some reparameterized income distribution defined in Table \ref{tab:gfuncions}, for a fixed (known) probability $
\tau \in (0, 1)$ associated with the quantile of interest. Then, in the formulation of the Singh-Maddala and Dagum quantile regression models, the parameter $\gamma$ of $Y_i$ assumes the following functional relation:
\begin{equation}\label{cs1}
g(\gamma_i) = \mathbf{x}^\top_{i}\bm{\beta}(\tau),
\end{equation}
where $\bm{\beta}(\tau) = (\beta_{0}(\tau), \ldots, \beta_{k}(\tau))^\top$ is the vector of the unknown regression coefficients, which are assumed to be functionally independent;
$\bm{\beta}(\tau) \in \mathbb{R}^{(k+1)}$, with $k +1  < n$;
and $\mathbf{x}_{i} = (x_{i1}, \ldots, x_{il})^\top$ is the observations of the $l$ known regressors, for $i = 1, \ldots, n$. In addition, we assume that the covariate matrices $\mathbf{X} = (\mathbf{x}_1, \ldots, \mathbf{x}_n)^\top$ has rank $l$. The link function $g: \mathbb{R}^+ \rightarrow \mathbb{R}$ in \eqref{cs1} must be strictly monotone, positive, and at least twice differentiable, with $g^{-1}(\cdot)$ being the inverse function of $g(\cdot)$. Here, we chose to work with $\log$ as link since it is widely used and more flexible when it comes to simulation studies.

Consider a sample of size $n$, $Y_1,\ldots,Y_n$ say, such that $Y_i \sim \text{QSM}(a, \gamma_i, q)$. Then, the corresponding likelihood function for $\bm{\theta} = (\bm{\beta}(\tau)^{\top},a,q)^{\top}$, is
\begin{equation}\label{sm_ll1}
    L(\bm{\theta}) = \prod_{i=1}^n \frac{aq { c_q} (y/\gamma_i)^{a-1}}{\gamma_i [1+{ c_q}(y/\gamma_i)^a]^{1+q}} \ ,
\end{equation}
where $\gamma_i$ is as in \eqref{cs1}. By applying the logarithm in \eqref{sm_ll1}, we obtain the log-likelihood function
\begin{equation}\label{sm_logll1}
    \ell(\bm{\theta}) = \sum_{i=1}^n \left\{ \big[(a-1)\log(aq { c_q}(y/\gamma_i))\big] - \big[\log(\gamma_i) + (1 + q) \log(1 + { c_q}(y/\gamma_i)^a)\big] \right\}.
\end{equation}

Now, consider a sample of size $n$, $Y_1,\ldots,Y_n$ say, such that $Y_i \sim \text{QDA}(a, \gamma_i, p)$. Then, the corresponding likelihood function for $\bm{\theta} = (\bm{\beta}(\tau)^{\top},a,p)^{\top}$, is
\begin{equation}\label{dg_ll1}
    L(\bm{\theta}) = \prod_{i=1}^n \frac{ap (y/\gamma_i)^{ap-1}}{\gamma_i
{ e_p}^{p}
[1+
{ e_p^{-1}}{(y/\gamma_i)^a}
]^{1+p}}
\end{equation}
where $\gamma_i$ is as in \eqref{cs1}. By applying the logarithm in \eqref{dg_ll1}, we obtain the log-likelihood function \begin{equation} \label{dg_ll2}
    \ell(\bm{\theta}) = \sum_{i=1}^n \left\{ \big[(ap-1)\log(ap(y/\gamma_i))\big] - \big[p\log({ e_p}\gamma_i) + (1 + p) \log(1 + { e_p^{-1}}(y/\gamma_i)^a)\big] \right\}.
\end{equation}

To obtain the ML estimate of $\bm{\theta}$, it is necessary to maximize the log-likelihood functions in \eqref{sm_logll1} and \eqref{dg_ll2}. Therefore, we need to differentiate the log-likelihood functions to find the score vector $\dot{\ell}(\bm{\theta})$ and then equate it to zero, providing the likelihood equations. They are solved using the Broyden-Fletcher-Goldfarb-Shanno (BFGS) quasi-Newton method, see \cite{mittelhammer2000}. The method is implemented and applied using the \texttt{R} software. Under some regularity conditions \citep{Cox1979} and when $n$ is large,
the asymptotic distribution of the ML estimator $\widehat{\bm{\theta}} = (\bm{\beta}(\tau)^{\top},a,q)^{\top}$ (QSM) or
$\widehat{\bm{\theta}} = (\bm{\beta}(\tau)^{\top},a,p)^{\top}$ (QDA) follows asymptotically a multivariate normal distribution
$$\widehat{\bm{\theta}} \,\dot{\sim}\, \textrm{N}_{k+3}(\bm{\theta}, {\bm{\Sigma}}^{-1}(\bm{\theta})),$$
where $\,\dot{\sim}\,$ means `approximately distributed' and ${\bm{\Sigma}}(\bm{\theta})$ is the expected Fisher information matrix, which is given by
$$\bm{\Sigma}(\bm{\theta})= \mathbb{E}\left[- \ {\partial \ell \left(\bm{\theta}\right)\over \partial \bm{\theta} \; \partial \bm{\theta}^\top} \right].$$
A consistent estimator of $\bm{\Sigma}(\bm{\theta})$ is the estimated observed Fisher information matrix, given by
$$\mathbf{K}(\widehat{\bm{\theta}})=- \ {\partial \ell \left(\bm{\theta}\right)\over \partial \bm{\theta} \;\partial \bm{\theta}^\top} \Big{|}_{\bm{\theta} = \widehat{\bm{\theta}}}.$$
Then, we can approximate $\bm{\Sigma}(\bm{\theta})$ by $\mathbf{K}(\widehat{\bm{\theta}})$.

Departures from regression models assumptions and goodness of fit are assessed by means of a residual analysis. Particularly, we use the generalized Cox-Snell (GCS) and randomized quantile (RQ) residuals:
\begin{equation*}
    \hat{r}_i^{\text{GCS}} = -\log(1-F_Y(y_i;\widehat{\bm{\theta}}))\,\, \text{and} \,\,
     \hat{r}_i^{\text{RQ}} = \Phi^{-1}(F_Y(y_i;\widehat{\bm{\theta}})), \ \  i = 1,\dots,n,
\end{equation*}
where $F_Y$ is quantile-based Singh-Maddala or Dagum CDF, and $\widehat{\bm{\theta}}$ is the ML estimate of ${\bm{\theta}}$.  If the model is correctly specified, the GCS is asymptotically standard exponential distributed, while the RQ is asymptotically standard normal distributed. With both residuals, graphical techquines, such as quantile-quantile (QQ) plots with simulated envelope, can be used to assess distributions assumptions.


\section{Monte Carlo simulation}\label{sec:4}
\noindent

In this section, we present Monte Carlo simulation studies for each reparameterized quantile model, considering different scenarios of parameters and sample sizes. The first part of the study consists in evaluating ML estimation performance, while the second evaluates the empirical distribution of the GCS and RQ residuals. Both studies consider simulated data generated from each one of the Singh-Maddala and Dagum quantile regression models according to
\begin{equation*}
\gamma_i = \exp\left( {\beta}_0(\tau) + {\beta}_1(\tau)x_{1i} + {\beta}_2(\tau)x_{2i} \right).
\end{equation*}

The
Monte Carlo simulation experiments were performed using the R environment; see
http://www.r-project.org.

\subsection{ML estimation}\label{ml_simulation_res}
\noindent

The simulation scenario considers the following settings: sample sizes $n \in {50,100,150,250,600}$, vector of betas $\bm{\beta}(\tau) = (1, 0.5, 1.5)^{\top}$, quantiles $\tau \in \{0.10, 0.25, 0.50, 0.75, 0.90\}$, $(a,q)=(5, 1)$ (Singh-Maddala), and $(a,p)=(1, 0.5)$ (Dagum), with 500 Monte Carlo replications for each sample size. Covariate values $x_{1i},x_{2i}$ are obtained from a uniform distribution in the interval (0,1). To study the ML estimators, we use compute the relative bias (RB), root mean square error (RMSE) and the coverage probability (CP). We expect that, as sample size increases, the RB and RMSE reduces, and the CP approaches the 95\% nominal level. The estimates of RB, RMSE and CP are computed from the Monte Carlo replicas as:
\begin{eqnarray*}
 \widehat{\textrm{RB}}(\widehat{\theta}) &=& \left| \frac{\frac{1}{m} \sum_{i = 1}^{m} \widehat{\theta}^{(i)} - \theta} {\theta} \right|,\\
\widehat{\mathrm{RMSE}}(\widehat{\theta}) &=& {\sqrt{\frac{1}{m} \sum_{i = 1}^{m} (\widehat{\theta}^{(i)} - \theta)^2}}, \\
\widehat{\mathrm{CP}}(\widehat{\theta}) &=& \frac{1}{m} \sum_{i = 1}^{m} \mathcal{I}(\theta \in [L^{(i)}_{\widehat{\theta}},U^{(i)}_{\widehat{\theta}}]),
\end{eqnarray*}
where $\theta$ and $\widehat{\theta}^{(i)}$ are the true parameter value and its respective $i$-th ML estimate, $m$ is the number of Monte Carlo replicas,  $\mathcal{I}$ is an indicator function taking the value 1 if $\theta\in [L^{(i)}_{\widehat{\theta}},U^{(i)}_{\widehat{\theta}}]$, and 0 otherwise, where $L^{(i)}_{\widehat{\theta}}$ and $U^{(i)}_{\widehat{\theta}}$ are the $i$-th upper and lower limit estimates of the 95\% confidence interval.

\begin{figure}[!ht]
\vspace{-0.25cm}
\centering
{\includegraphics[height=3.5cm,width=3.5cm]{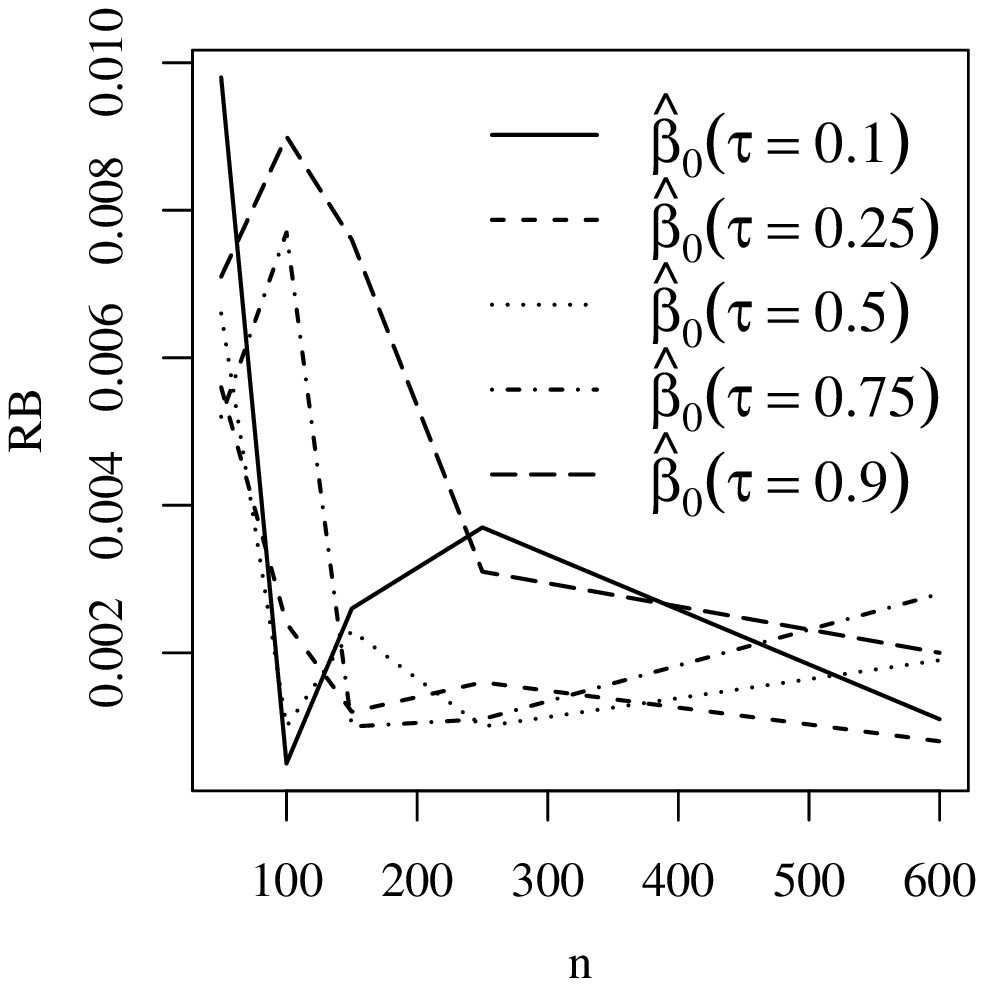}}\hspace{-0.25cm}
{\includegraphics[height=3.5cm,width=3.5cm]{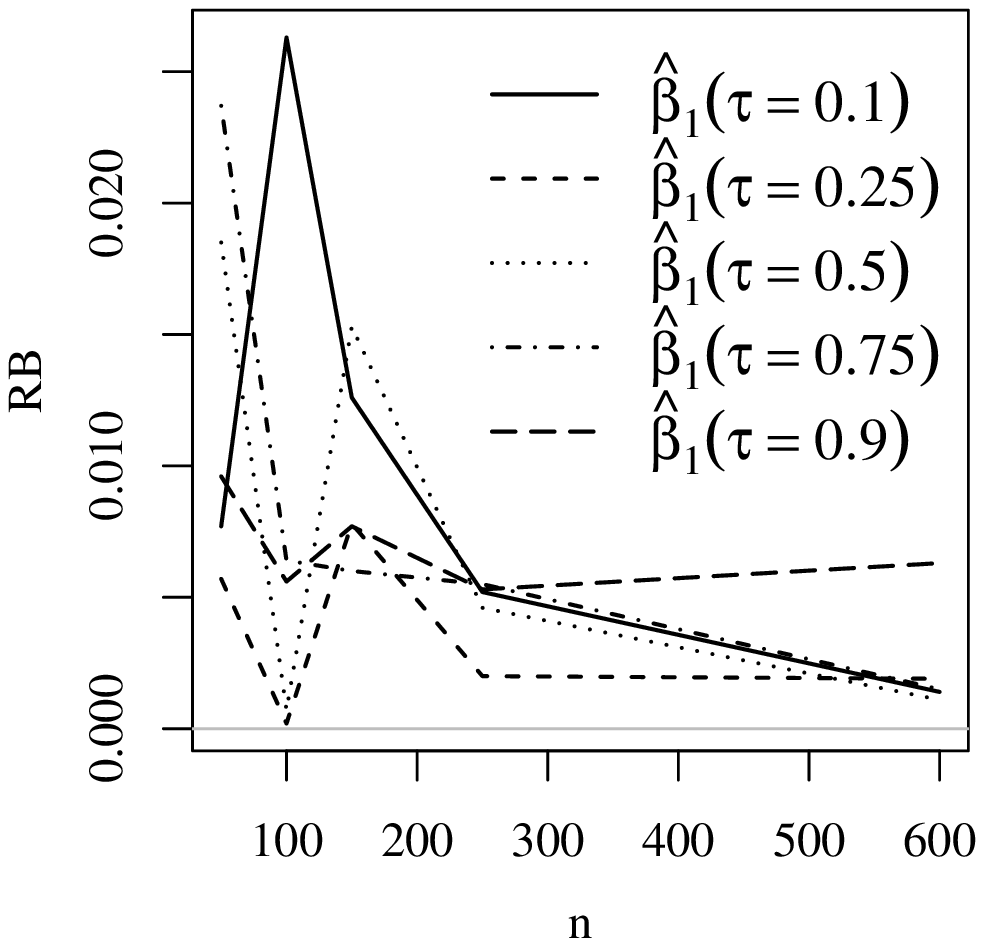}}\hspace{-0.25cm}
{\includegraphics[height=3.5cm,width=3.5cm]{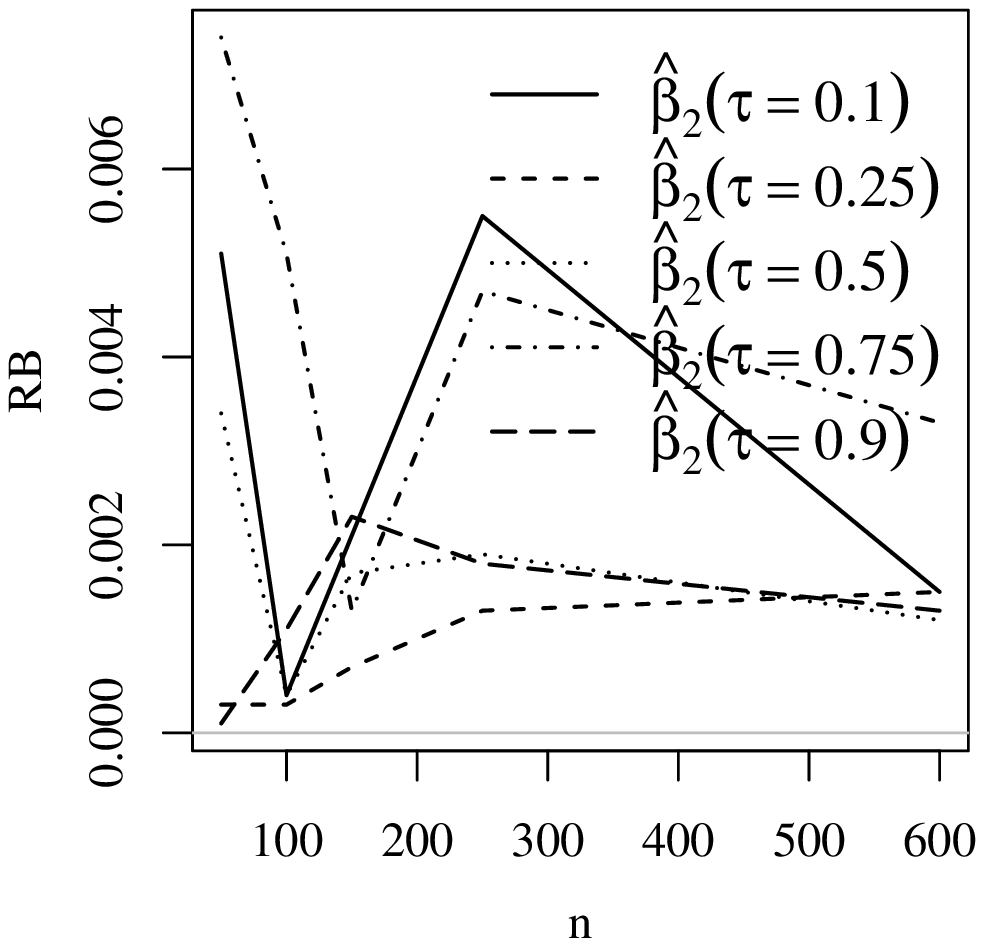}}\hspace{-0.25cm}
{\includegraphics[height=3.5cm,width=3.5cm]{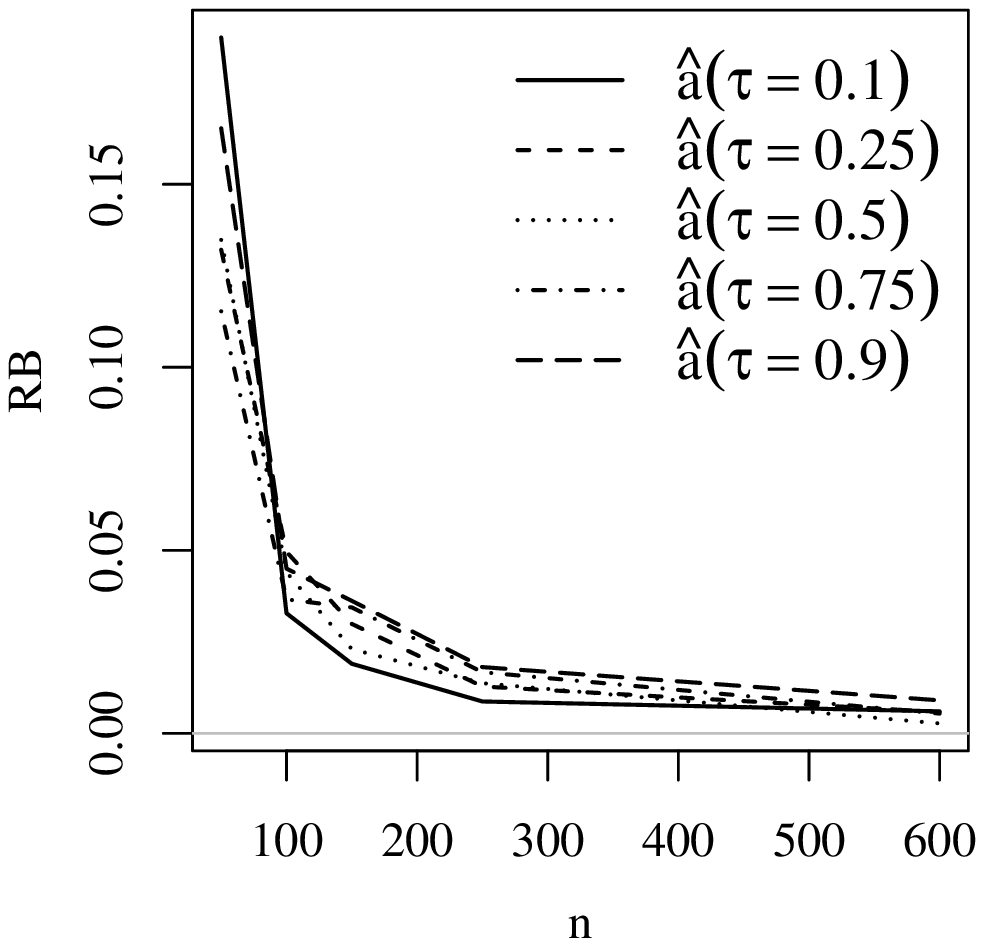}}\hspace{-0.25cm}
{\includegraphics[height=3.5cm,width=3.5cm]{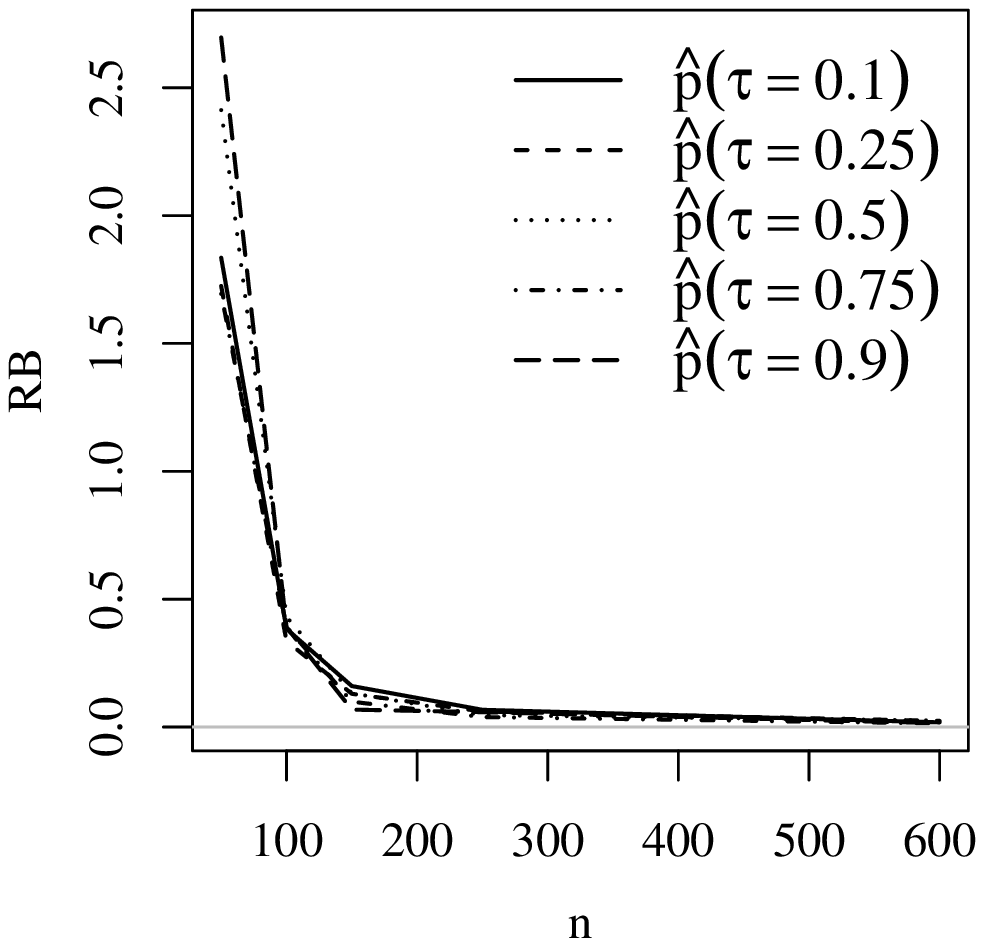}}
{\includegraphics[height=3.5cm,width=3.5cm]{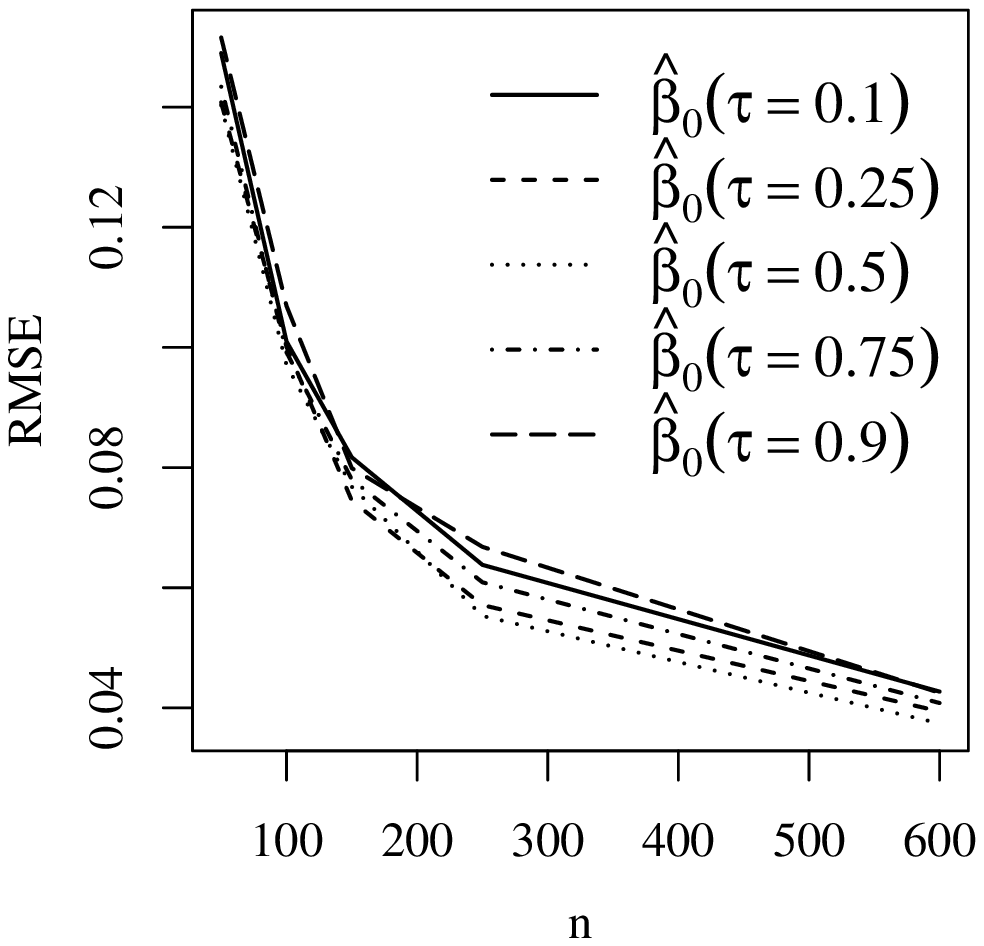}}\hspace{-0.25cm}
{\includegraphics[height=3.5cm,width=3.5cm]{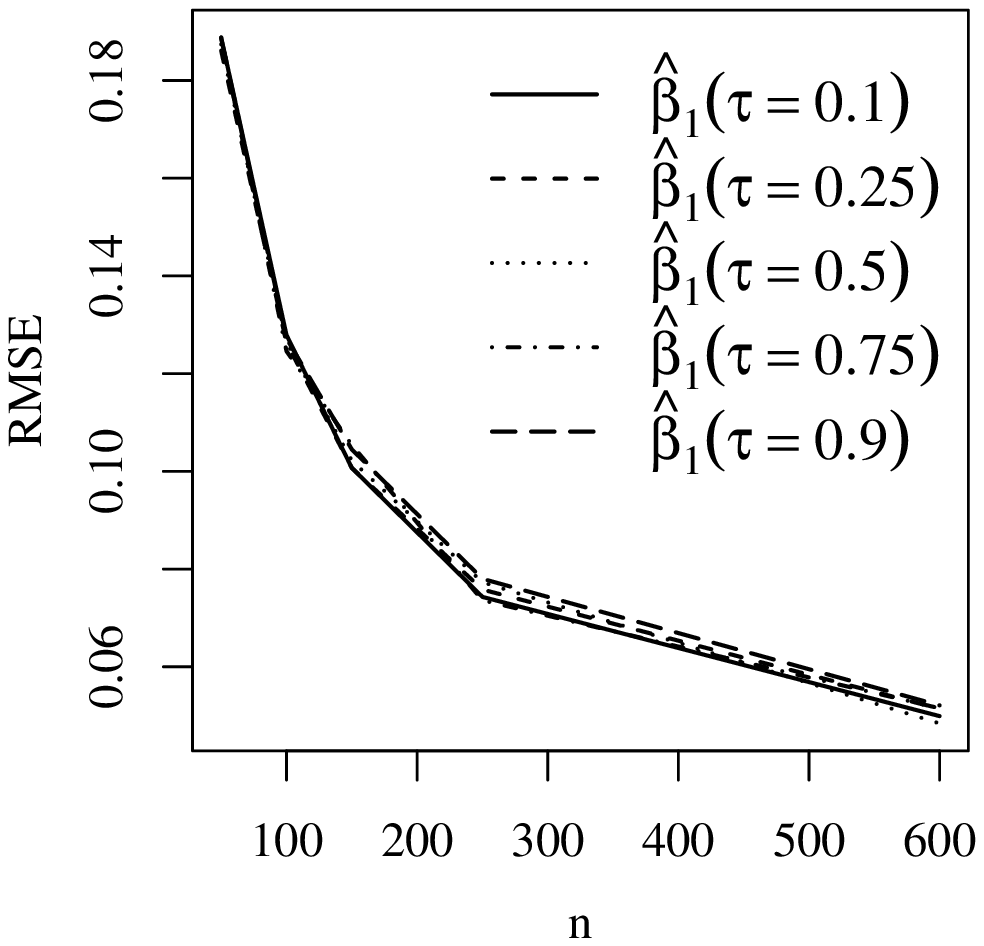}}\hspace{-0.25cm}
{\includegraphics[height=3.5cm,width=3.5cm]{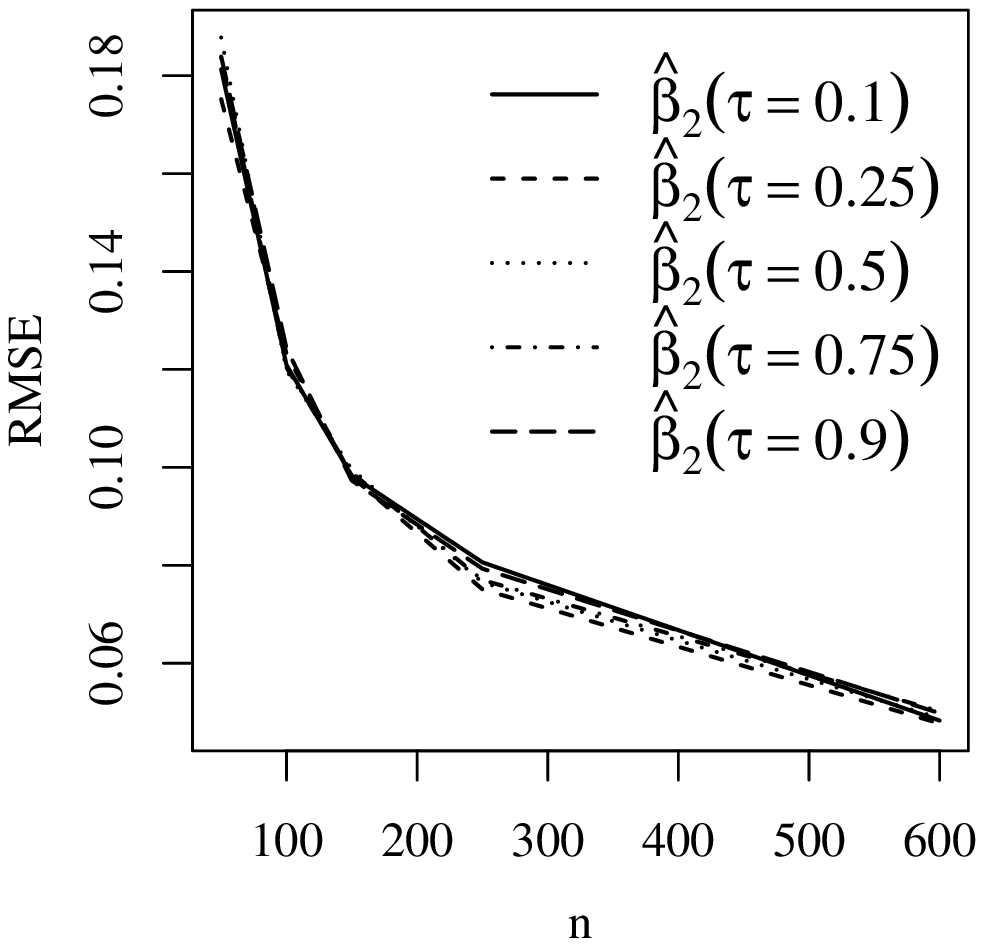}}\hspace{-0.25cm}
{\includegraphics[height=3.5cm,width=3.5cm]{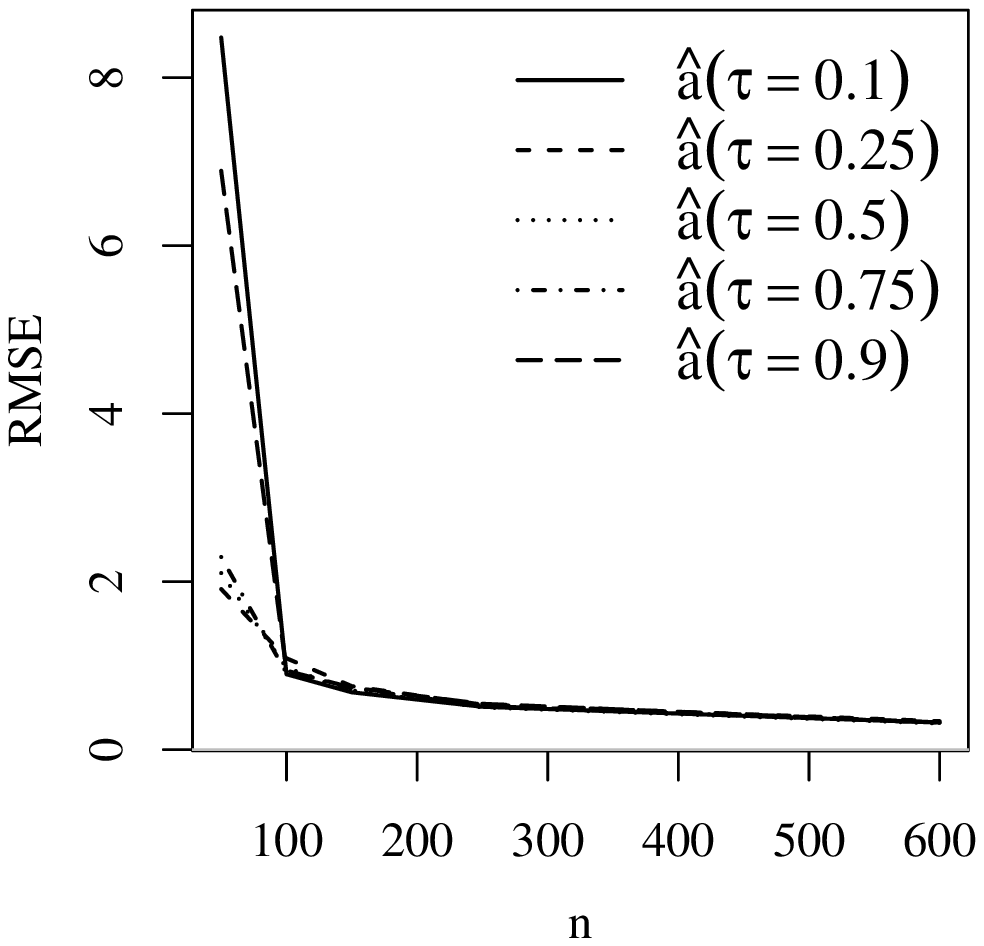}}\hspace{-0.25cm}
{\includegraphics[height=3.5cm,width=3.5cm]{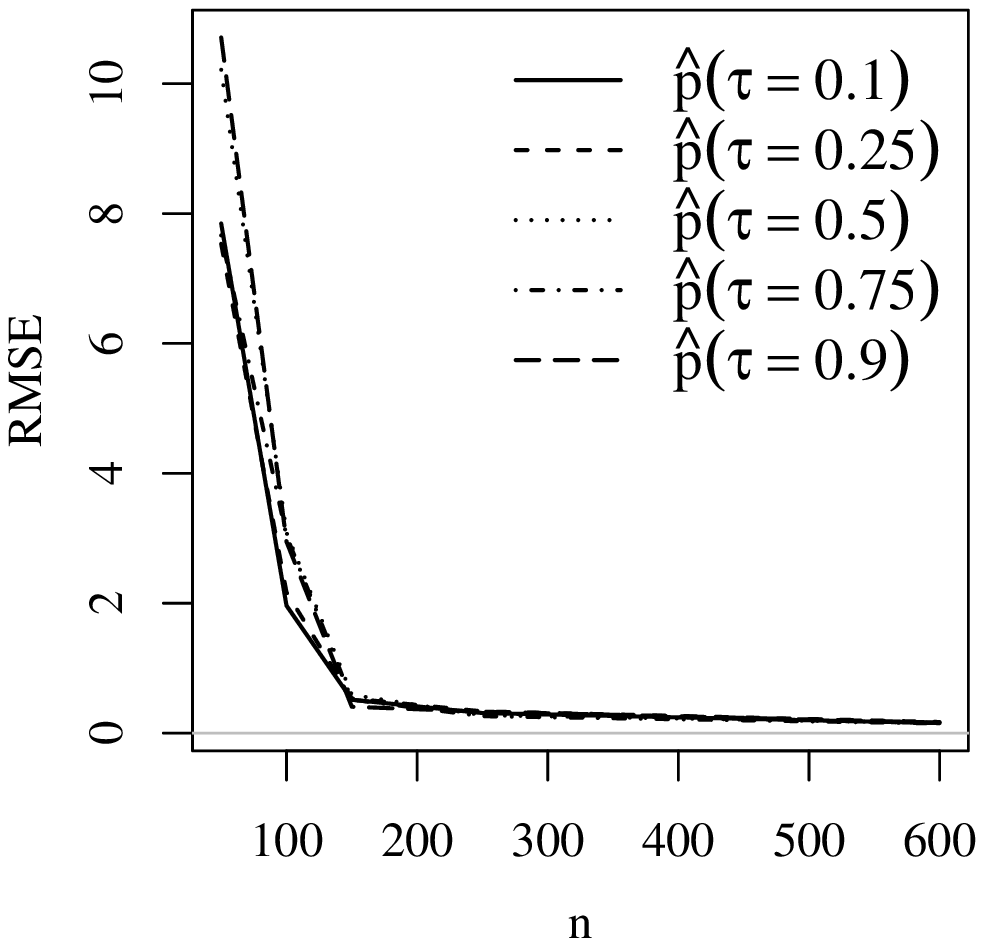}}
{\includegraphics[height=3.5cm,width=3.5cm]{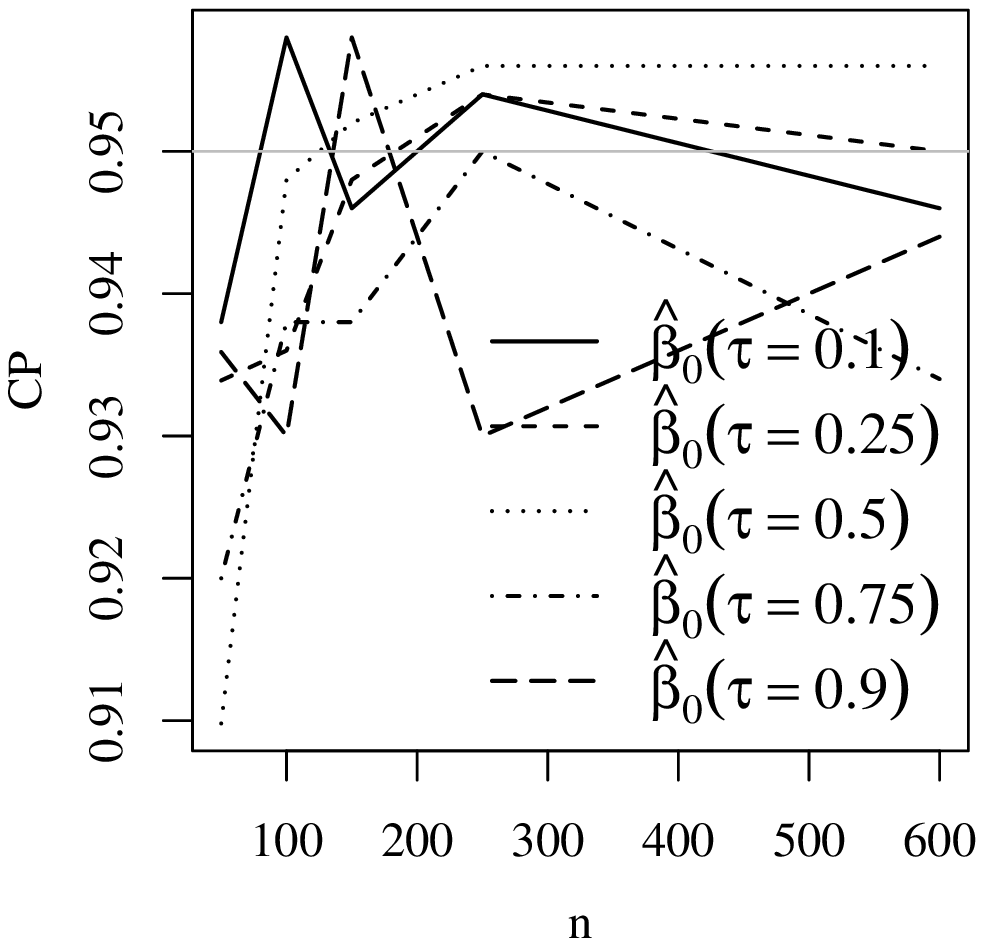}}\hspace{-0.25cm}
{\includegraphics[height=3.5cm,width=3.5cm]{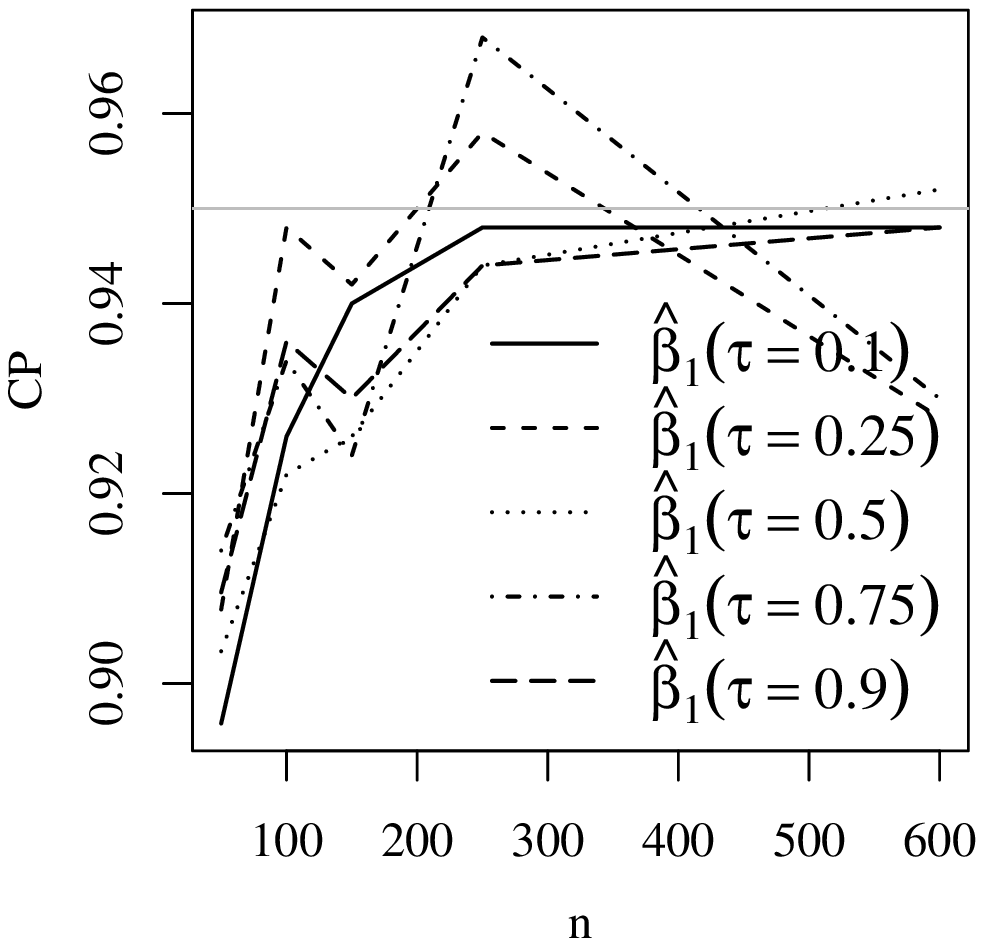}}\hspace{-0.25cm}
{\includegraphics[height=3.5cm,width=3.5cm]{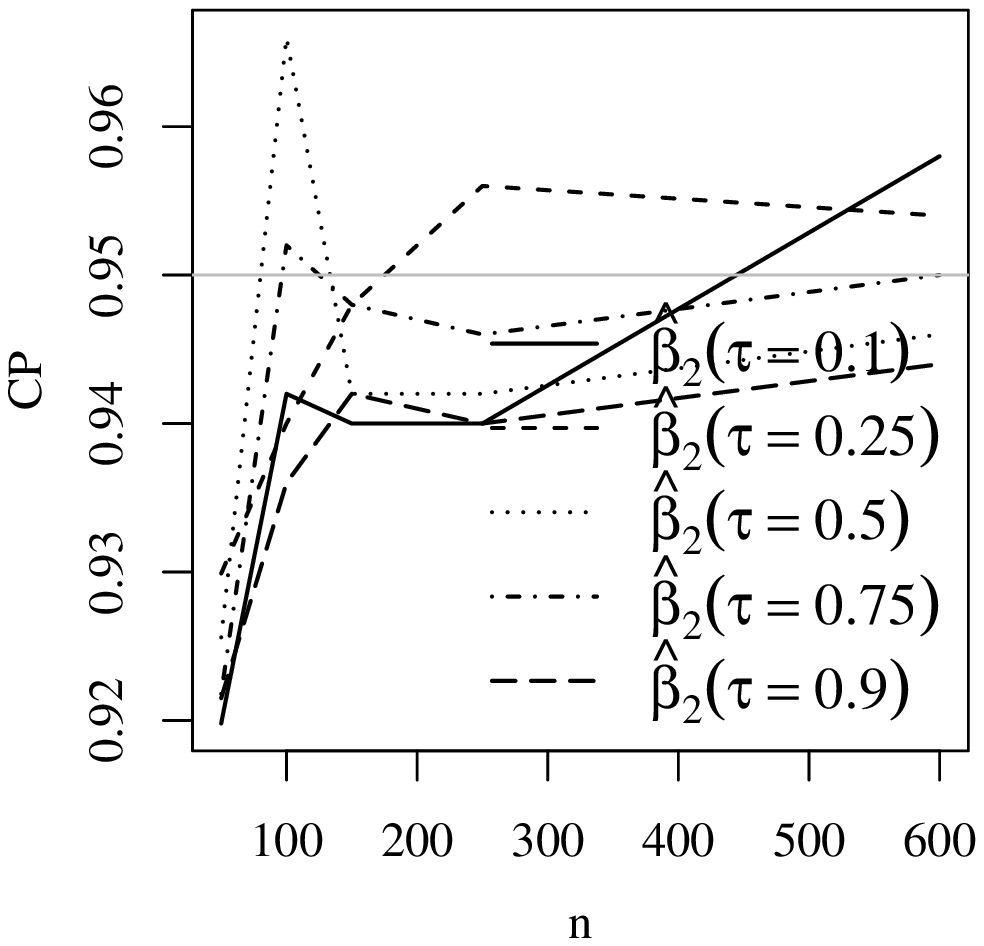}}\hspace{-0.25cm}
{\includegraphics[height=3.5cm,width=3.5cm]{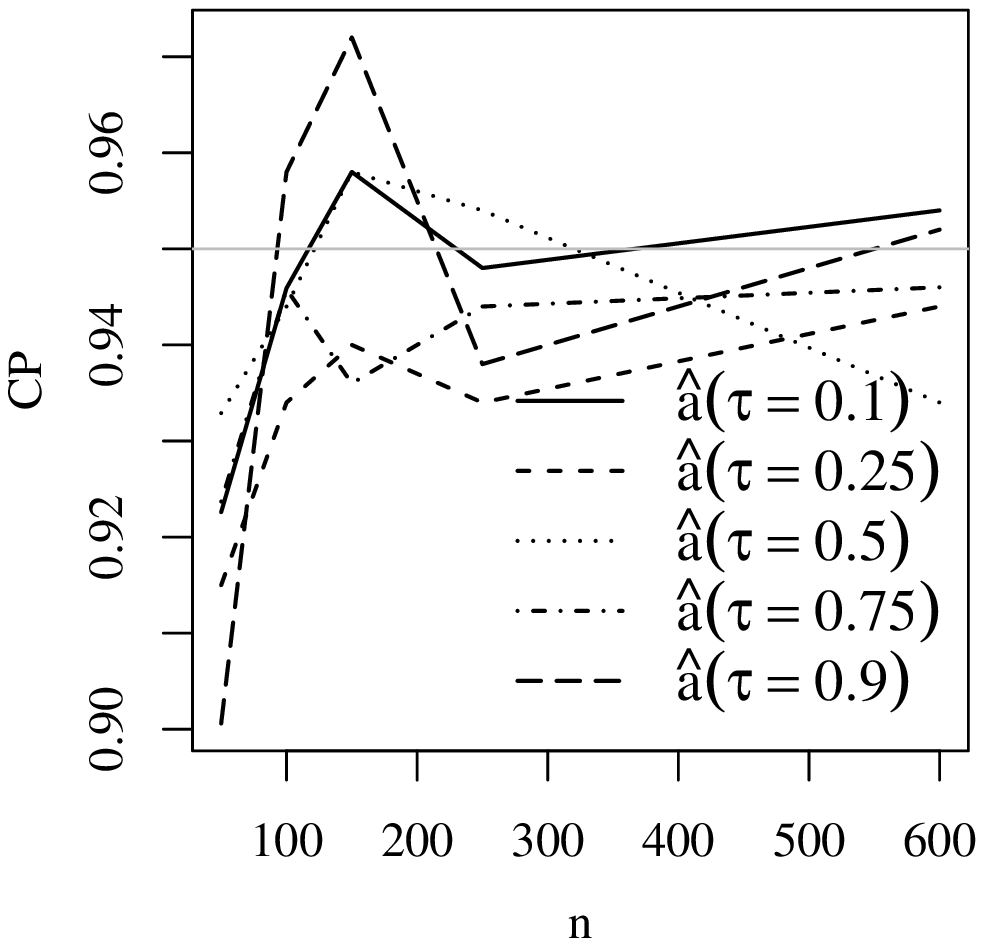}}\hspace{-0.25cm}
{\includegraphics[height=3.5cm,width=3.5cm]{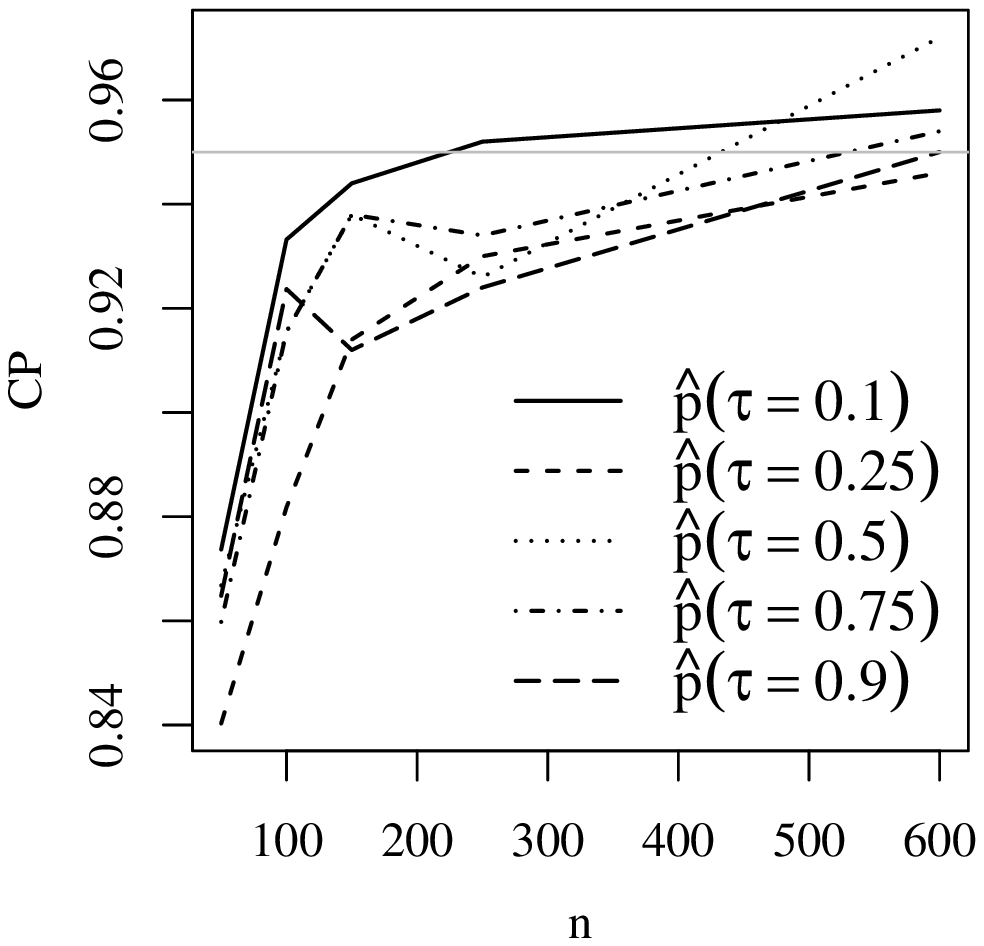}}
\vspace{-0.2cm}
\caption{Monte Carlo simulation results for the Singh-Maddala model with $a = 5$ and $q = 1$.}
\label{fig_singh_maddala_MC_BRC}
\end{figure}

\begin{figure}[!ht]
\vspace{-0.25cm}
\centering
{\includegraphics[height=3.5cm,width=3.5cm]{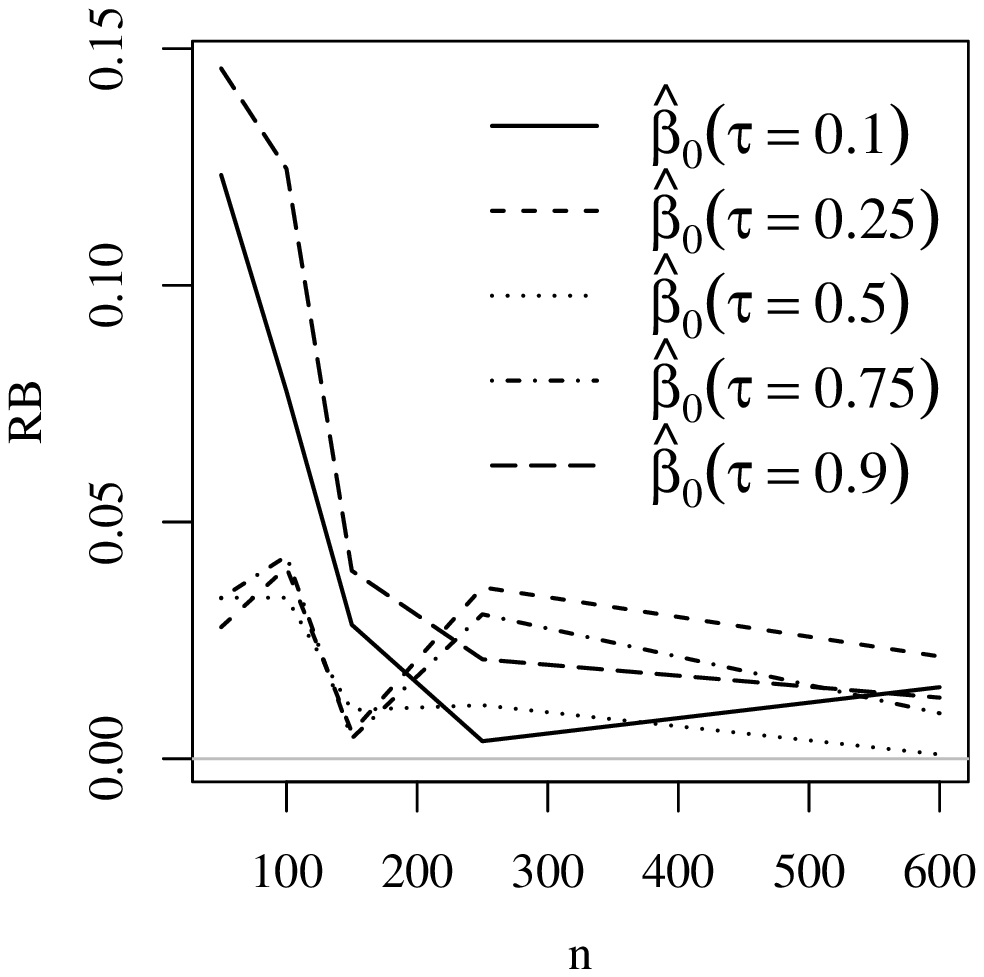}}\hspace{-0.25cm}
{\includegraphics[height=3.5cm,width=3.5cm]{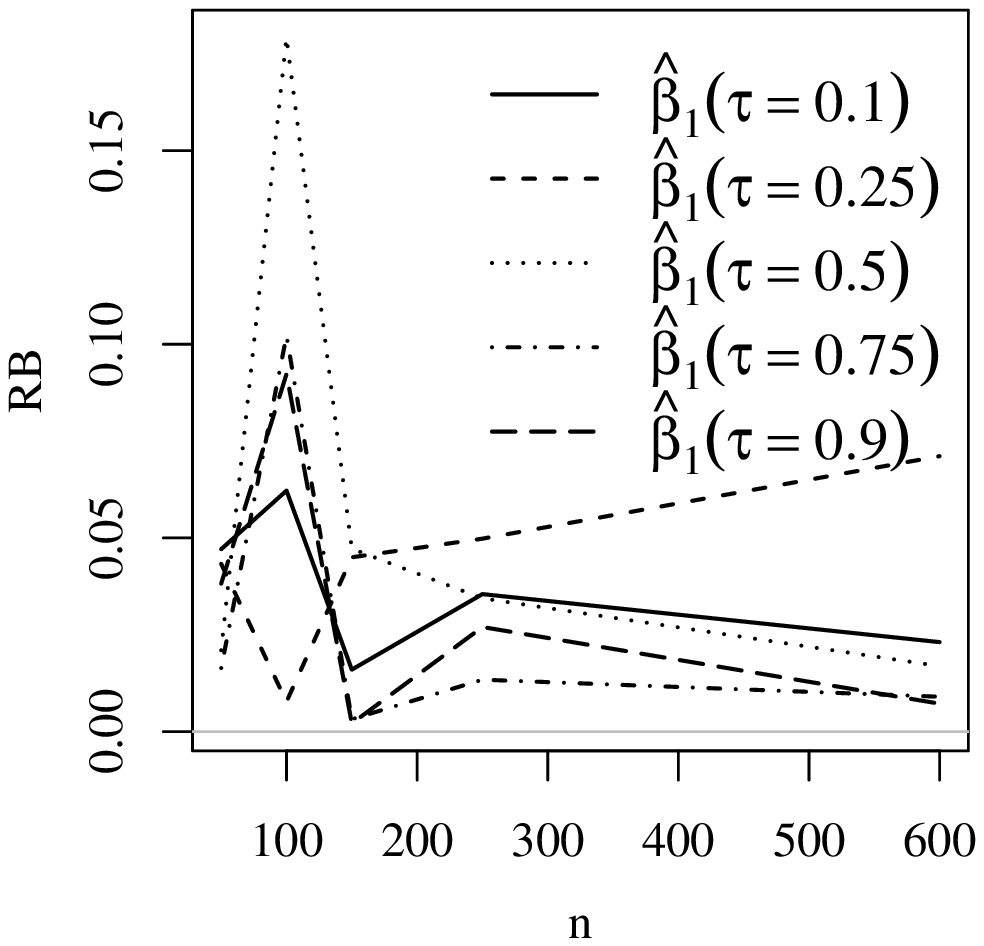}}\hspace{-0.25cm}
{\includegraphics[height=3.5cm,width=3.5cm]{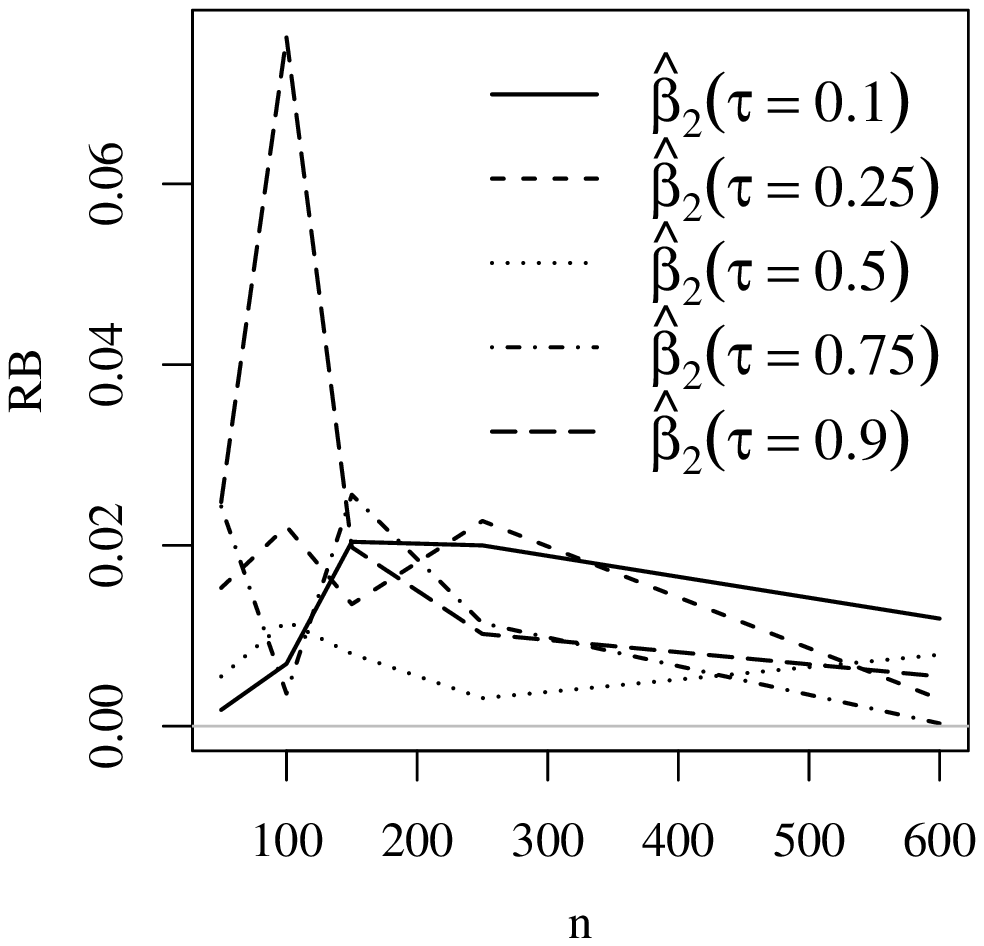}}\hspace{-0.25cm}
{\includegraphics[height=3.5cm,width=3.5cm]{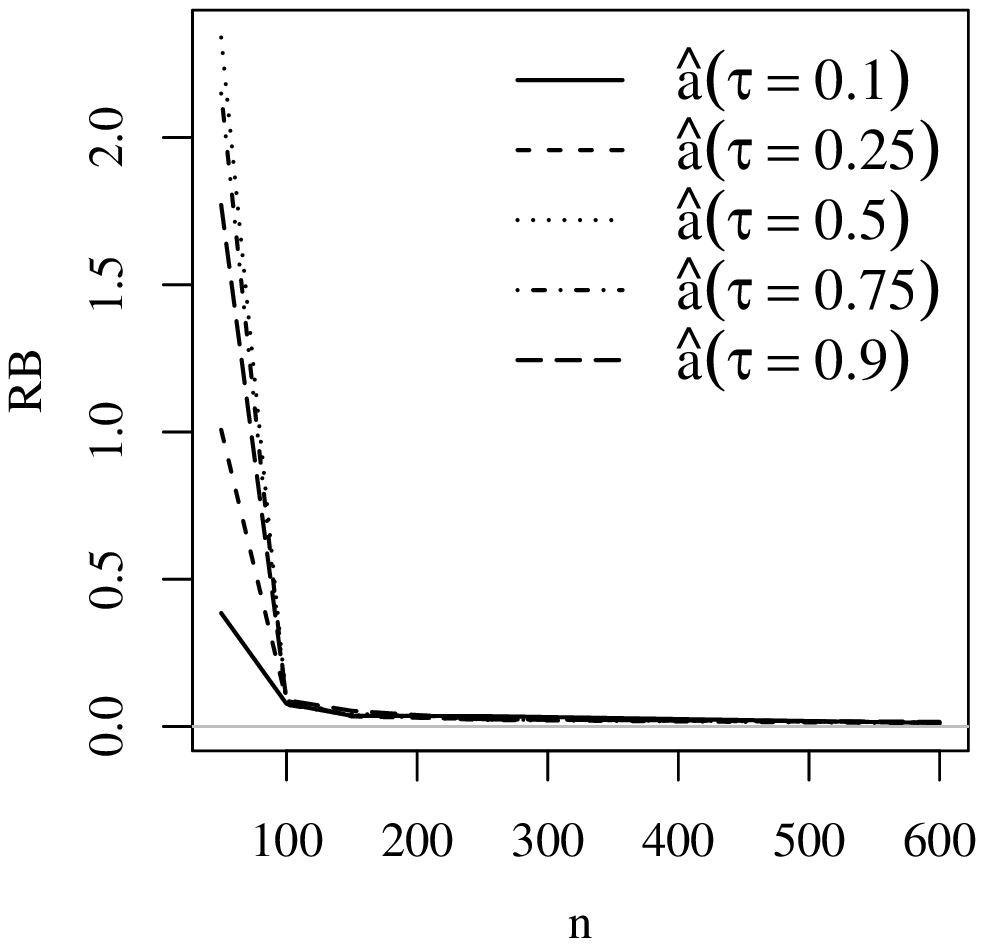}}\hspace{-0.25cm}
{\includegraphics[height=3.5cm,width=3.5cm]{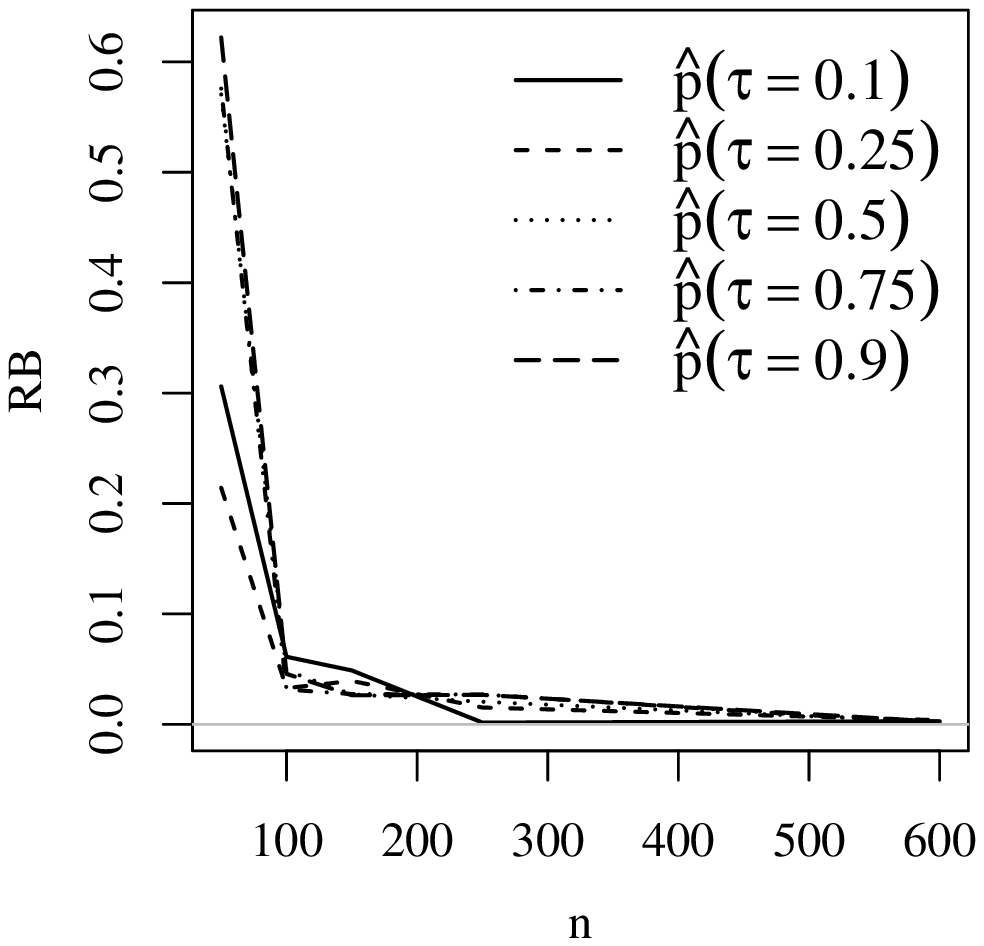}}
{\includegraphics[height=3.5cm,width=3.5cm]{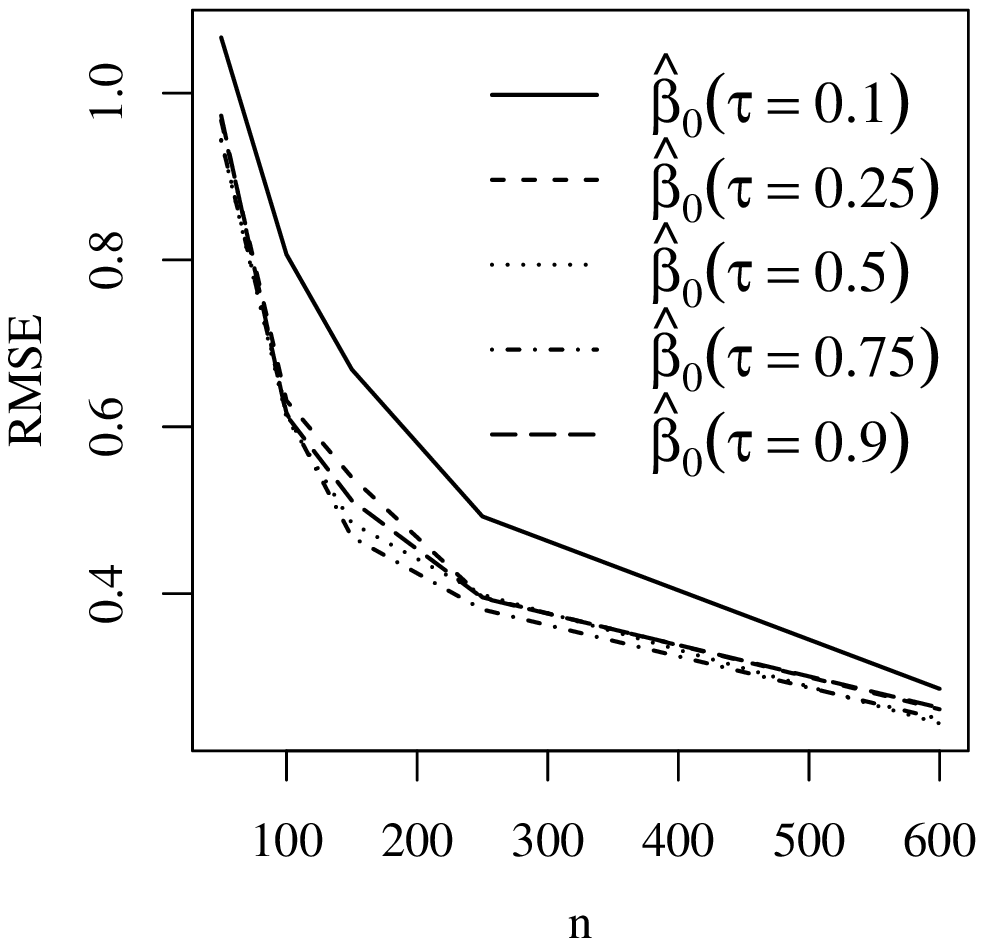}}\hspace{-0.25cm}
{\includegraphics[height=3.5cm,width=3.5cm]{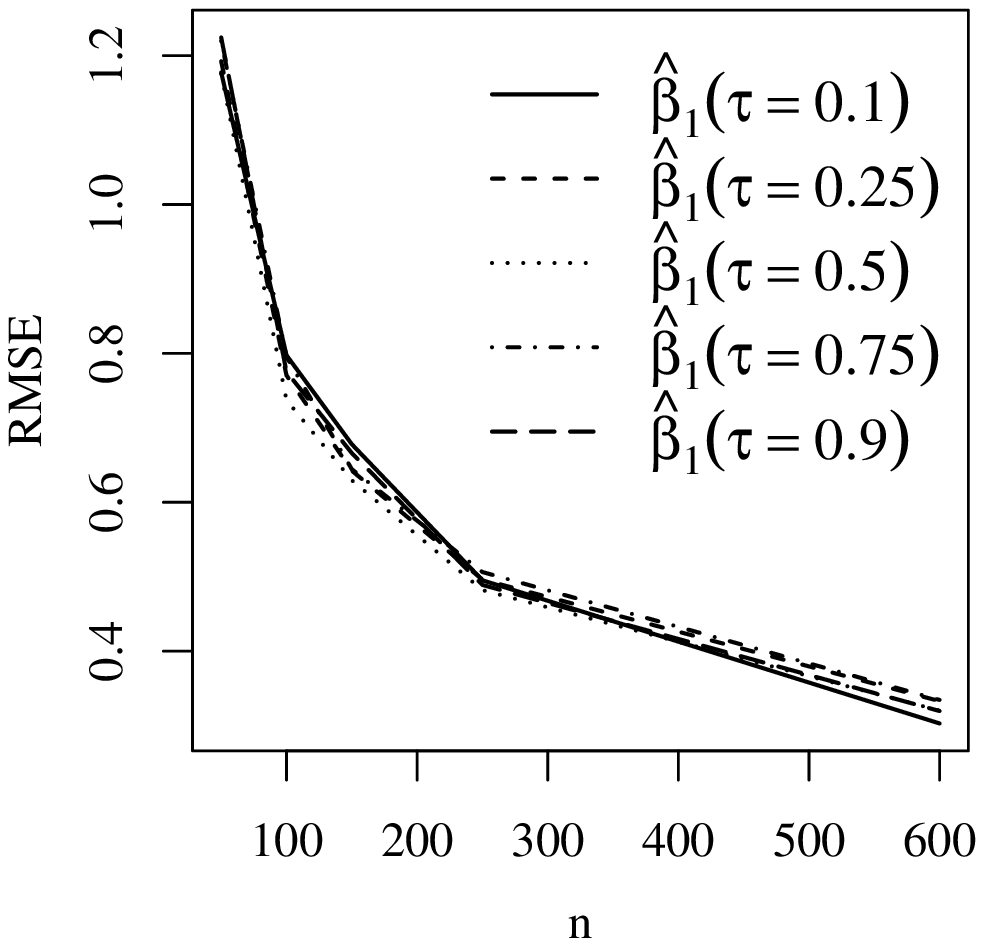}}\hspace{-0.25cm}
{\includegraphics[height=3.5cm,width=3.5cm]{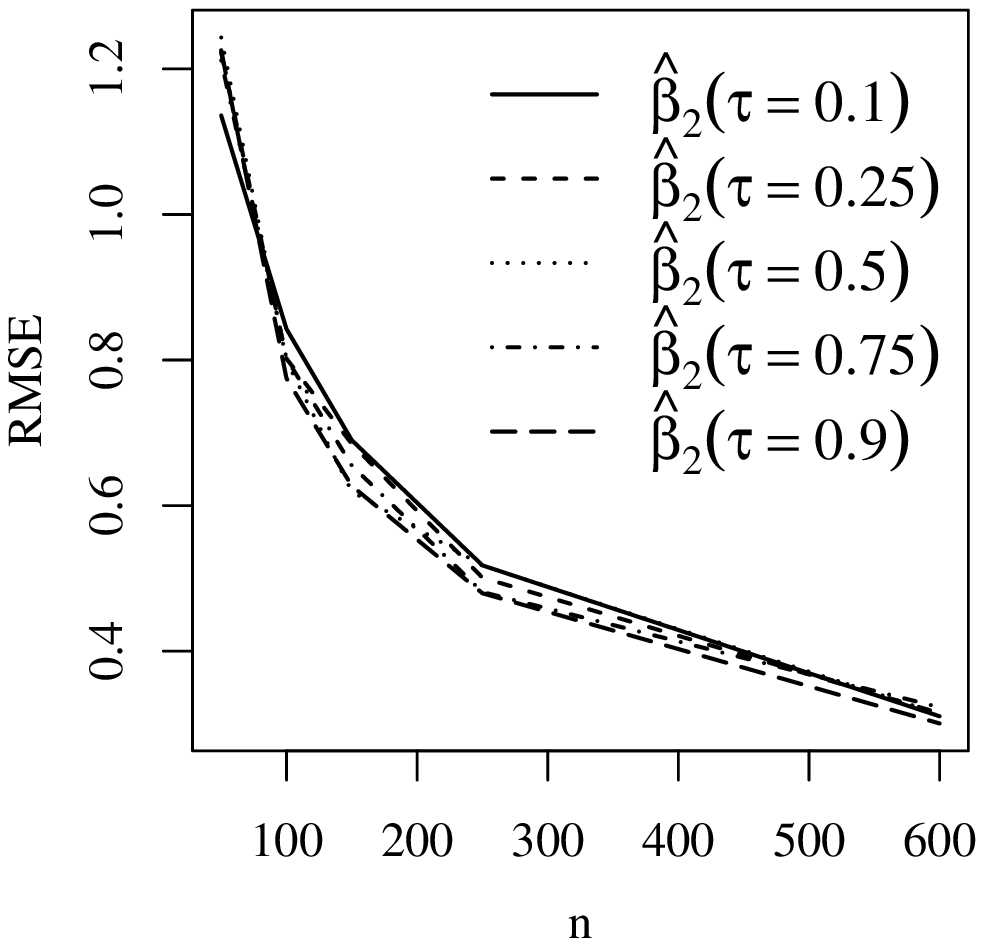}}\hspace{-0.25cm}
{\includegraphics[height=3.5cm,width=3.5cm]{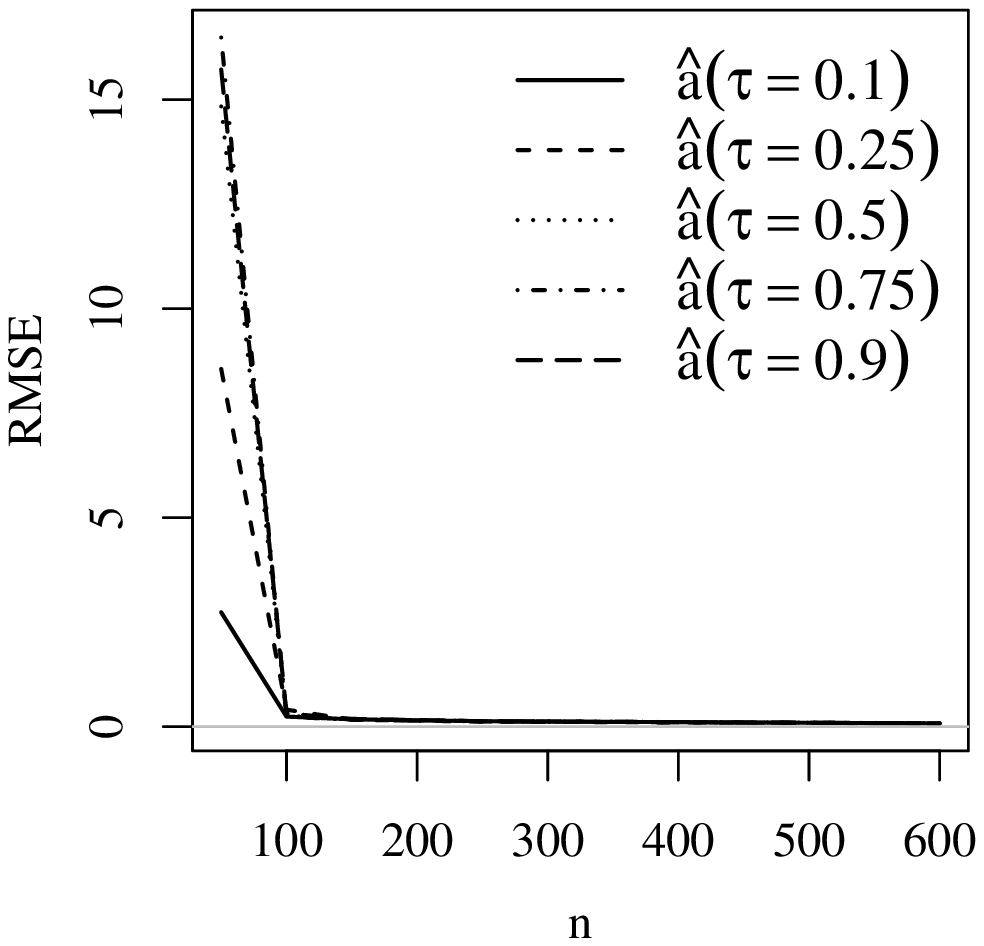}}\hspace{-0.25cm}
{\includegraphics[height=3.5cm,width=3.5cm]{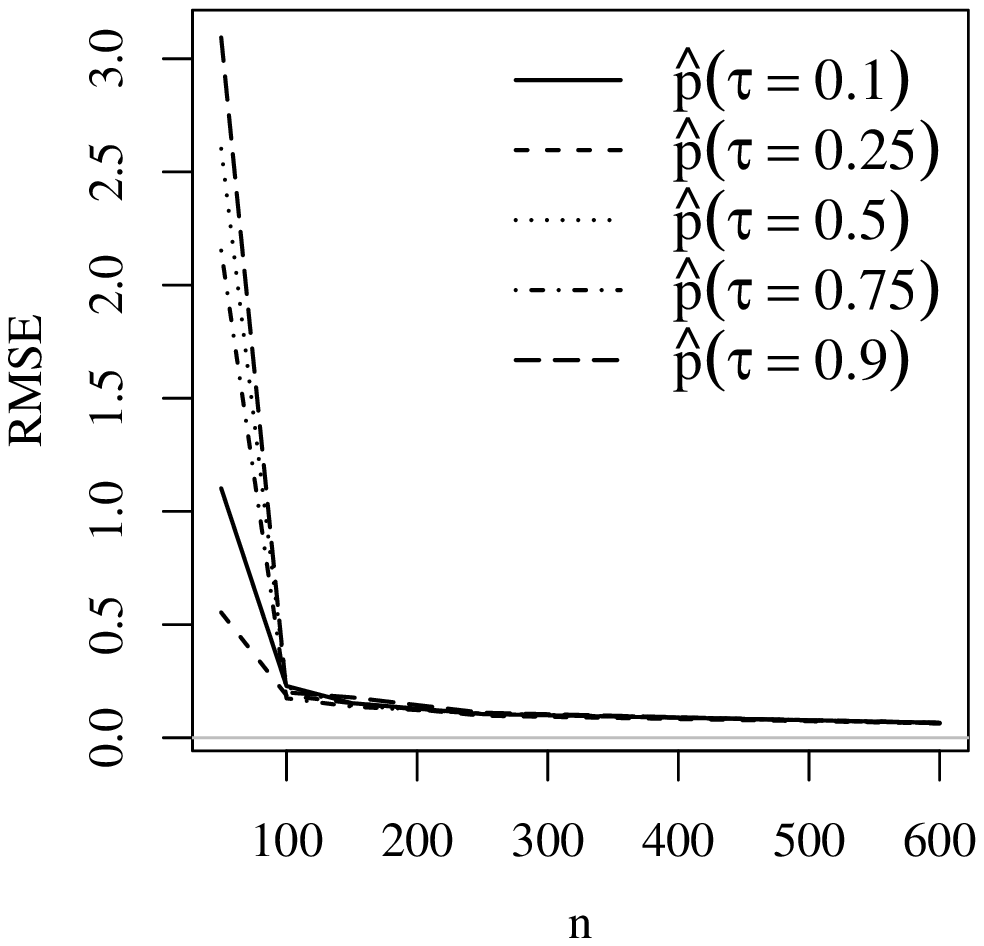}}
{\includegraphics[height=3.5cm,width=3.5cm]{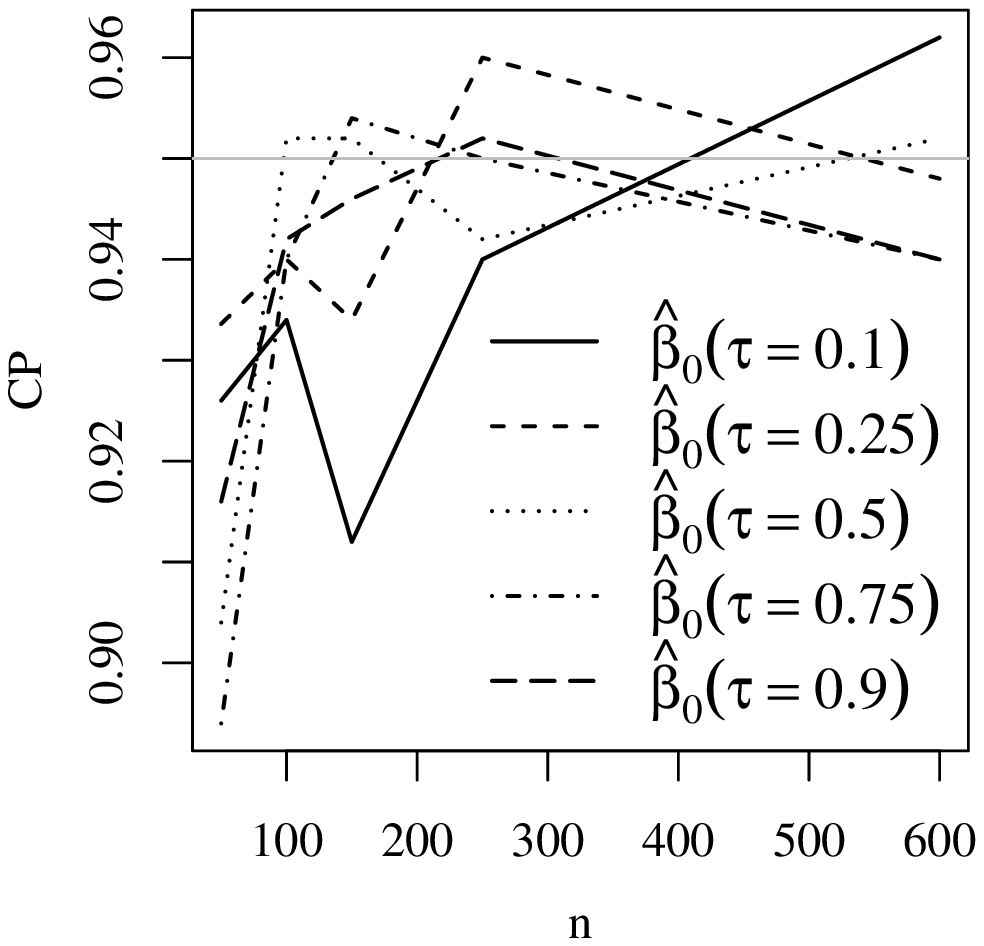}}\hspace{-0.25cm}
{\includegraphics[height=3.5cm,width=3.5cm]{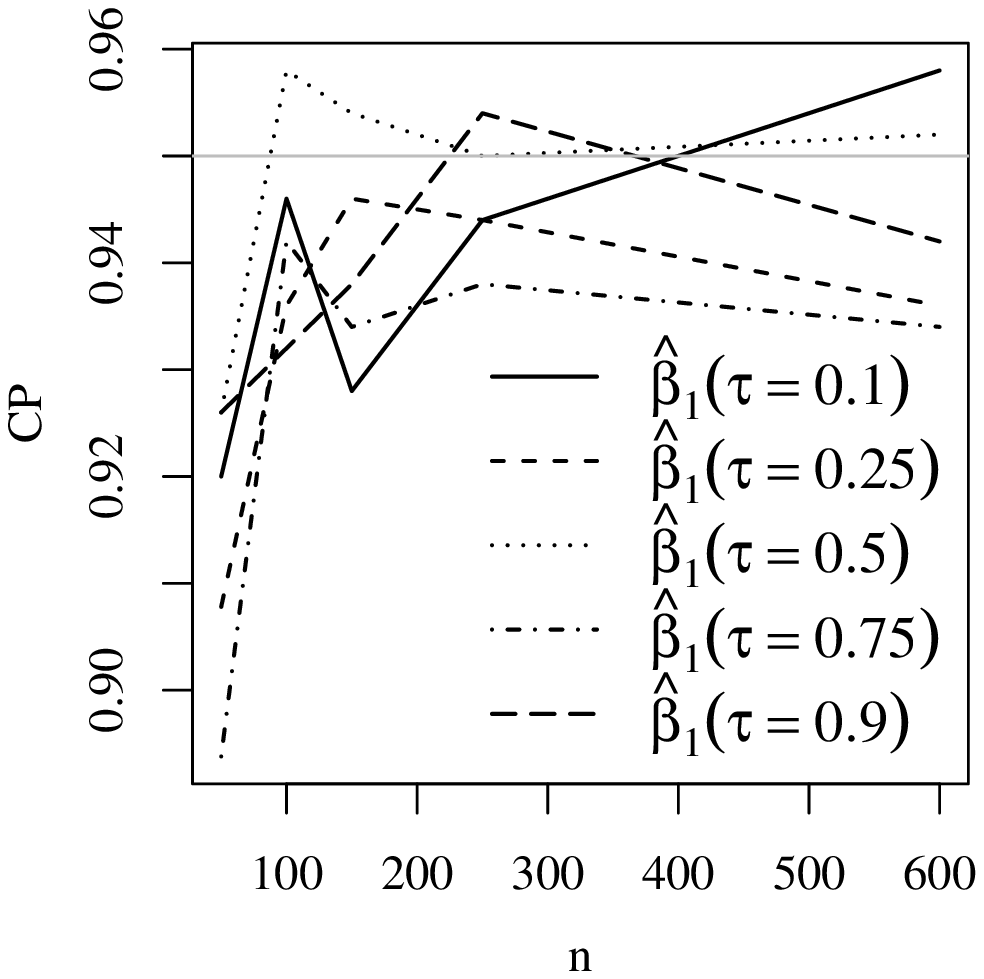}}\hspace{-0.25cm}
{\includegraphics[height=3.5cm,width=3.5cm]{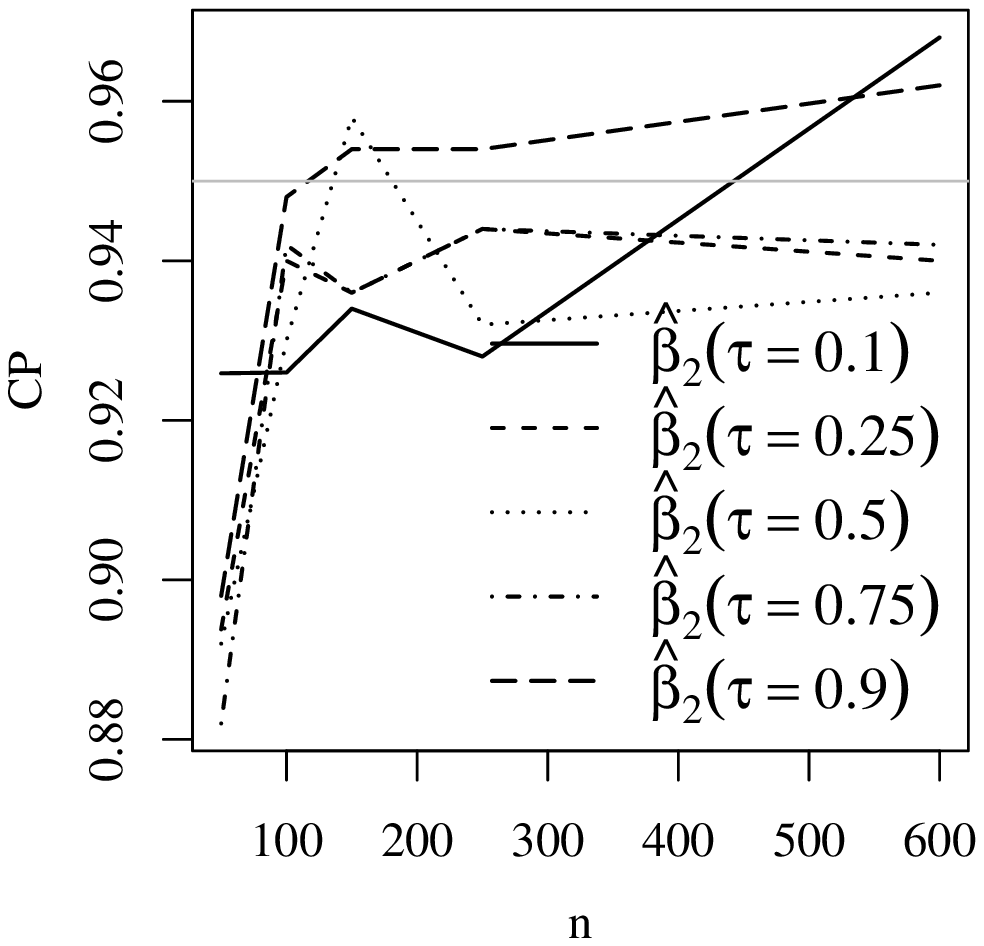}}\hspace{-0.25cm}
{\includegraphics[height=3.5cm,width=3.5cm]{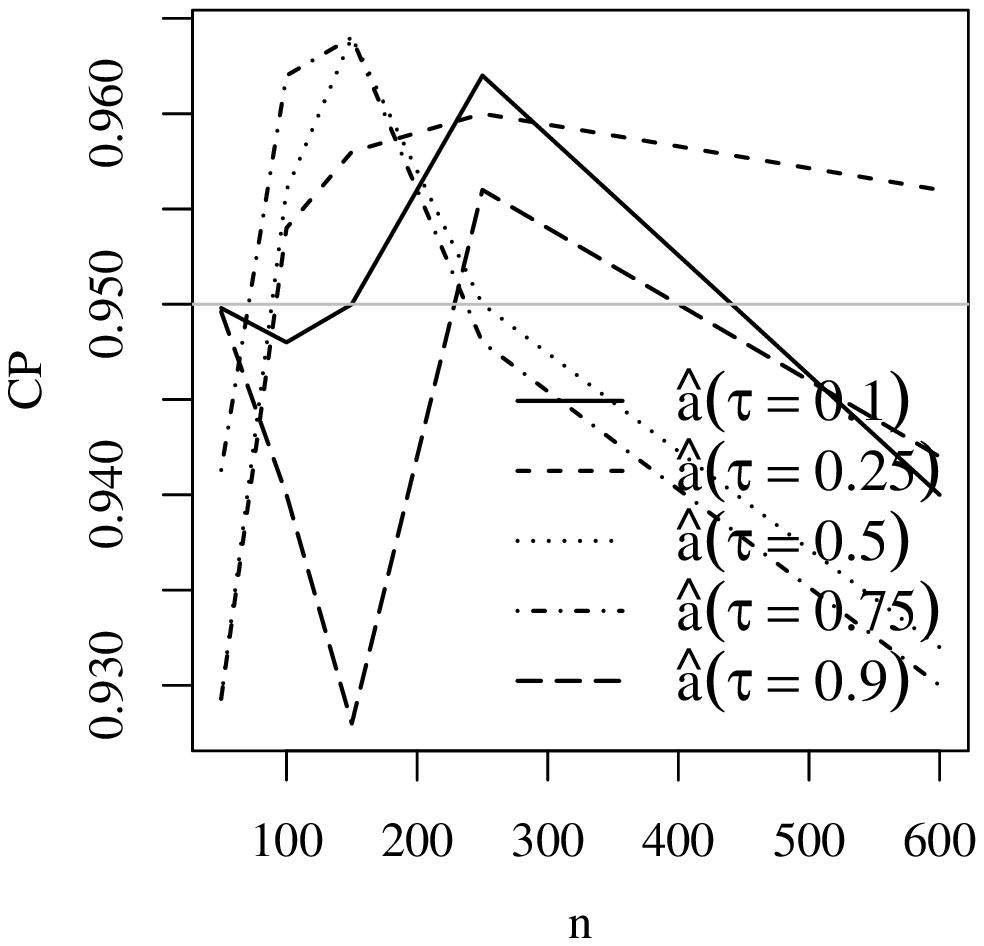}}\hspace{-0.25cm}
{\includegraphics[height=3.5cm,width=3.5cm]{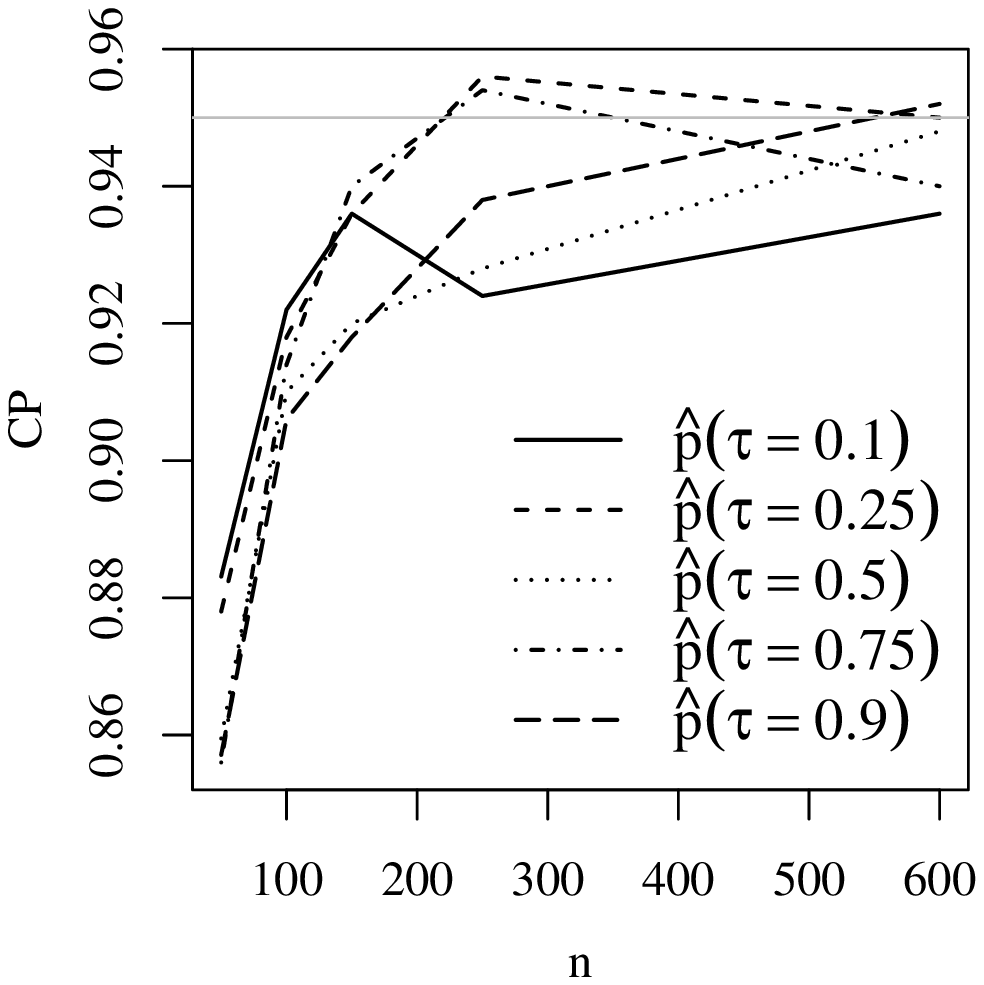}}
\vspace{-0.2cm}
\caption{Monte Carlo simulation results for the Dagum model with $a = 1$ and $p = 0.5$.}
\label{fig_dagum_MC_BRC}
\end{figure}

The results for Singh-Maddala models are shown in Figure \ref{fig_singh_maddala_MC_BRC}. It is possible to see that the simulations produced the expected outcomes. As the sample size increases, the RB and RMSE both decrease, and the CP tends to 95\%. The results for the Dagum model are shown in Figure \ref{fig_dagum_MC_BRC}. This figure presents results similar to those found for the Singh-Maddala model.

\subsection{Empirical distribution of residuals}\label{residuals_simulation_res}
\noindent

\begin{figure}[!ht]
\vspace{-0.25cm}
\centering
{\includegraphics[height=3.5cm,width=3.5cm]{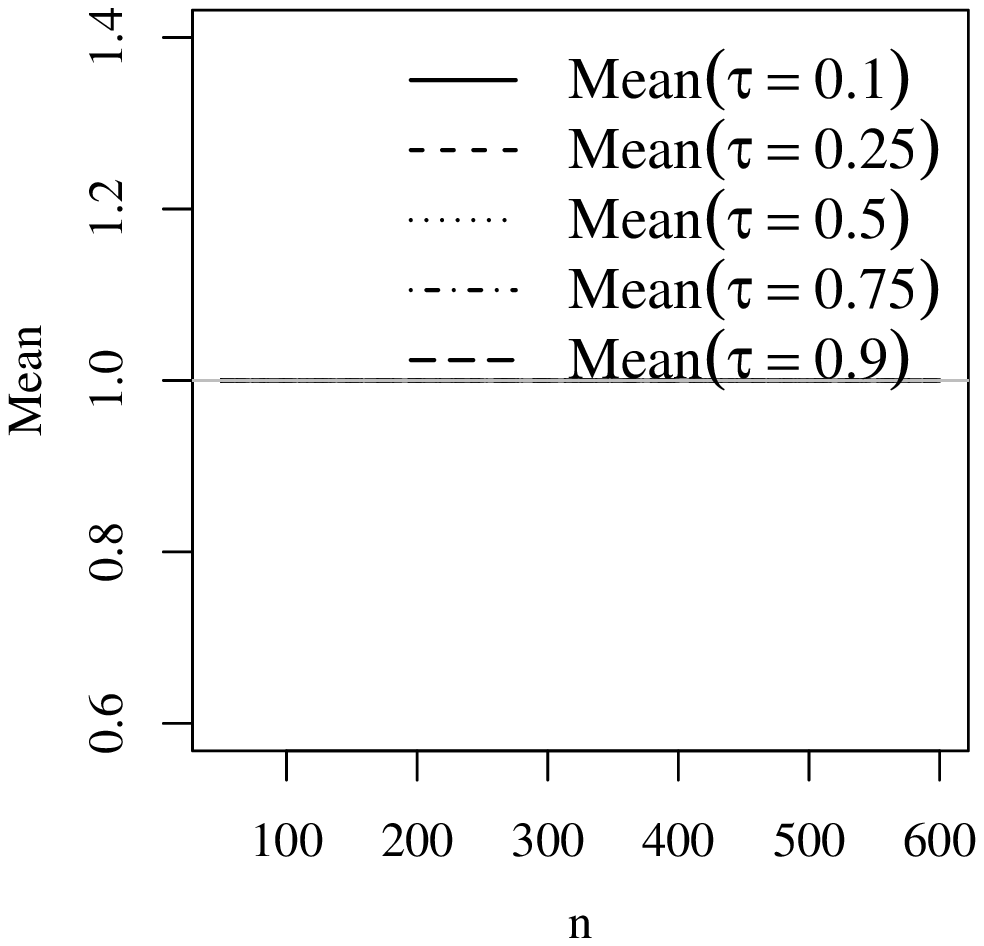}}\hspace{-0.25cm}
{\includegraphics[height=3.5cm,width=3.5cm]{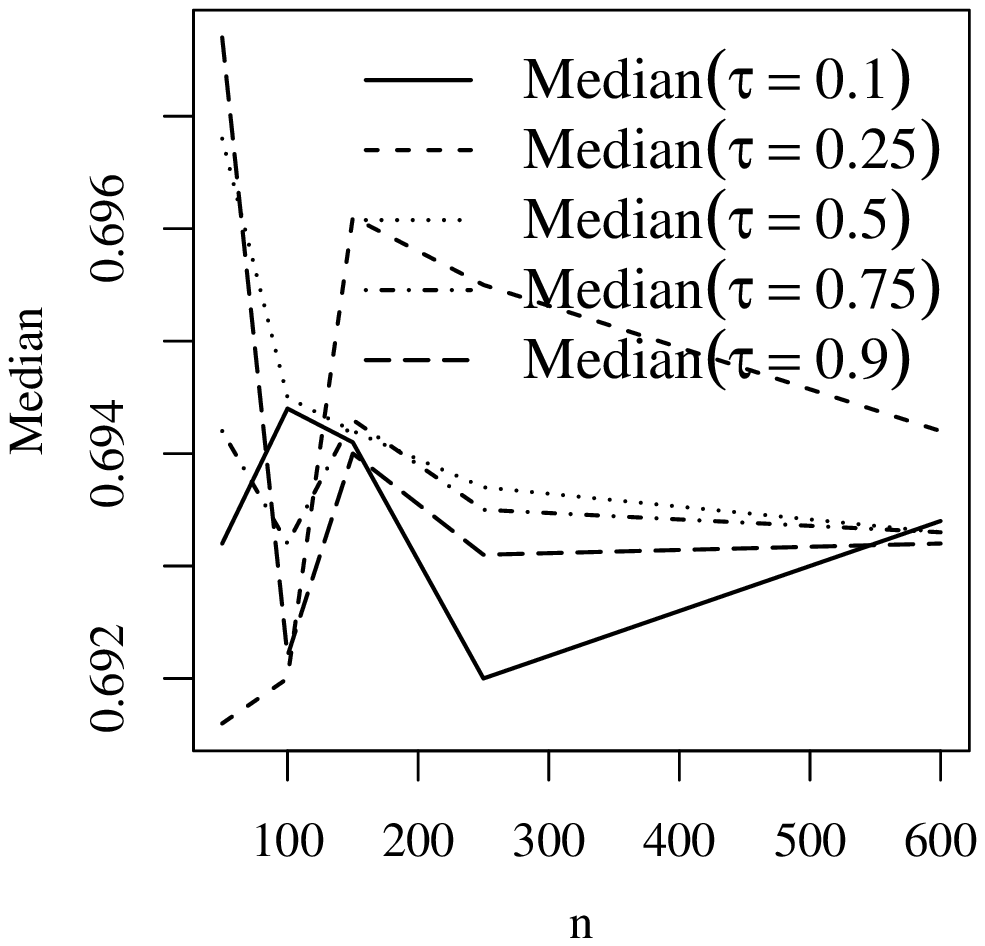}}\hspace{-0.25cm}
{\includegraphics[height=3.5cm,width=3.5cm]{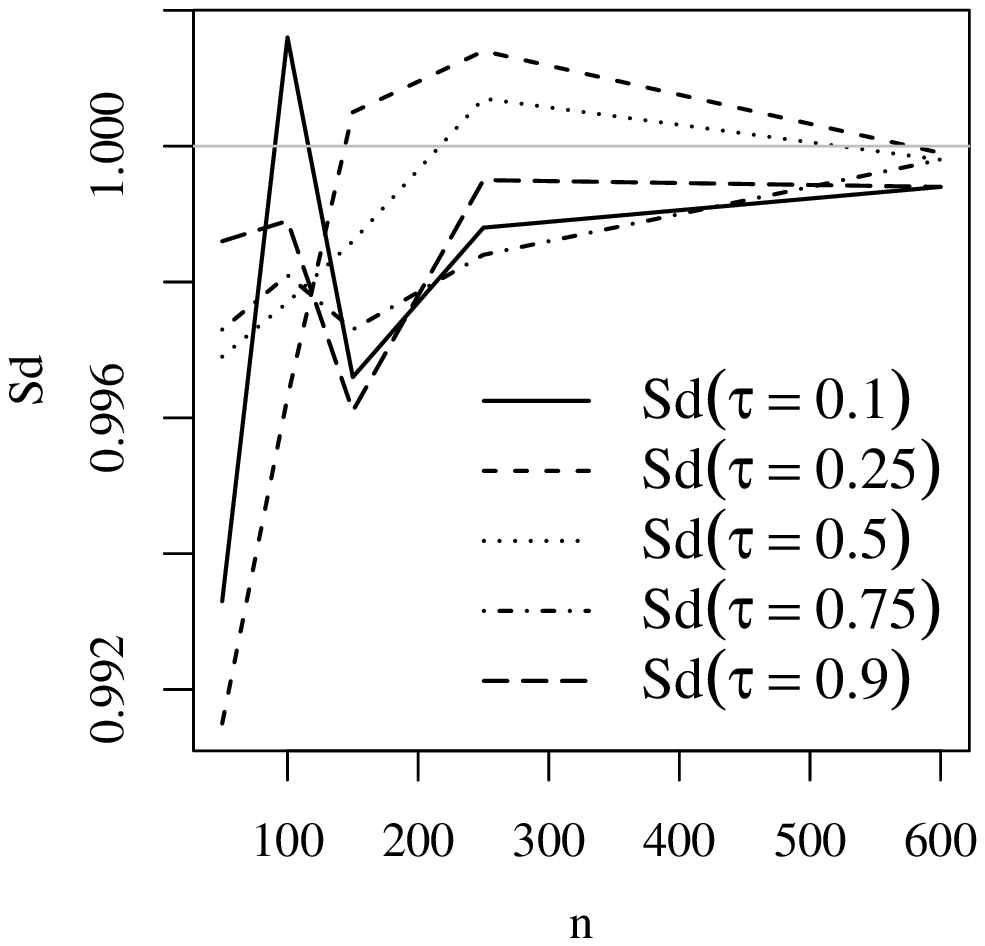}}\hspace{-0.25cm}
{\includegraphics[height=3.5cm,width=3.5cm]{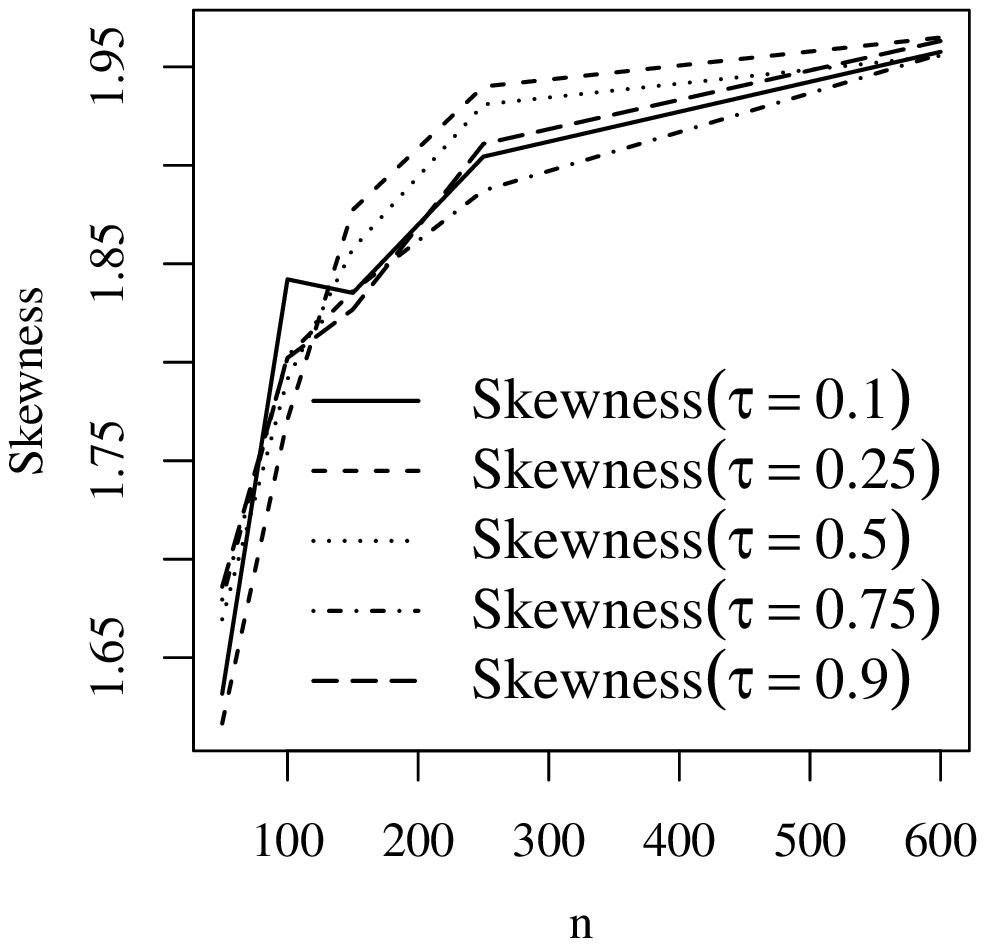}}\hspace{-0.25cm}
{\includegraphics[height=3.5cm,width=3.5cm]{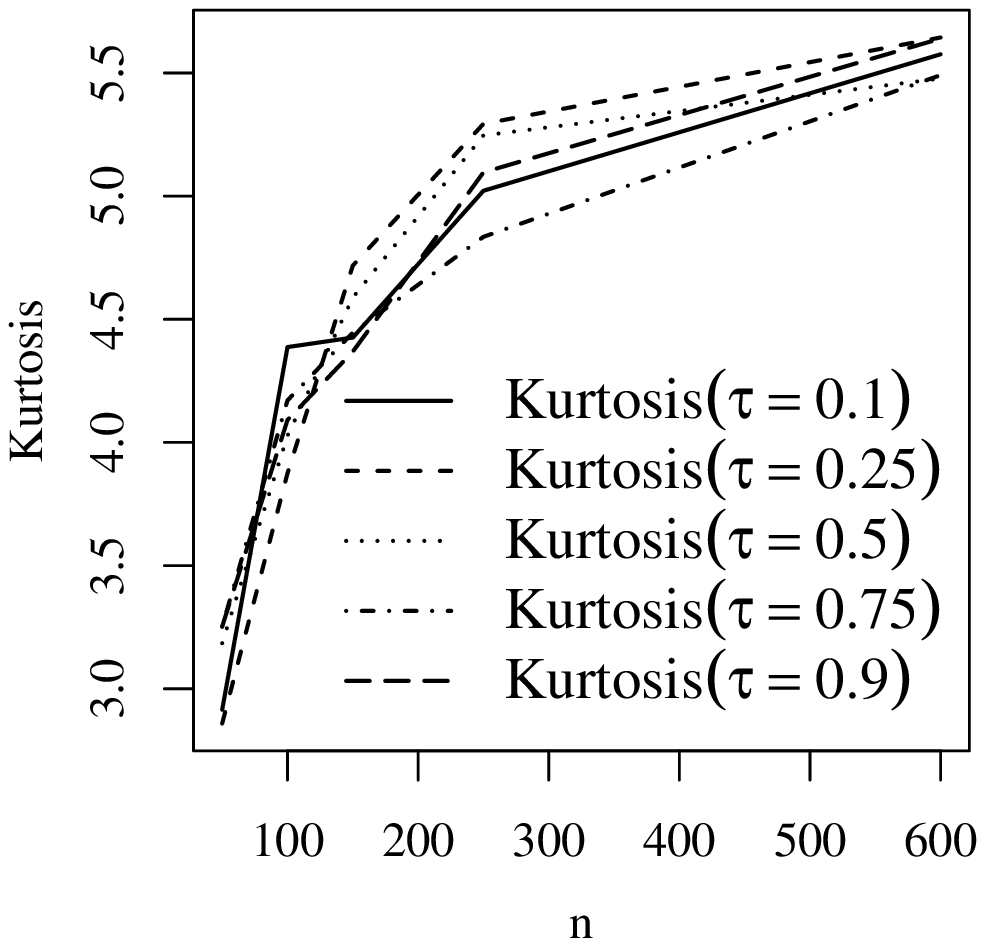}}
{\includegraphics[height=3.5cm,width=3.5cm]{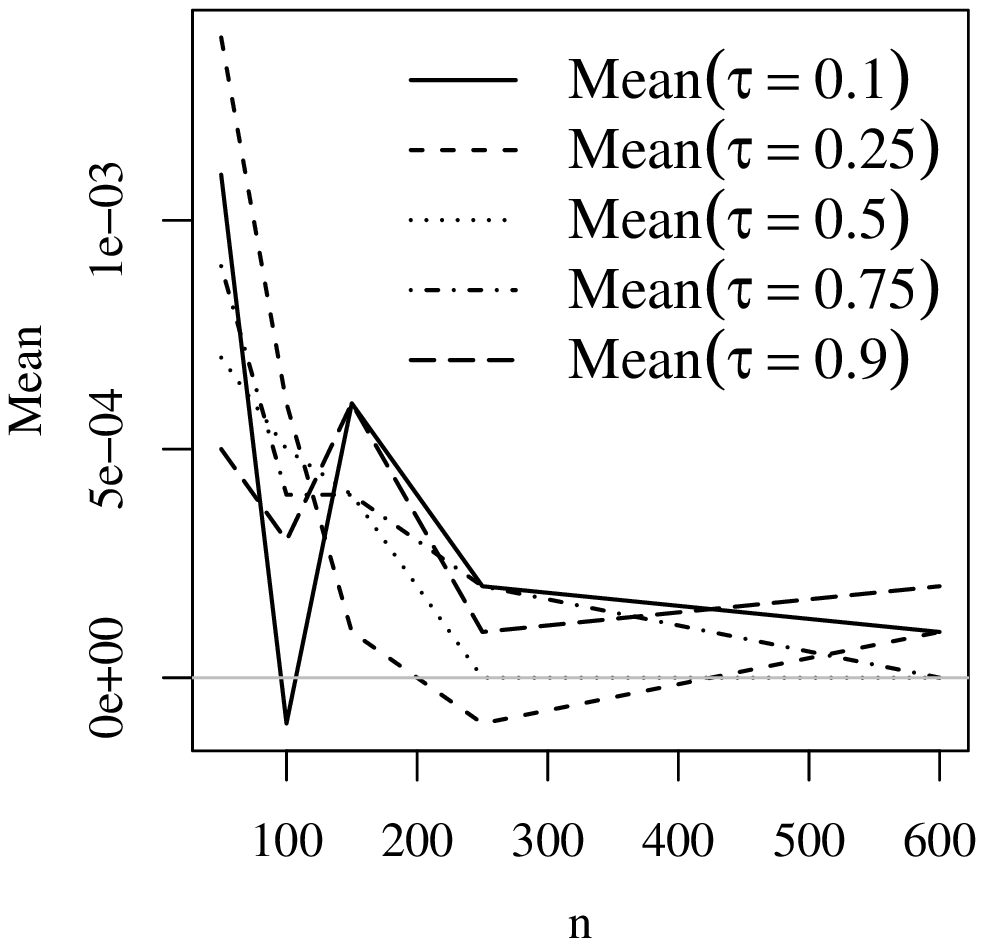}}\hspace{-0.25cm}
{\includegraphics[height=3.5cm,width=3.5cm]{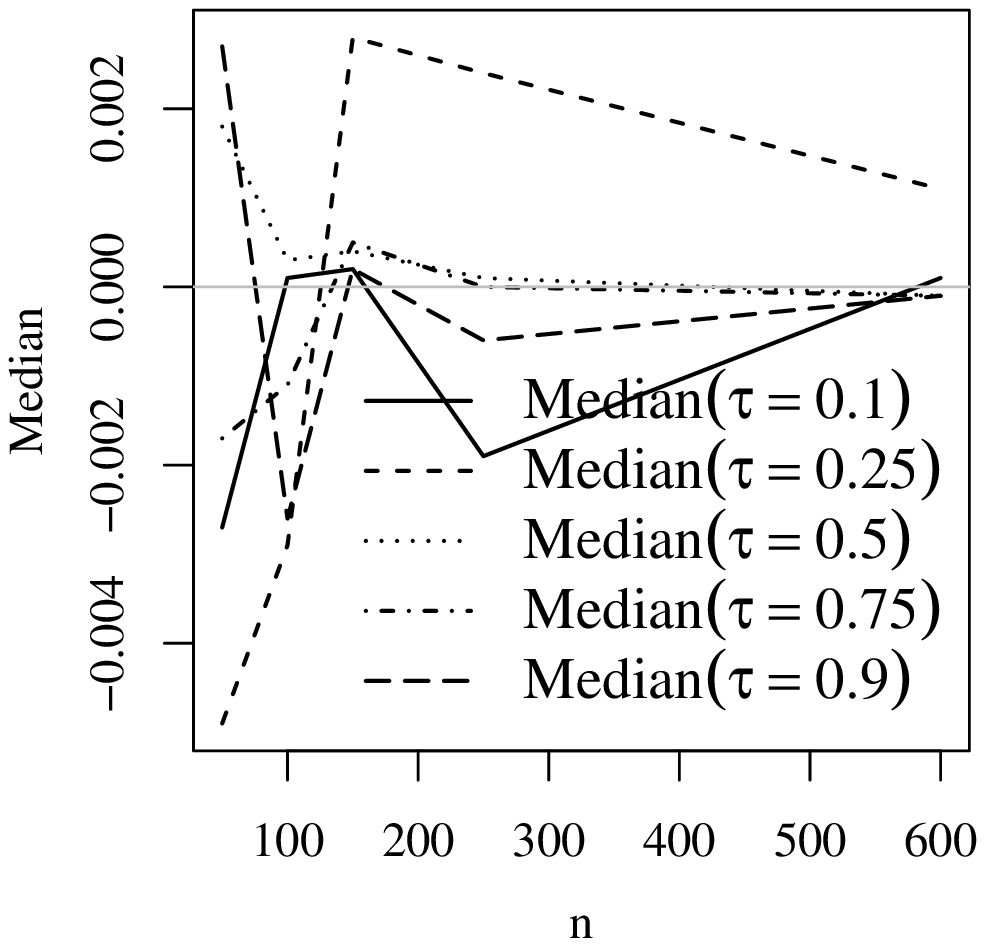}}\hspace{-0.25cm}
{\includegraphics[height=3.5cm,width=3.5cm]{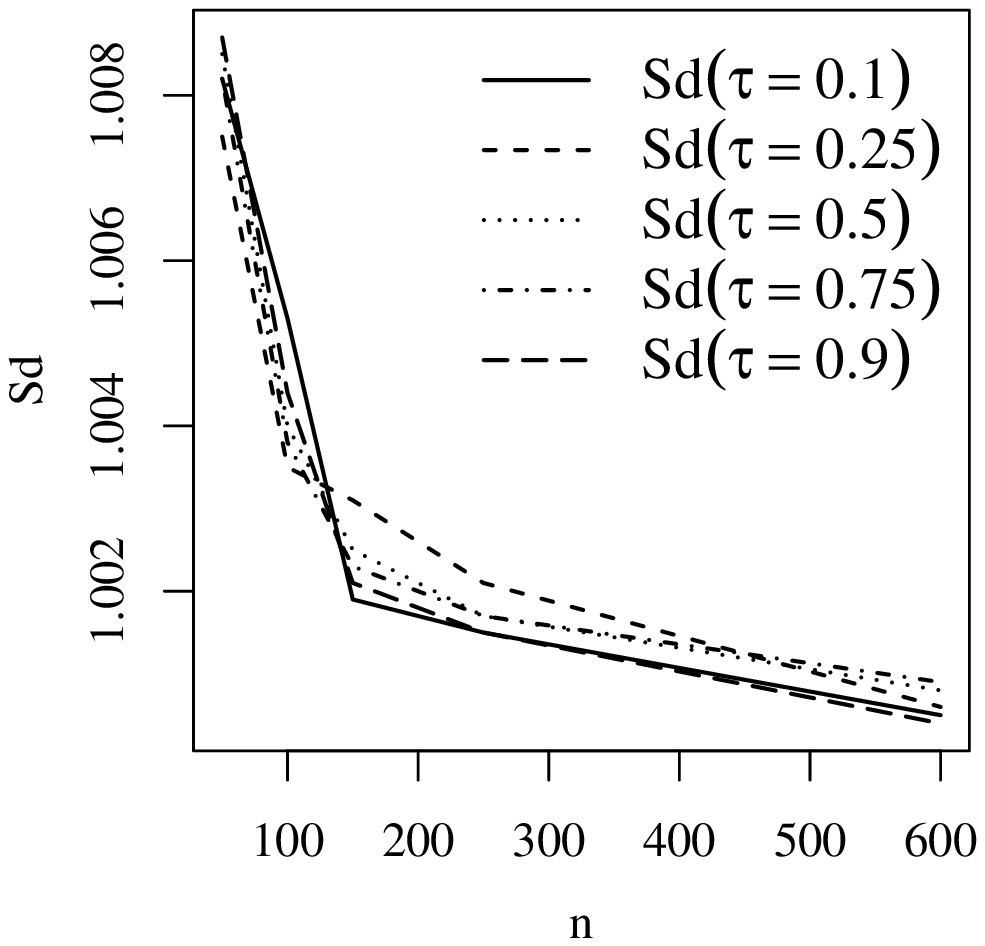}}\hspace{-0.25cm}
{\includegraphics[height=3.5cm,width=3.5cm]{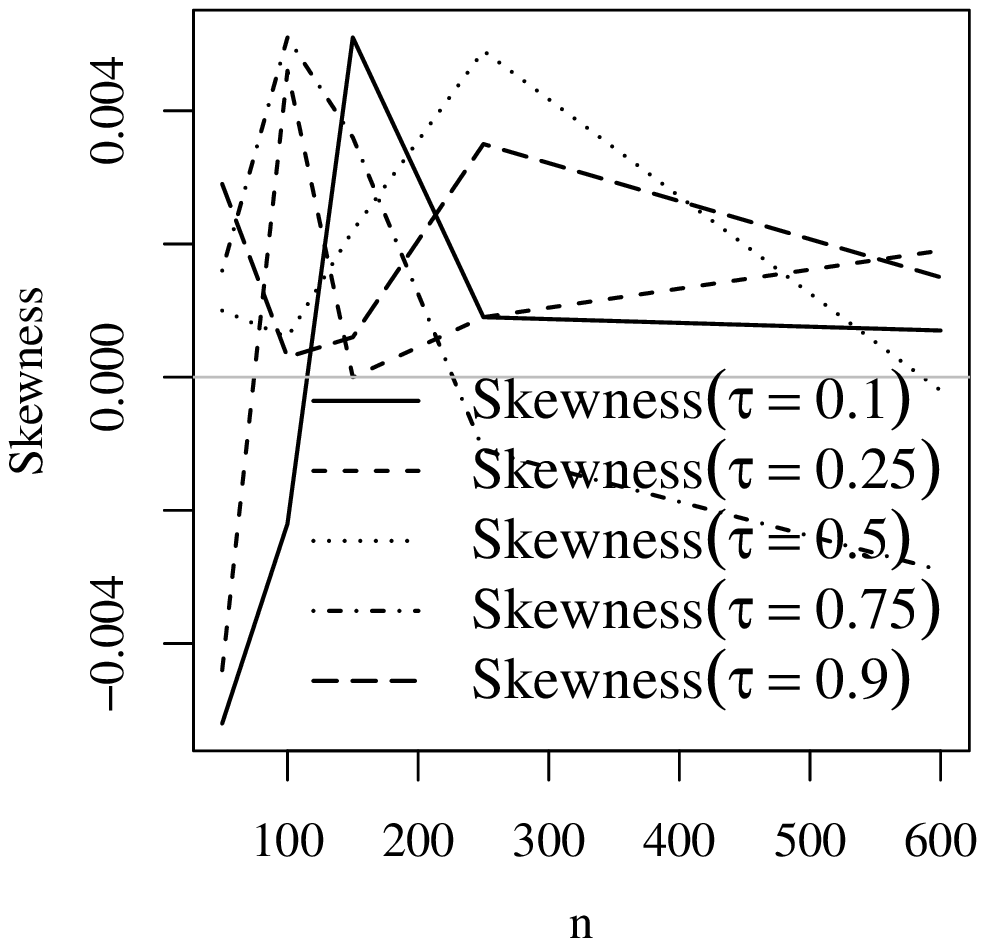}}\hspace{-0.25cm}
{\includegraphics[height=3.5cm,width=3.5cm]{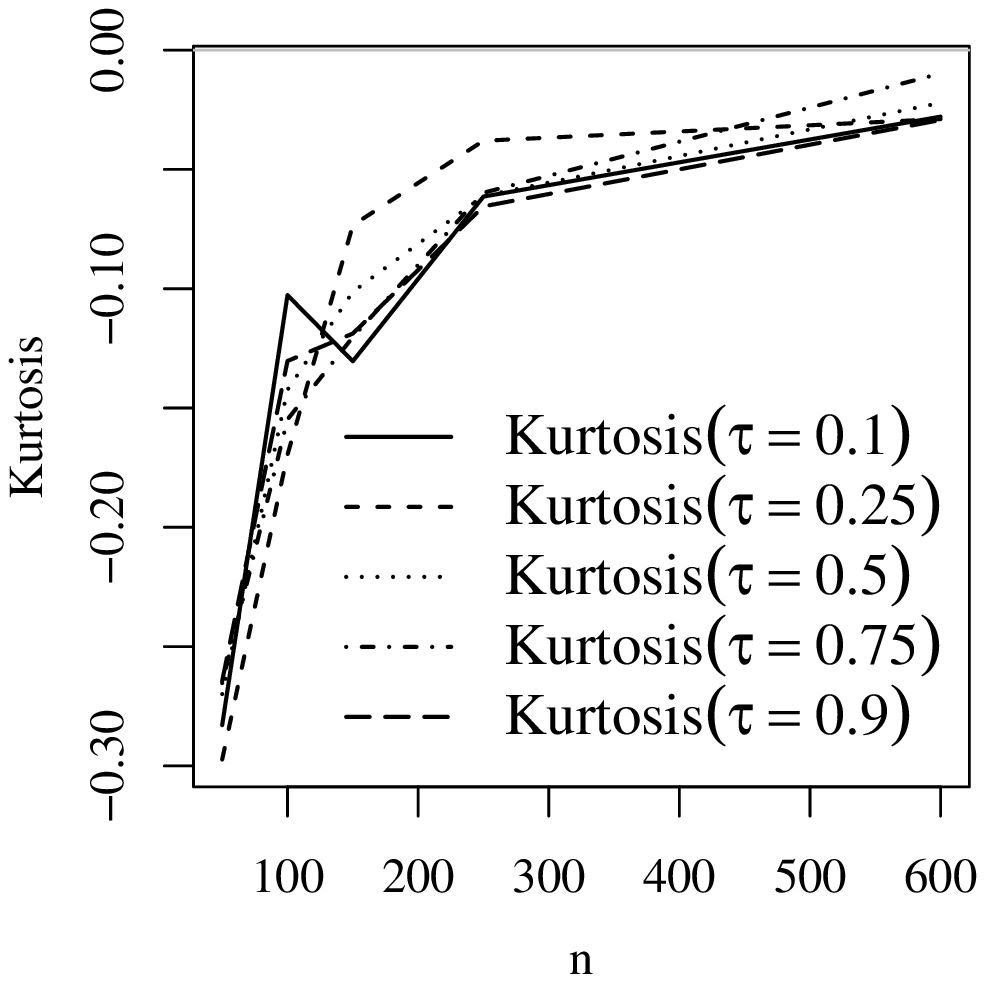}}
\vspace{-0.2cm}
\caption{Monte Carlo simulation results of the GCS (top) and RQ (bottom) residuals for the Singh-Maddala model with $a = 5$ and $p = 1$.}
\label{fig_singh_maddala_MC_GCS_RQ}
\end{figure}

\begin{figure}[!ht]
\vspace{-0.25cm}
\centering
{\includegraphics[height=3.5cm,width=3.5cm]{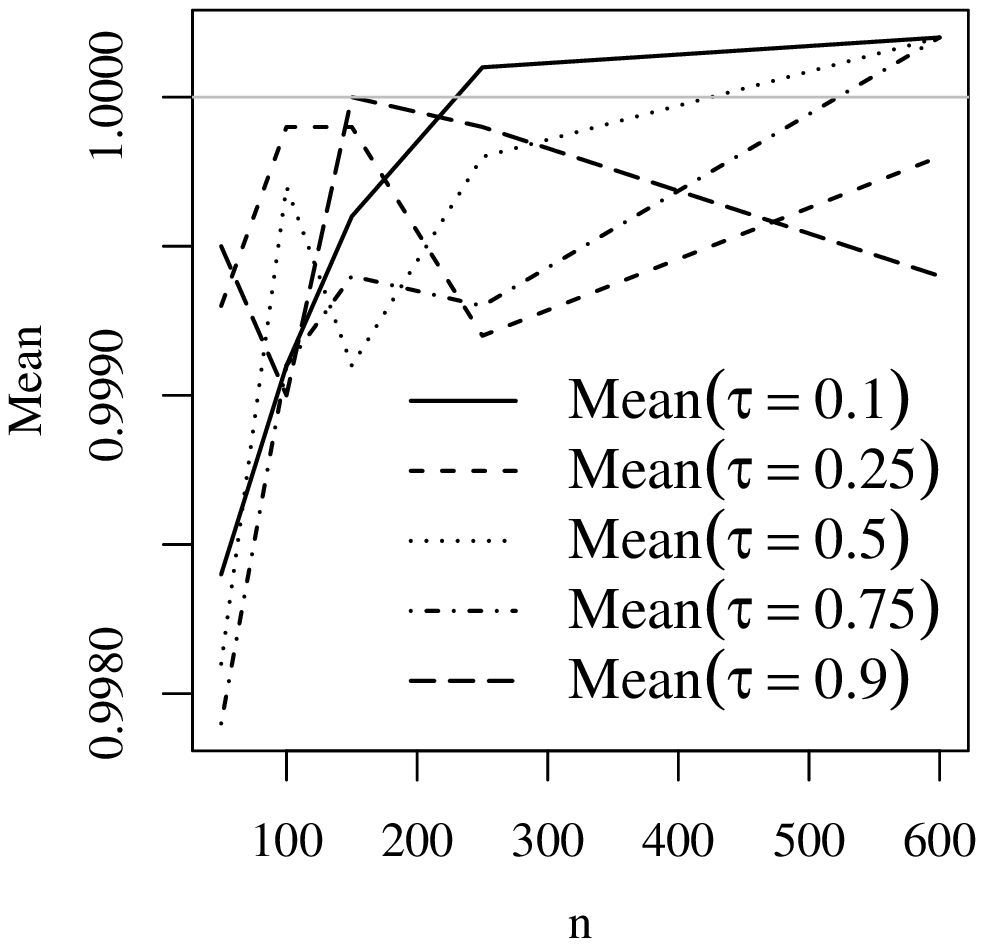}}\hspace{-0.25cm}
{\includegraphics[height=3.5cm,width=3.5cm]{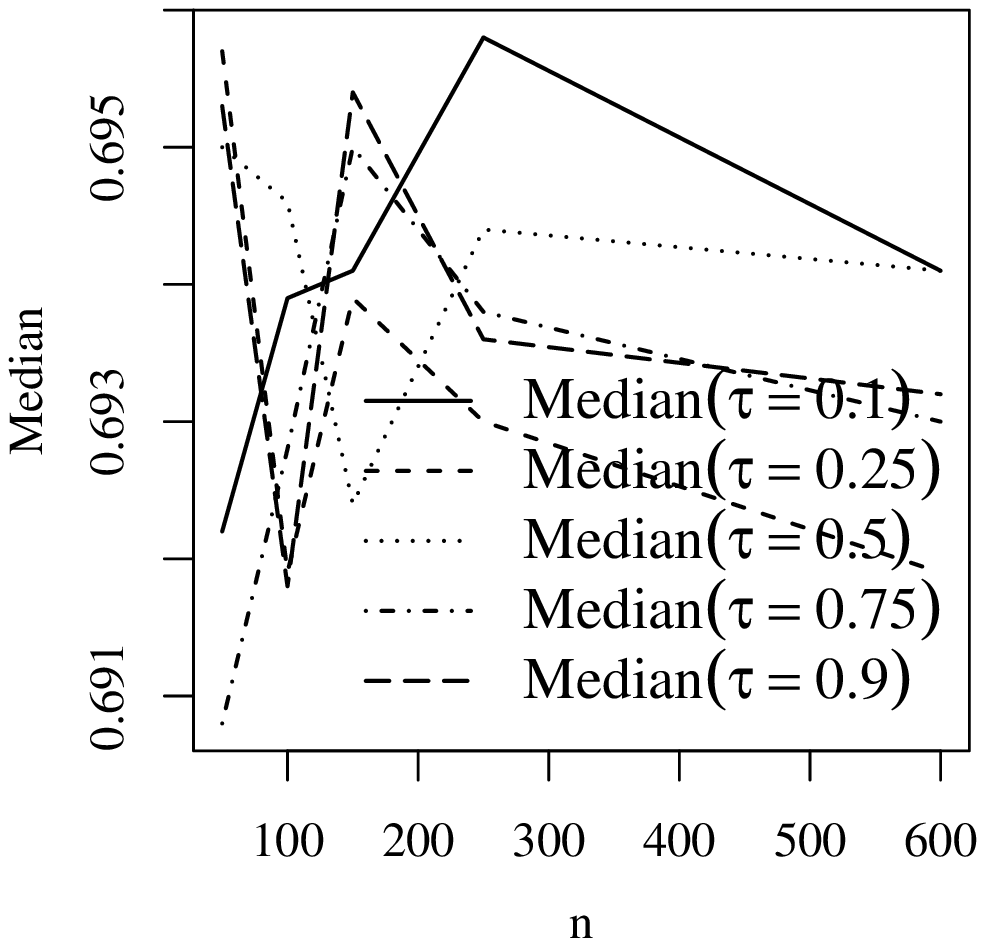}}\hspace{-0.25cm}
{\includegraphics[height=3.5cm,width=3.5cm]{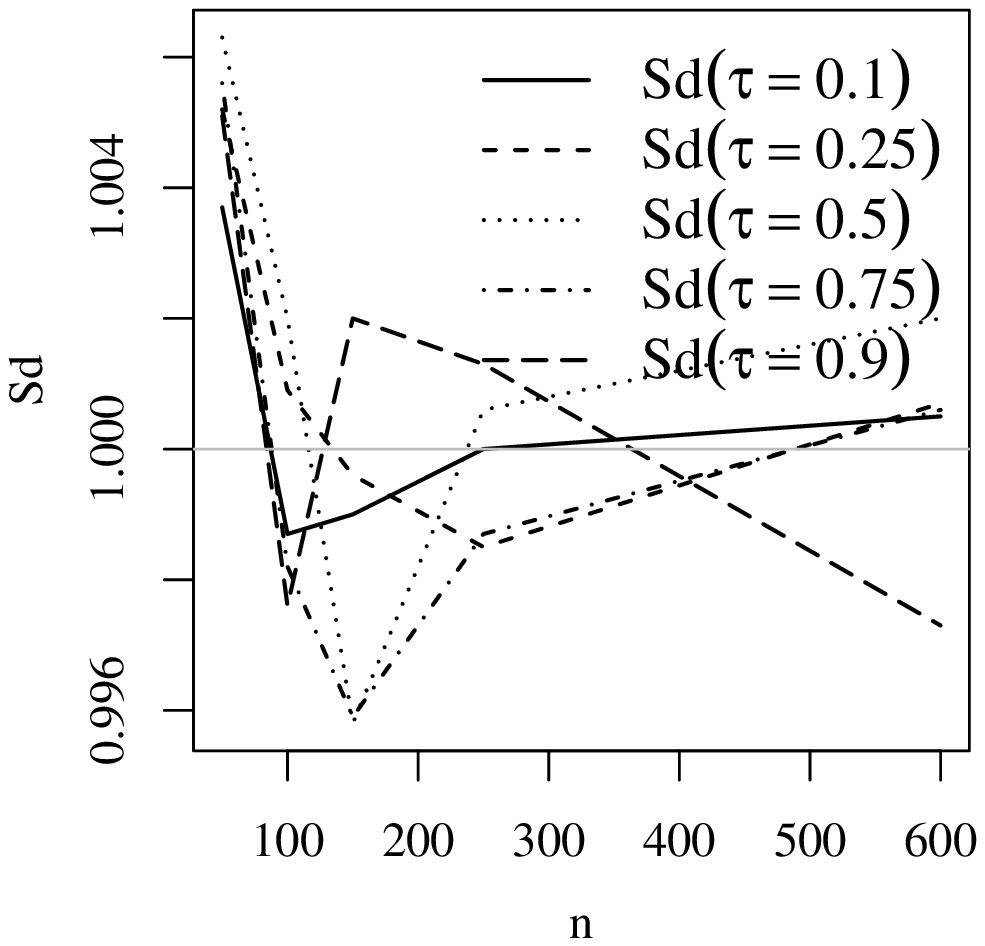}}\hspace{-0.25cm}
{\includegraphics[height=3.5cm,width=3.5cm]{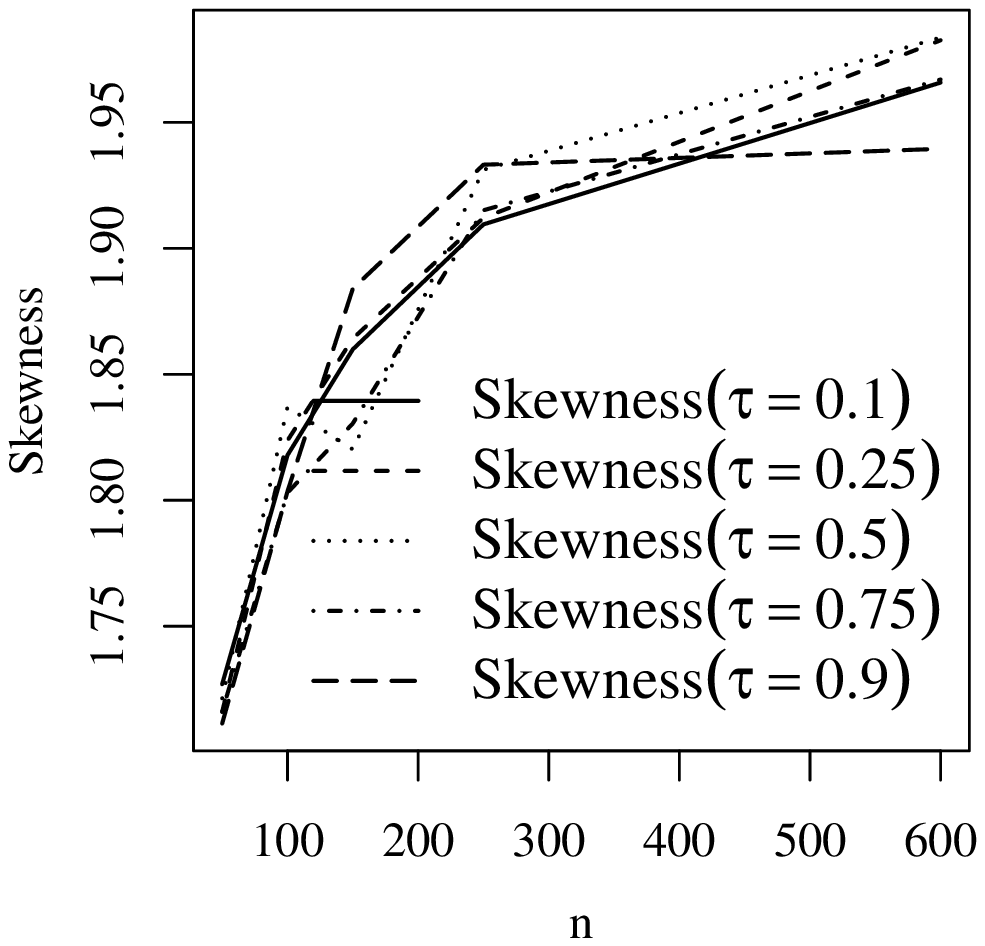}}\hspace{-0.25cm}
{\includegraphics[height=3.5cm,width=3.5cm]{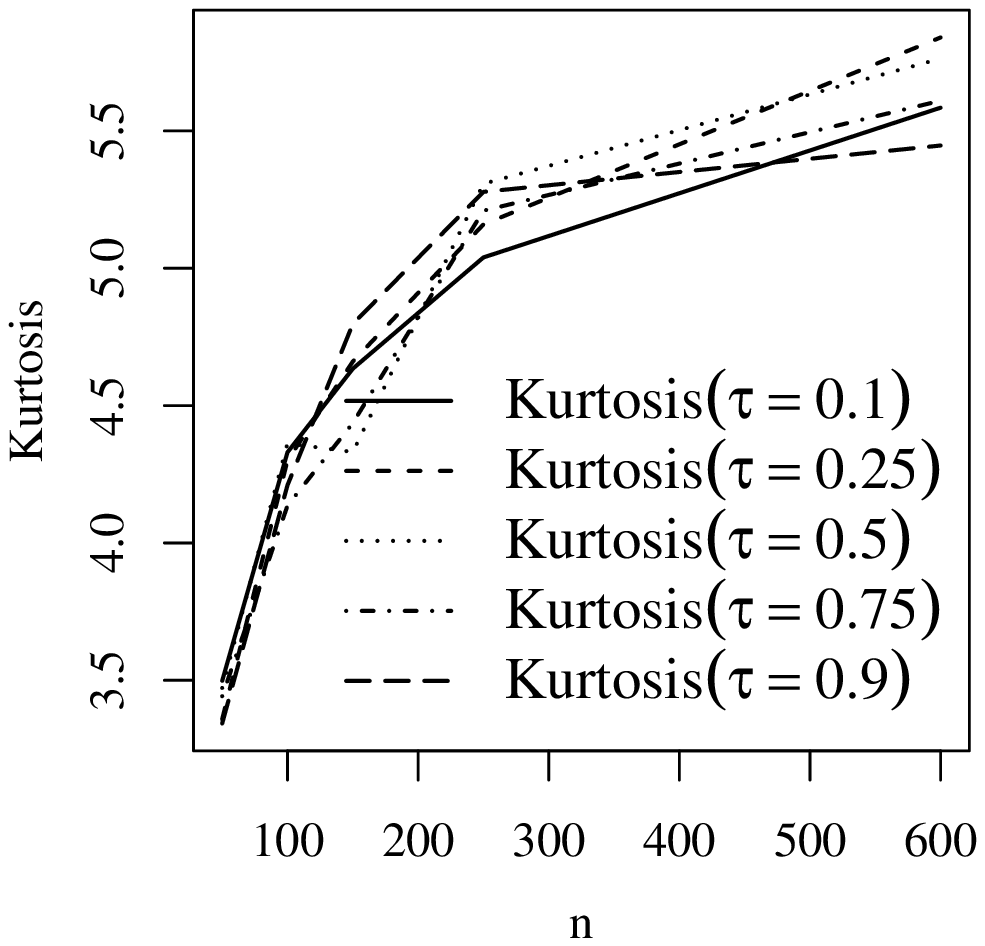}}
{\includegraphics[height=3.5cm,width=3.5cm]{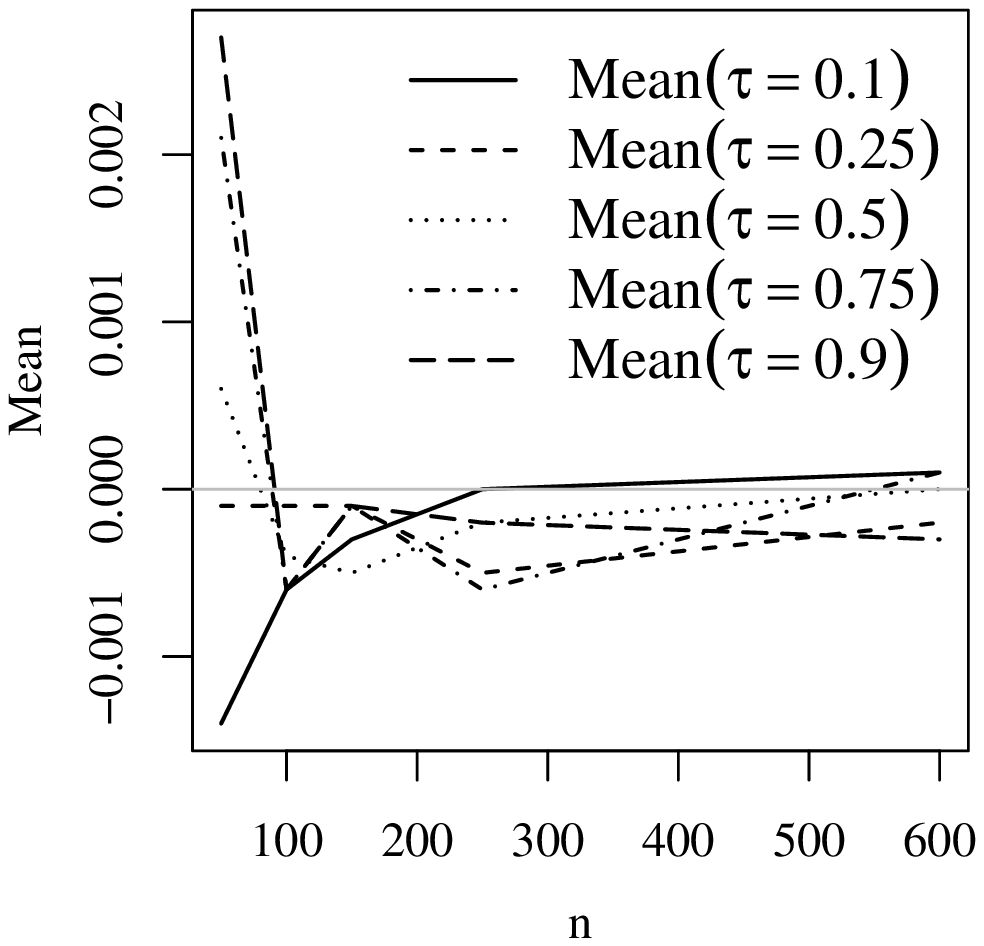}}\hspace{-0.25cm}
{\includegraphics[height=3.5cm,width=3.5cm]{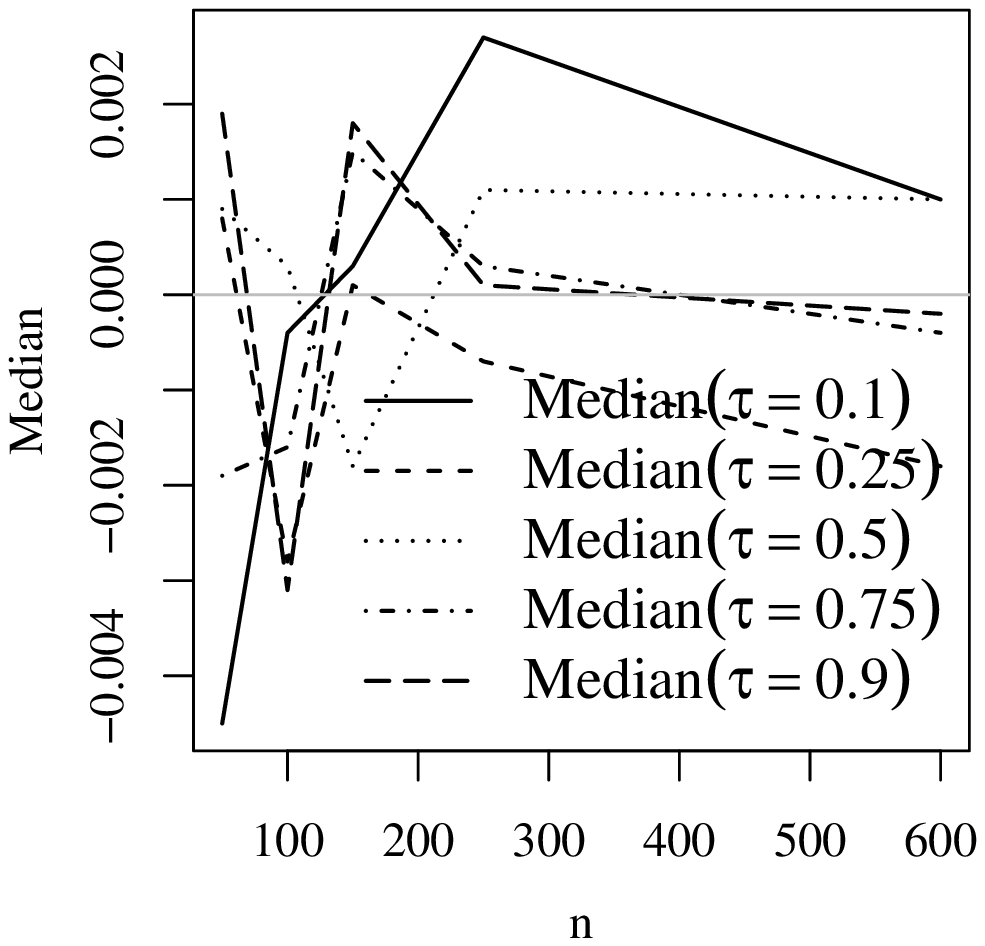}}\hspace{-0.25cm}
{\includegraphics[height=3.5cm,width=3.5cm]{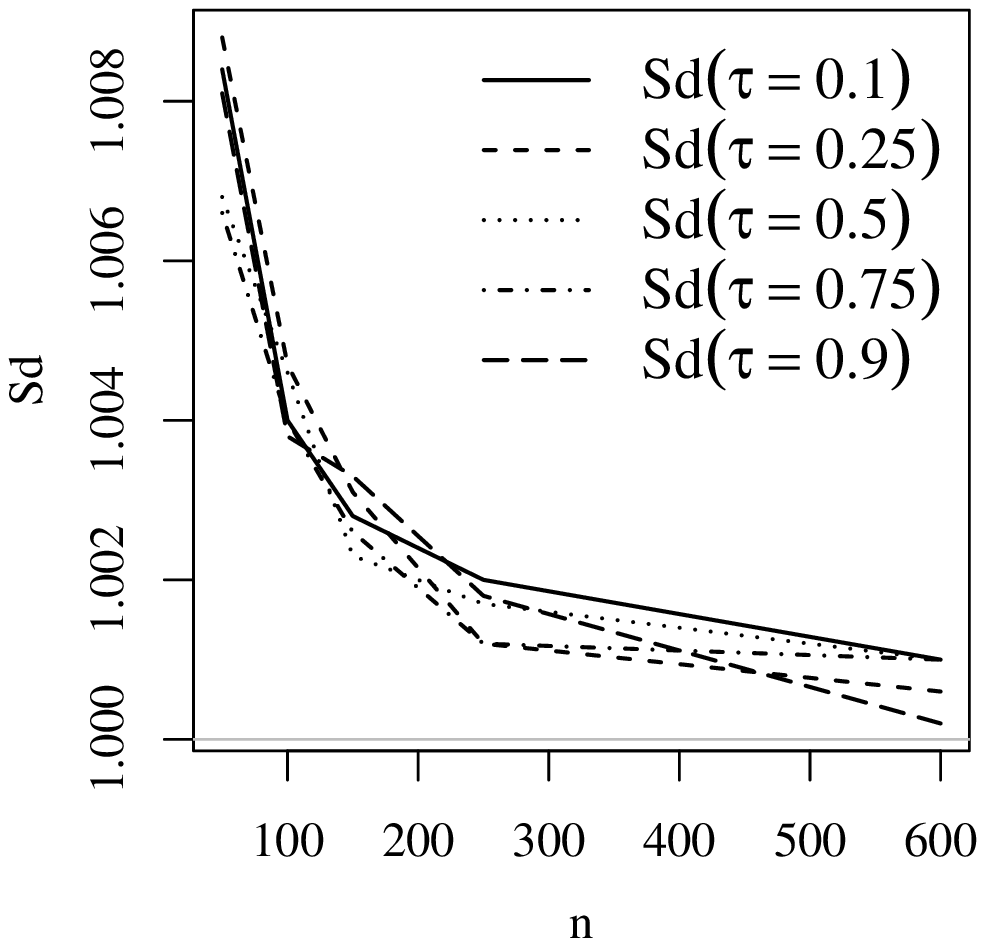}}\hspace{-0.25cm}
{\includegraphics[height=3.5cm,width=3.5cm]{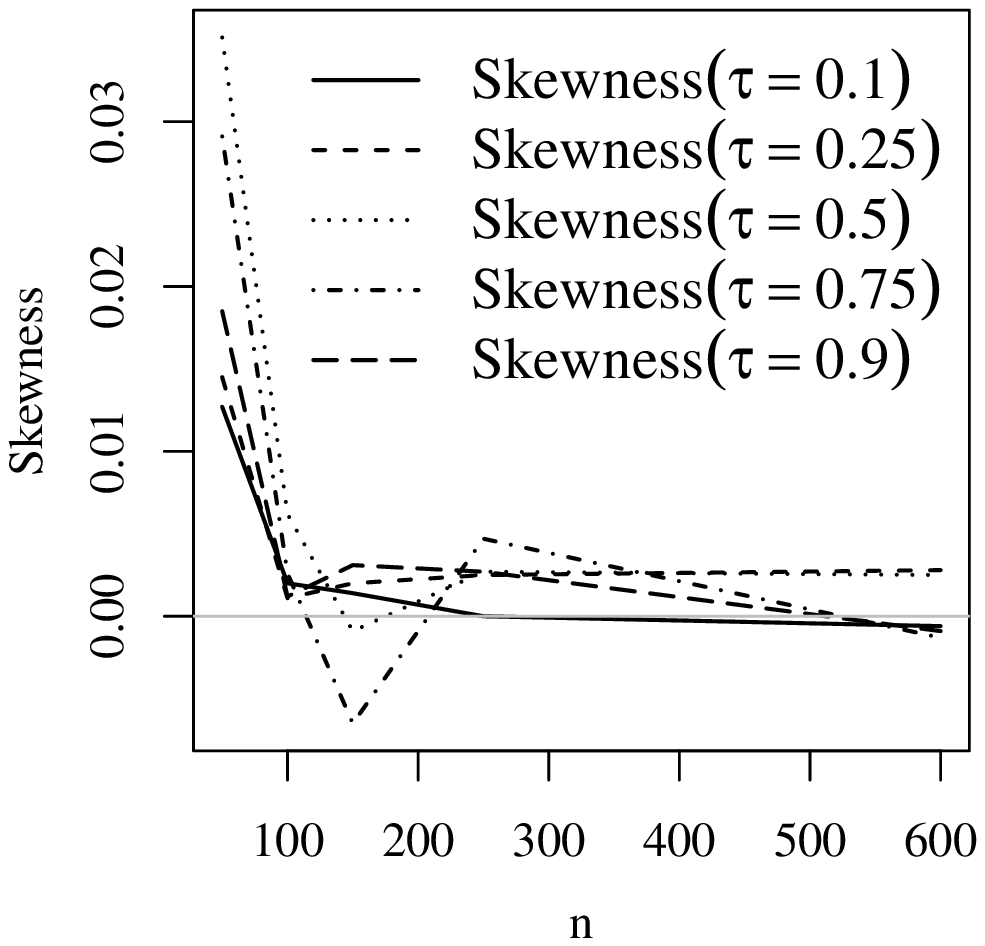}}\hspace{-0.25cm}
{\includegraphics[height=3.5cm,width=3.5cm]{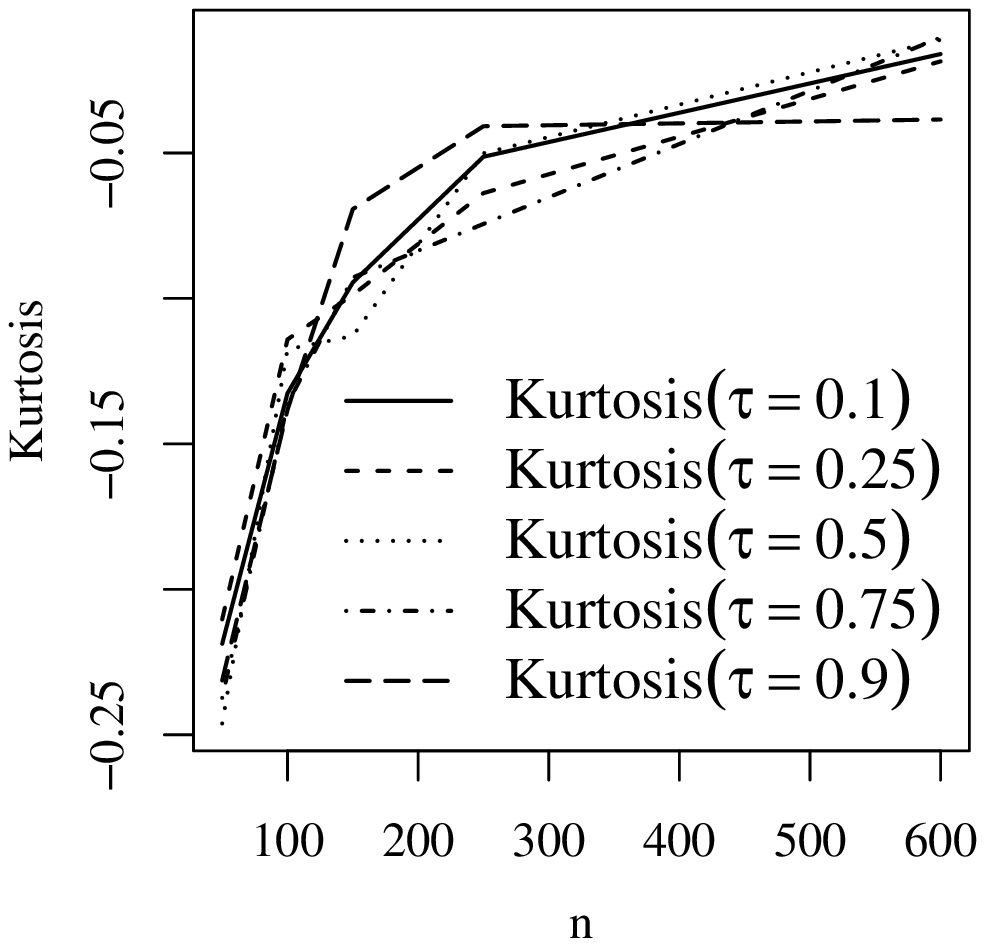}}
\vspace{-0.2cm}
\caption{Monte Carlo simulation results of the GCS (top) and RQ (bottom) residuals for the Dagum model with $a = 1$ and $p = 0.5$.}
\label{fig_dagum_MC_GCS_RQ}
\end{figure}

Here we show the performance of GCS and RQ residuals. We analyse the results with descriptive statistics (mean, median, standard deviation, coefficient of skewness and coefficient of kurtosis). The simulation scenario is
exactly the same as in Subsection \ref{ml_simulation_res}. Figures \ref{fig_singh_maddala_MC_GCS_RQ} and \ref{fig_dagum_MC_GCS_RQ} show the simulation results of the Singh-Maddala and Dagum models, respectively.

The reference values of mean, median, Sd, skewness and kurtosis are 1, 0.69, 1, 2 and 6, respectively, for GCS residual, and 0, 0, 1, 0 and 0, respectively, for RQ residual. From Figures \ref{fig_singh_maddala_MC_GCS_RQ} and \ref{fig_dagum_MC_GCS_RQ}, it is possible to verify that, as the sample size increases, the values tend to the expected results for each $\tau$. Therefore, we can use the both residuals to verify the fit of the proposed models.

\clearpage

\section{Application to income data}\label{sec:5}
\noindent

In this section, we use the 2016 Chilean household income data set, provided by the National Institute of Statistics in Chile\footnote{Available at \url{https://www.ine.cl/estadisticas/sociales/ingresos-y-gastos/encuesta-suplementaria-de-ingresos}.} to illustrate the proposed parametric quantile regression models. This data set was also used by \cite{sanchez2021b}, who introduced the Birnbaum-Saunders quantile regression model. While the Birnbaum-Saunders is not a distribution commonly used for income data, Singh-Maddala and Dagum are, so we assess if these models can produce better fits than the BS model.

The household income is the response variable ($Y$), whereas the covariates are the total income due to salaries ($X_1$), the total income due to independent work ($X_2$) and the total income due to retirements ($X_3$). The original dataset contains 107 variables, including the aforementioned, but these were selected based on economic and statistical criteria in relation to the response variable and descriptive analysis conducted by \cite{sanchez2021b}. Moreover, all incomes are expressed in thousands of Chilean pesos\footnote{\ See \url{http://www.bancocentral.cl} for their equivalence in American dollars.}.

\begin{figure}[!ht]
\vspace{-0.25cm}
\centering
\subfigure{\includegraphics[height=6.8cm,width=6.8cm]{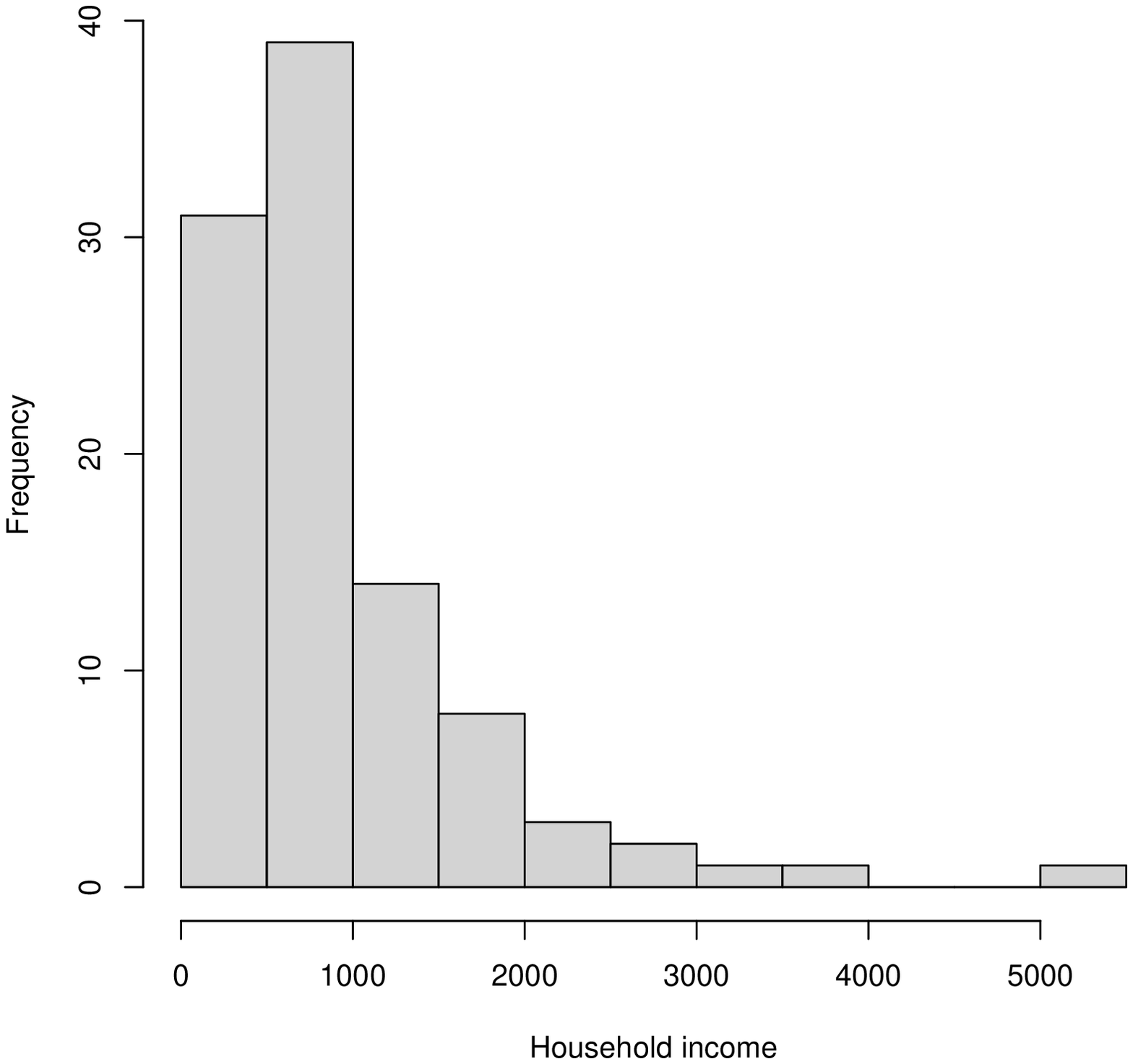}}
\subfigure{\includegraphics[height=6.8cm,width=6.8cm]{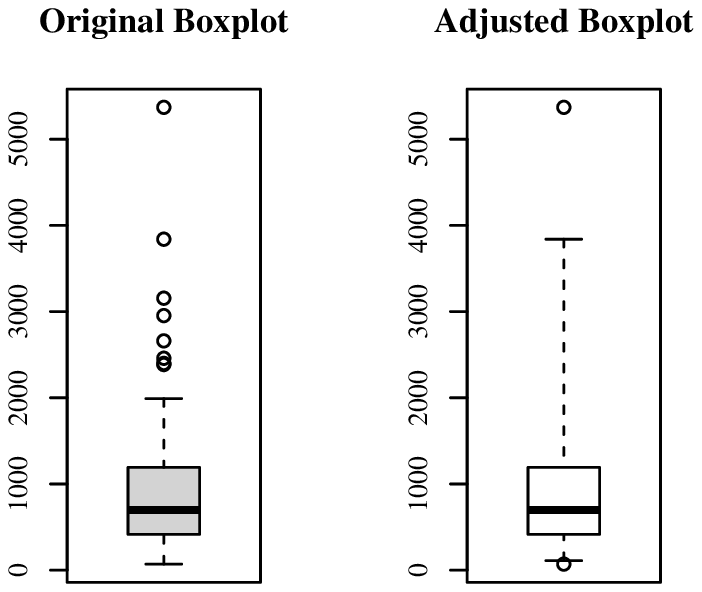}}\\
\vspace{-0.2cm}
\caption{Histogram and boxplots for the household income data (in thousands of Chilean pesos).}
\label{fig:stats}
\end{figure}

\begin{figure}[!ht]
\vspace{-0.25cm}
\centering
\subfigure{\includegraphics[height=12cm,width=12cm]{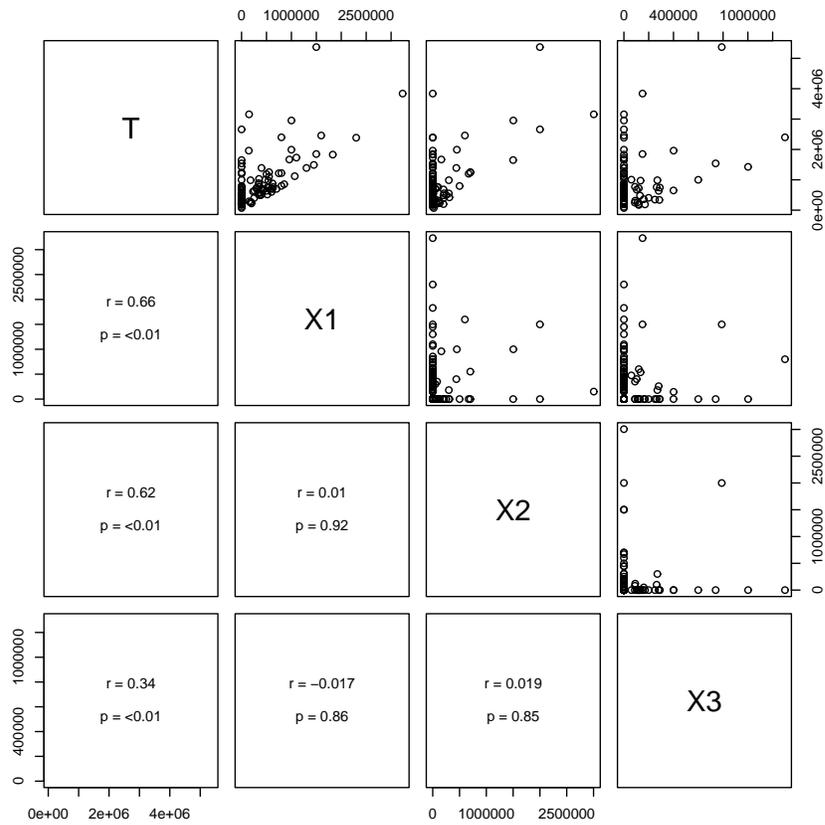}}\\
\vspace{-0.2cm}
\caption{Scatterplots and correlations between variables $Y, X_1, X_2$ and $X_3$.}
\label{fig:corr}
\end{figure}

We report in Table \ref{tab:stats} descriptive statistics for the household income ($Y$). Figure \ref{fig:stats} shows the histogram along with usual and adjusted box plots \citep{rousseeuw2016}. We observe that the household income data have a unimodal and right-skewed behavior, which i the precise needed scenario to uphold the usage of asymmetric distribution. Figure \ref{fig:corr} shows scatterplots (with correlation) between the household income ($Y$) and the covariates ($X_1$, $X_2$ and $X_3$). We observe that correlations are reasonable and significant, meanwhile the covariates have almost no linear correlation between each other.

\begin{table}[!ht]
\centering
\footnotesize
\caption{Descriptive statistics for the household income data (in thousands of Chilean pesos).}
\label{tab:stats}
\begin{tabular}{ccccccccc}
\hline
Mean & Median & Sd & Coef. Variation & Coef. Skewness & Coef. Kurtosis & minimum  & maximum & $n$ \\ \hline
\multicolumn{1}{r}{698.80} & \multicolumn{1}{r}{938.10} & \multicolumn{1}{r}{837.52} & \multicolumn{1}{r}{0.89} & \multicolumn{1}{r}{2.45} & \multicolumn{1}{r}{11.03} & \multicolumn{1}{r}{70} & \multicolumn{1}{r}{5369.90} & \multicolumn{1}{r}{100} \\ \hline
\end{tabular}
\end{table}

We then analyze the household income data using the Singh-Maddala and Dagum quantile regression models, with regression structure expressed as\footnote{We use this specification in order to compare the results of the proposed models with those of the Birnbaum-Saunders quantile regression model.}
\begin{equation*}
\gamma_i =  {\beta}_0(\tau) + {\beta}_1(\tau)x_{1i} + {\beta}_2(\tau)x_{2i} + {\beta}_3(\tau)x_{3i} ,
\end{equation*}
for $i,1,\ldots,100$.  The proposed models are fitted using the function \texttt{IncomeReg.fit}, implemented in the \texttt{R} software \citep{R} by the authors. The codes are available upon request.

\begin{figure}[!ht]
\centering
\subfigure[$\widehat{\beta}_0$]{\includegraphics[height=5.5cm,width=5.5cm]{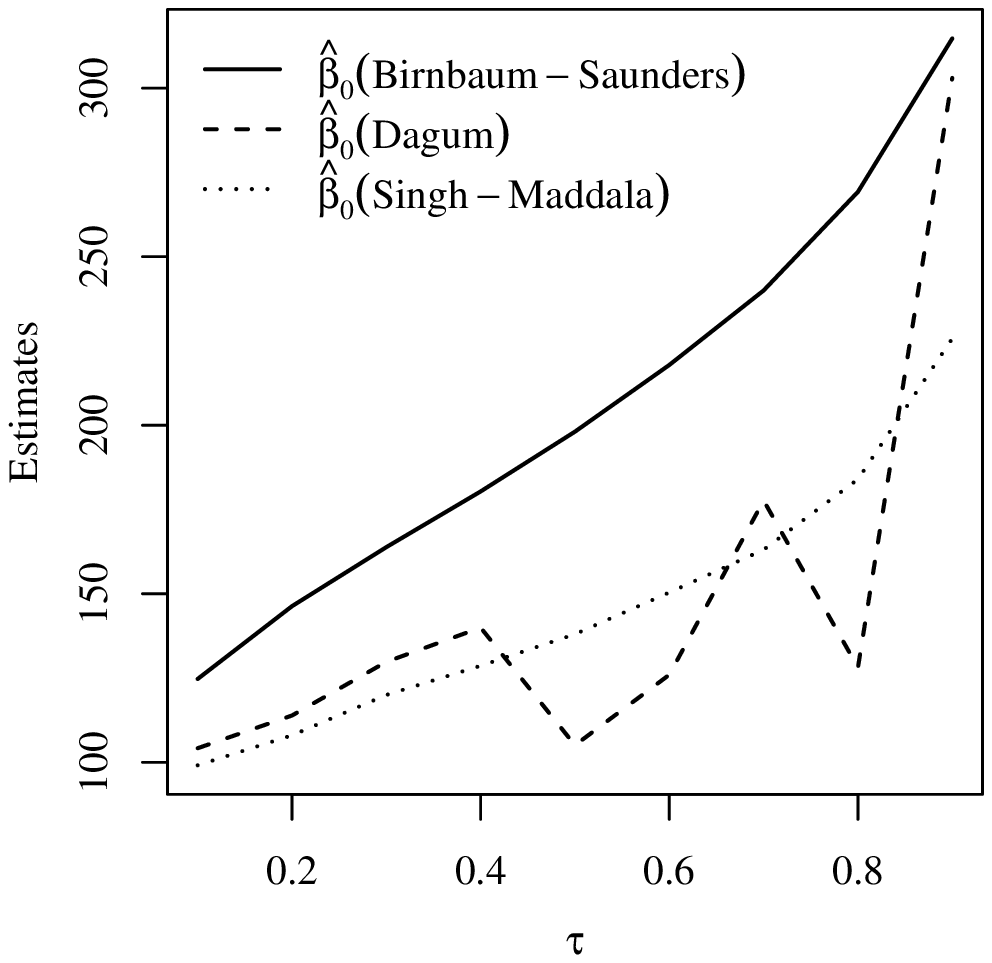}}
\subfigure[$\widehat{\beta}_1$]{\includegraphics[height=5.5cm,width=5.5cm]{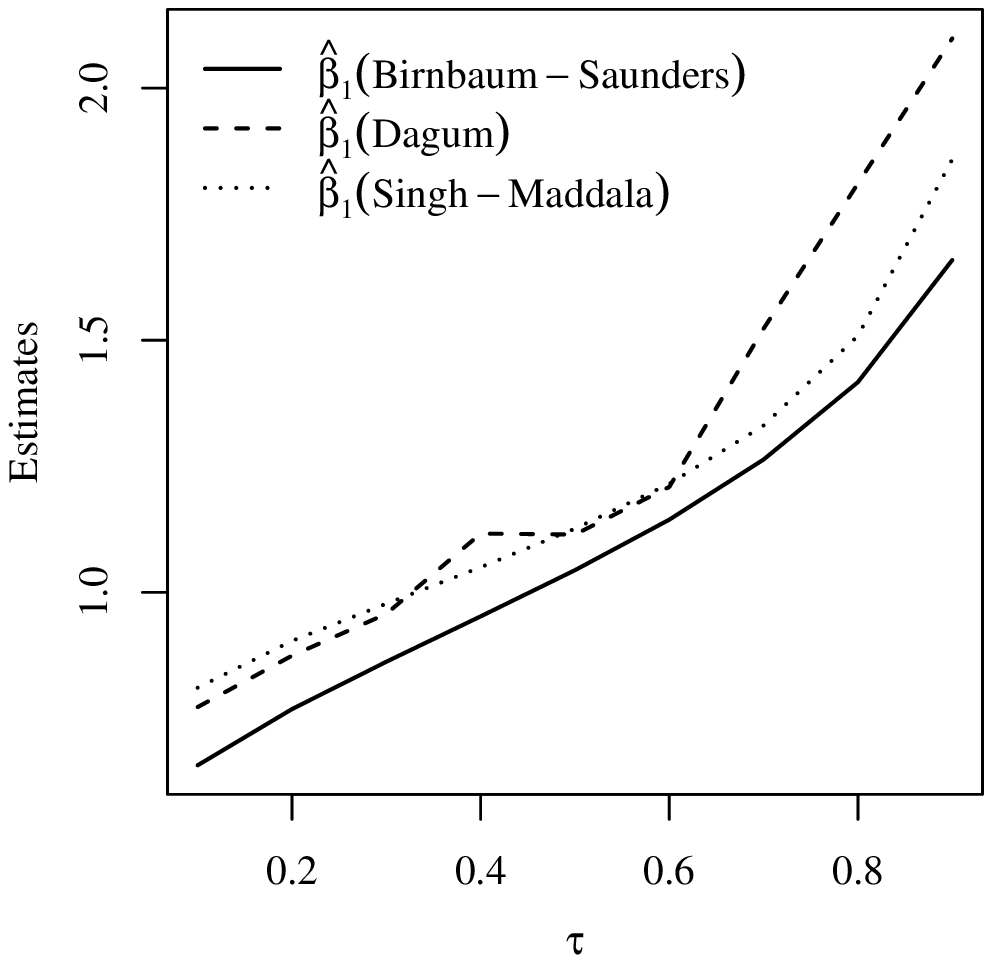}}\\
\subfigure[$\widehat{\beta}_2$]{\includegraphics[height=5.5cm,width=5.5cm]{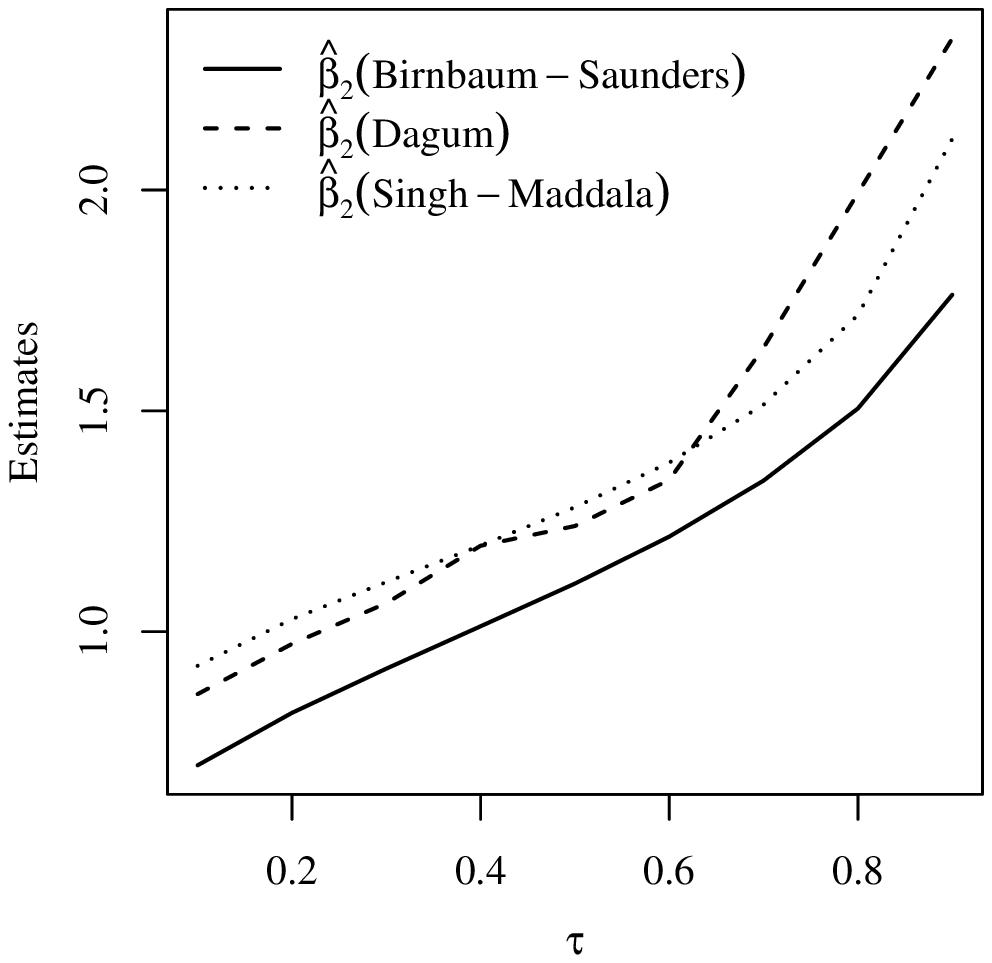}}
\subfigure[$\widehat{\beta}_3$]{\includegraphics[height=5.5cm,width=5.5cm]{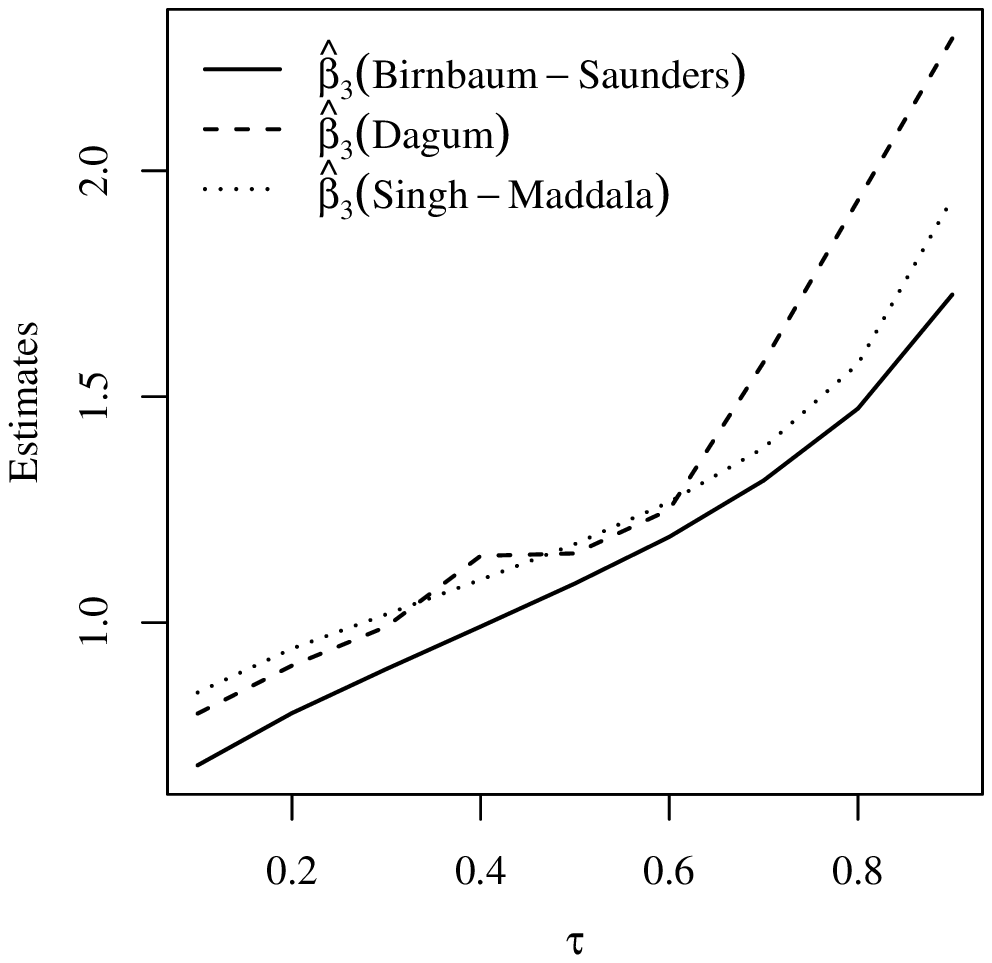}}
 \caption{\small {Estimated parameters of the indicated model across $\tau$ for the income data.}}
\label{fig:estimates_models}
\end{figure}

Table \ref{tab:resulall_ex2} presents the ML estimates, computed by the BFGS quasi-Newton method, standard errors (SEs) and Akaike (AIC) and Bayesian information (BIC) criteria values, for the Singh-Maddala and Dagum quantile regression models with $\tau=0.50$. As mentioned earlier, the results of the Birnbaum-Saunders quantile regression are presented as well. The results of Table \ref{tab:resulall_ex2} reveal that the proposed Singh-Maddala and Dagum models provide better adjustments than the Birnbaum-Saunders model based on the values of log-likelihood, AIC and BIC. Particularly, the Singh-Maddala model has the lowest AIC and BIC values. The estimated parameters of the Birnbaum-Saunders, Dagum and Singh-Maddala models across $\tau$ are shown in Figure \ref{fig:estimates_models}. From this figure, we observe that the estimates associated with all the covariates tend to increase as $\tau$ increases, as expected.

\begin{figure}[!ht]
\centering
\subfigure[Birnbaum-Saunders (GCS)]{\includegraphics[height=5.5cm,width=5.5cm]{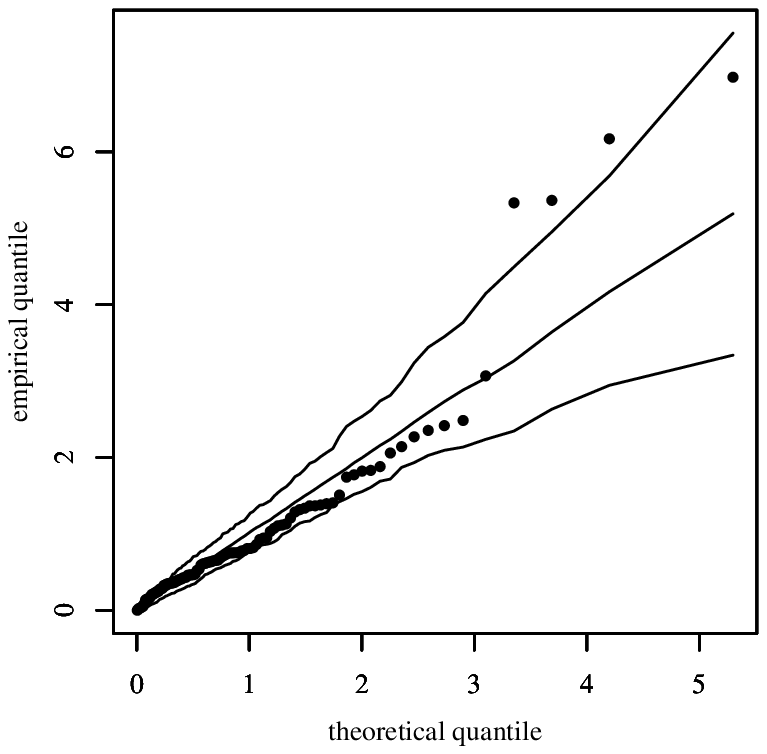}}
\subfigure[Birnbaum-Saunders (RQ)]{\includegraphics[height=5.5cm,width=5.5cm]{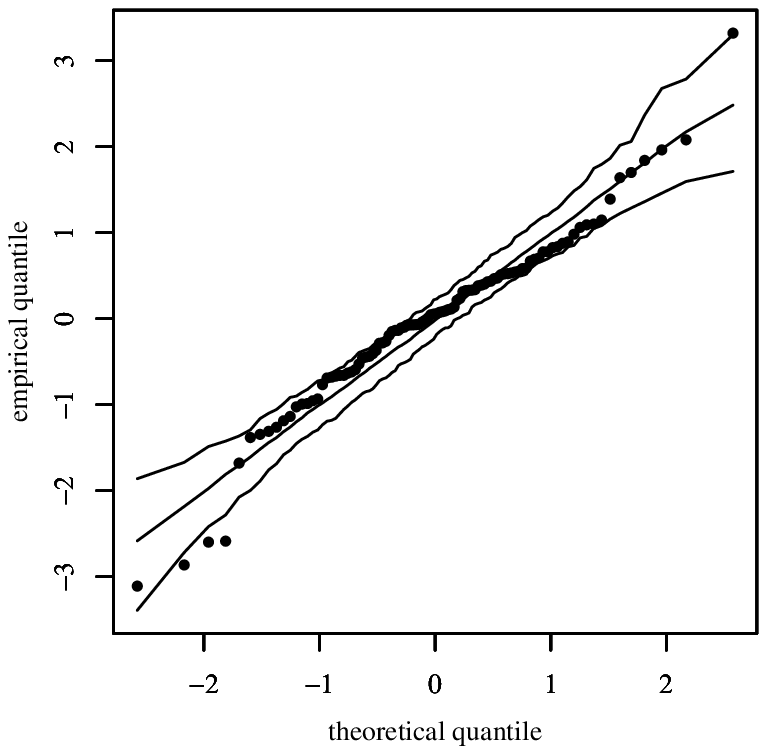}}\\
\subfigure[Dagum (GCS)]{\includegraphics[height=5.5cm,width=5.5cm]{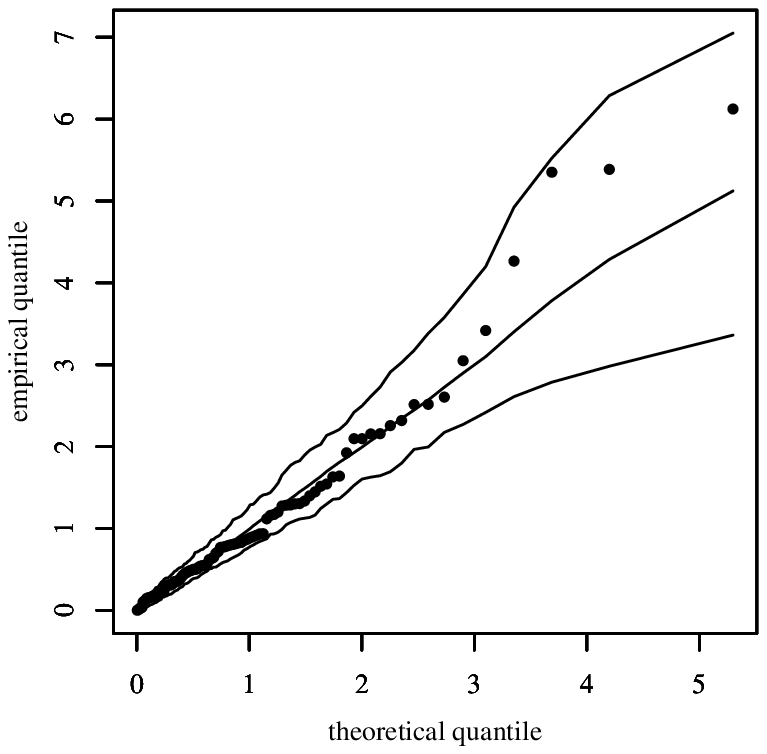}}
\subfigure[Dagum (RQ)]{\includegraphics[height=5.5cm,width=5.5cm]{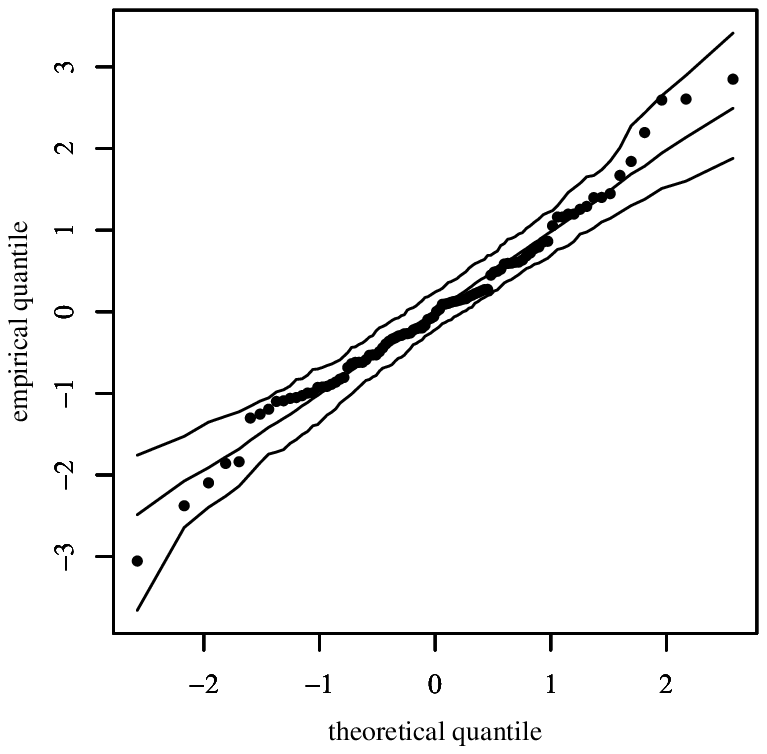}}\\
\subfigure[Singh-Maddala (GCS)]{\includegraphics[height=5.5cm,width=5.5cm]{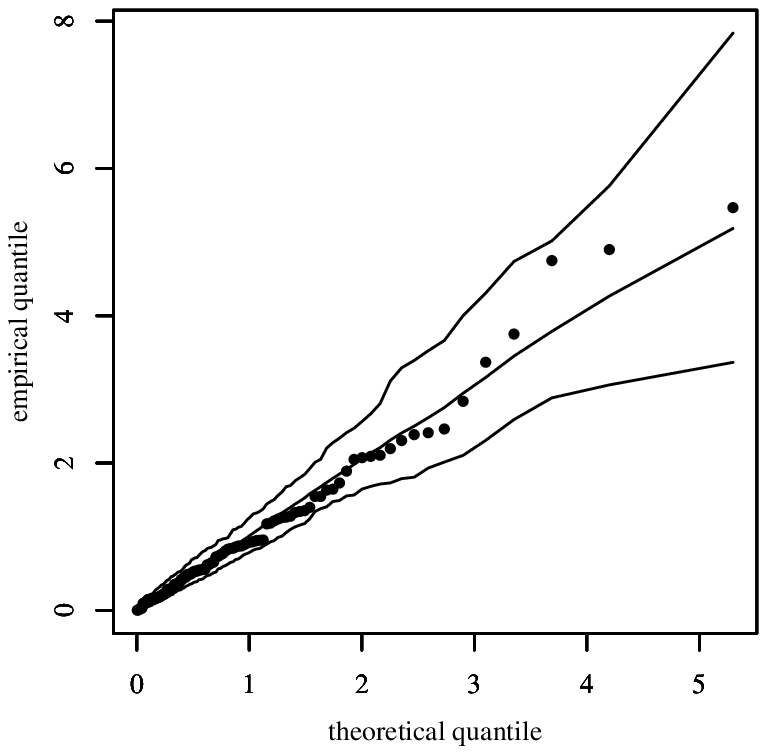}}
\subfigure[Singh-Maddala (RQ)]{\includegraphics[height=5.5cm,width=5.5cm]{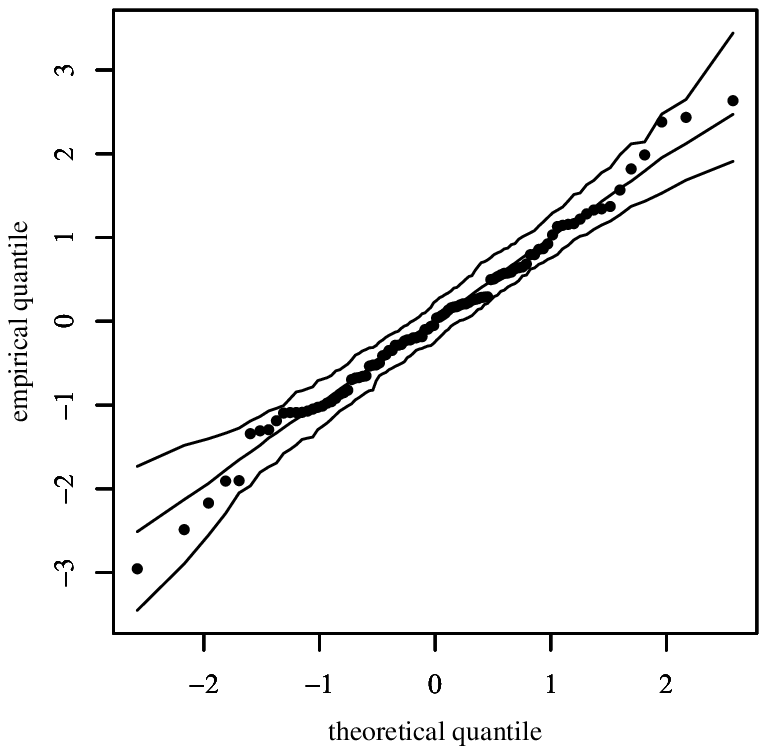}}
 \caption{\small QQ plot and its envelope for the GSC and RQ residuals in the indicated models for the income data ($\tau=0.50$).}
\label{fig:qqplots_ex2}
\end{figure}

The QQ plots with simulated envelope of the GCS and RQ residuals for the models considered in Table \ref{tab:resulall_ex2} confirm the results presented in Table \ref{tab:resulall_ex2}; see Figure \ref{fig:qqplots_ex2}. Similar results are obtained when considering $\tau = \{0.10,\ldots, 0.90\}$.

\begin{table}[!ht]
\centering
\begin{small}
 \caption{{ML estimates (with standard errors in parentheses) and model selection measures for the income data.}}
 \begin{tabular}{l rrrrrrrrrrrrrrrrrrrrr}
\hline
          & Birnbaum-Saunders ($\tau=0.50$)   &  Dagum ($\tau=0.50$)         & Singh-Maddala ($\tau=0.50$) \\
\hline
$\beta_0$  & 198.0903*          & 150.8307*       & 137.8478*  \\[-0.1cm]
           & (22.3166)          &  (3.0771)       & ( 3.2826)   \\
$\beta_1$  & 1.0440*            & 1.1173*         & 1.1252*  \\[-0.1cm]
           & (0.0870)           &  (0.0636)       & (0.0569)   \\
$\beta_2$  & 1.1090*            & 1.2424*         & 1.2805*  \\[-0.1cm]
           & (0.1502)           &  (0.1172)       & (0.1103)   \\
$\beta_3$  & 1.0865*            & 1.1562*         & 1.1730*  \\[-0.1cm]
           & (0.1759)           &  (0.1395)       & (0.1382)   \\
$\alpha$   & 0.3646*            &                 &          \\[-0.1cm]
           & (0.0087)           &                 &          \\
$a$        &                    & 4.3720          & 8.3380  \\[-0.1cm]
           &                    &  (0.5692)       & (1.4720)   \\
$p\,\text{or}\,q$ &             & 2.2100          & 0.4034 \\[-0.1cm]
           &                    &  (1.0290)       & (0.1144)   \\[0.25cm]
Log-lik.  & $-$692.8373         &  $-$686.9182      &  $-$685.2337\\
AIC       & 1395.675            &  1385.836       &  1382.467\\
BIC       & 1408.701            &  1386.740       & 1383.371 \\
\hline
\end{tabular}
\label{tab:resulall_ex2}
\end{small}
\begin{flushleft}
 {\footnotesize{* significant at 5\% level. ** significant at 10\% level.}}
\end{flushleft}
\end{table}

\begin{figure}[!ht]
\vspace{-0.25cm}
\centering
{\includegraphics[scale=0.7]{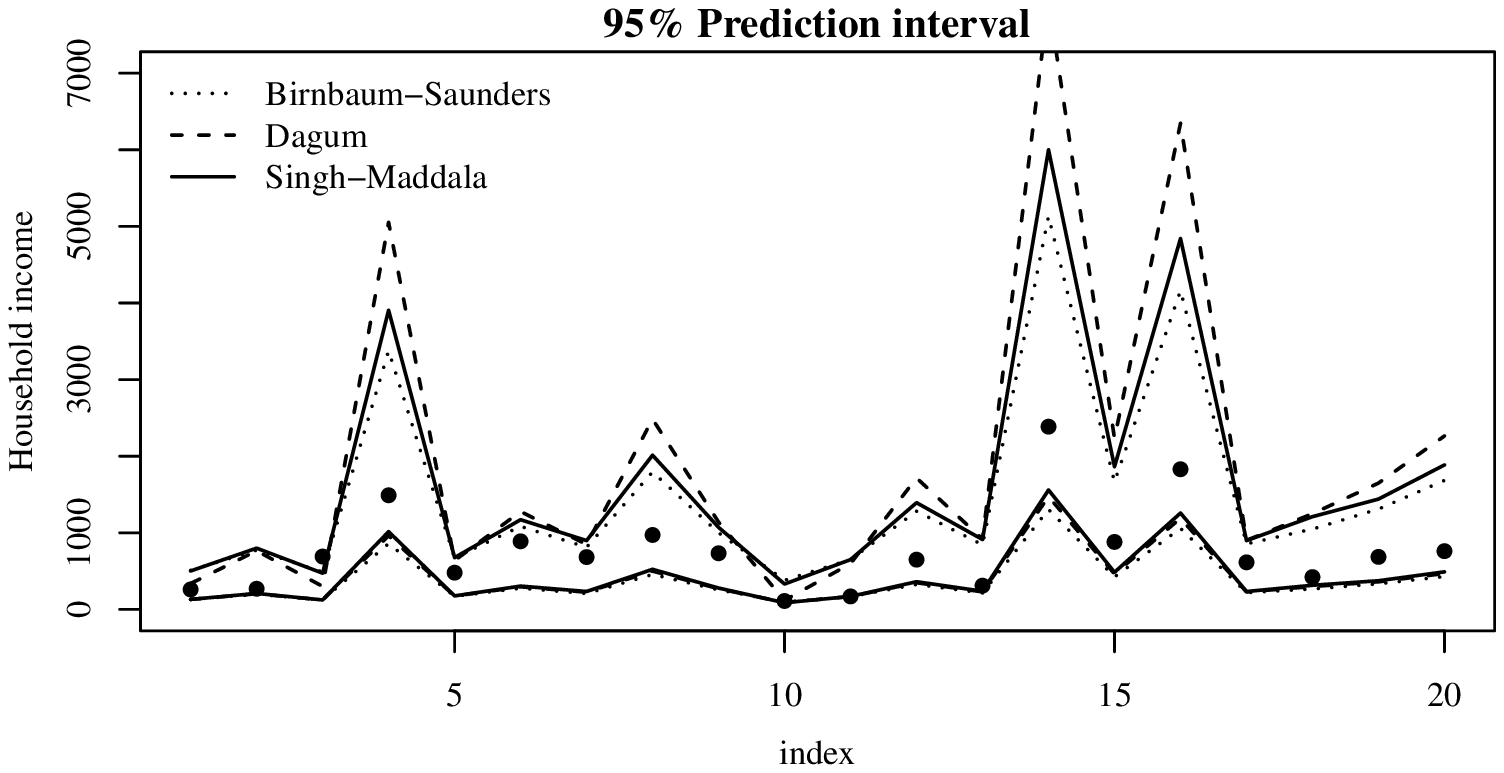}}
\vspace{-0.2cm}
\caption{{95\% prediction intervals ($20$-steps-ahead) from the indicated models for the household income data.}}\label{figfore:1}
\end{figure}

Figure \ref{figfore:1} shows 95\% prediction intervals from the Birnbaum-Saunders, Dagum and Singh-Maddala quantile regression models for the household income data. The predictions were performed $20$-steps-ahead, namely, $20$-observations were not included in the estimation. From Figure \ref{figfore:1}, we observe that 95\%, 95\% and 95\% of the observations are within the limits of the prediction interval for the Birnbaum-Saunders, Dagum and Singh-Maddala models, respectively. Therefore, all the models provide values closer to the nominal 95\% level.

\section{Concluding remarks}\label{sec:6}
\noindent

In this paper, we have proposed parametric quantile regression models based on the Singh-Maddala and Dagum distributions. The proposed models are based on reparametrizations of the original distributions, by including the quantile as a parameter. The maximum likelihood method was used to estimate model parameters and Monte Carlo simulation studies were conducted in order to evaluate the performance of the estimators and the empirical distribution of the generalized Cox-Snell and random quantile residuals. Results showed that the estimates had good performance, and the residuals presented good agreement with their reference distributions. We applied the proposed models to a real data set, where we have modeled the household income as a function of the following covariates: total income due to salaries, total income due to independent work and total income due to retirements. The results were compared to those obtained by \cite{sanchez2021b}, who proposed the Birnbaum-Saunders quantile regression model. We showed that both Singh-Maddala and Dagum models have better fit to data than Birnbaum-Saunders model, with Singh-Maddala also showing a slight superior performance than Dagum. Therefore, results were favorable to the usage of Singh-Maddala and Dagum quantile regression models. As part of future research, influence diagnostic tools can be investigated and also multivariate models can be studied.

\normalsize


\appendix

\section{Proof}
\label{Apendix-A}

\thispagestyle{empty}

\begin{proof}[Proof of Property (QDA3)]
If $Y\sim {\rm QDA}(a,\gamma,p)$ then
\begin{align*}
\mathbb{E}(Y^r\mathds{1}_{\{Y>x\}})
=
\int_{x}^{\infty}
y^r
	\,
\frac{a\,p (y/\gamma)^{ap-1}}{
	\gamma
	e_p^{p}
	[1+e_p^{-1}{(y/\gamma)^a}]^{1+p}}
\, {\rm d}y.
\end{align*}
Taking the  change of variables $z=e_p^{-1} (y/\gamma)^a$ and ${\rm d} z=ae_p^{-1}(y/\gamma)^{a-1} {\rm d}y/\gamma$ we get
\begin{align*}
=
p\,\gamma^r \,e_p^{ar+p-1}
\int_{e_p^{-1} (x/\gamma)^a}^{\infty}
\frac{z^{ar+p-1} }{
	(1+z)^{1+p}}
\, {\rm d}z.
\end{align*}
By using the identity:
$\int_{u}^{\infty} x^{a-1} (1+bx)^{-\nu} {\rm d}x
=
u^{a-\nu} b^{-\nu}(\nu-a)^{-1}
\, _2F_1(\nu,\nu-a;\nu-a+1;-1/(bu))$, $\nu>a$; see Eq. (3.194.2) in \citet{gr:15};
the last integral is
\begin{align*}
=
{p\gamma^r
 (\gamma/x)^{a(1-ar)} \over   (1-ar)e_p^{-p}}
\, _2F_1\biggl(1+p,1-ar;2-ar;-{(\gamma/x)^a\over e_p^{-1}}\biggr), \quad ar<1.
\end{align*}
Then the proof follows.
\end{proof}

\end{document}